\documentclass[longauth,doublecolumn]{aa} 

\usepackage{graphicx}

\usepackage{txfonts}
\usepackage{savesym}
\savesymbol{rotate}
\usepackage{deluxetable} 
\restoresymbol{DTBL}{rotate}

\usepackage[english]{babel}
\usepackage{times,epsfig}
\usepackage{longtable}
\usepackage{url}

\usepackage{mathrsfs}
\usepackage[normalem]{ulem}

\usepackage{hyperref}
\hypersetup{
    colorlinks = true,
    linkcolor = blue,
    anchorcolor = blue,
    citecolor = blue,
    filecolor = blue,
    urlcolor = blue
    }
    
\usepackage{placeins}

\newcommand{\orcid}[1]{\href{https://orcid.org/#1}{\includegraphics[width=10pt]{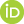}}}

\authorrunning{Stritzinger, Baron, Taddia et al.}
\titlerunning{SN~2016adj in Centaurus A.}
\begin{document} 

\title{The carbon-rich type Ic supernova 2016adj in the iconic dust lane of Centaurus A: signatures of interaction with circumstellar hydrogen?}

\author{
M. D. Stritzinger\inst{1}\orcid{0000-0002-5571-1833}
\and
 E. Baron\inst{2,3}\orcid{0000-0001-5393-1608}
 \and 
 F. Taddia\inst{1}\orcid{0000-0002-2387-6801}
 \and
 C. R. Burns\inst{4}\orcid{0000-0003-4625-6629}
 \and
 Morgan Fraser\inst{5}\orcid{0000-0003-2191-1674}
\and
  L. Galbany\inst{6,7}\orcid{0000-0002-1296-6887}
 \and 
 S. Holmbo\inst{1}\orcid{0000-0002-3415-322X}
 \and
 P. Hoeflich\inst{8}\orcid{0000-0002-4338-6586}
 \and 
  N. Morrell\inst{9}\orcid{0000-0003-2535-3091}
  \and
  E. Y. Hsiao\inst{8}\orcid{0000-0003-1039-2928}
 \and
 Joel P. Johansson\inst{10}\orcid{0000-0001-5975-290X}
 \and
 E. Karamehmetoglu\inst{1}\orcid{0000-0001-6209-838X}
 \and 
 Hanindyo Kuncarayakti\inst{11}\orcid{0000-0002-1132-1366}
 \and 
 Joe Lyman\inst{12}\orcid{0000-0002-3464-0642}
 \and 
 Takashi J. Moriya\inst{13,14}\orcid{0000-0003-1169-1954}
 \and 
 Kim Phan\inst{1,6}\orcid{0000-0001-6383-860X}
 \and 
 M. M. Phillips\inst{9}\orcid{0000-0003-2734-0796}
 \and 
 Joseph P. Anderson\inst{15}\orcid{0000-0003-0227-3451}
 \and 
  C. Ashall\inst{16}\orcid{0000-0002-5221-7557}
  \and 
  Peter J. Brown\inst{17}\orcid{0000-0001-6272-5507}
  \and 
  Sergio Castell\'{o}n\inst{9}
  \and 
  Massimo Della Valle\inst{18,19}\orcid{0000-0003-3142-5020}
  \and 
  Santiago Gonz\'{a}lez-Gait\'{a}n\inst{20}\orcid{0000-0001-9541-0317}
  \and 
  Mariusz Gromadzki\inst{21}\orcid{0000-0002-1650-1518}
  \and 
  Rasmus Handberg\inst{1}\orcid{0000-0001-8725-4502}
  \and 
  Jing Lu\inst{22}\orcid{0000-0002-3900-1452}
  \and 
 Matt Nicholl\inst{23}\orcid{0000-0002-2555-3192}
 \and 
 Melissa Shahbandeh\inst{24,25}\orcid{0000-0002-9301-5302}
 }

\institute{
 Department of Physics and Astronomy, Aarhus University, Ny Munkegade 120, DK-8000 Aarhus C, Denmark (\email{max@phys.au.dk})
 \and
 Planetary Science Institute, 1700 E Fort Lowell Rd., Ste 106, Tucson, AZ 85719 USA
 \and
 Hamburger Sternwarte, Gojensbergweg 112, 21029 Hamburg, Germany
 \and
 Observatories of the Carnegie Institution for Science, 813 Santa Barbara St., Pasadena, CA 91101, USA
 \and
 School of Physics, O'Brien Centre for Science North, University College Dublin, Belfield, Dublin 4, Ireland
 \and 
 Institute of Space Sciences (ICE, CSIC), Campus UAB, Carrer de Can Magrans, s/n, E-08193 Barcelona, Spain
 \and
 Institut d’Estudis Espacials de Catalunya (IEEC), E-08034 Barcelona, Spain
 \and
 Department of Physics, Florida State University, 77 Chieftain Way, Tallahassee, FL, 32306, USA 
 \and
Carnegie Observatories, Las Campanas Observatory, Casilla 601, La Serena, Chile
 \and 
 The Oskar Klein Centre, Department of Physics, Stockholm University, AlbaNova, 10691 Stockholm, Sweden
 \and
 Tuorla Observatory, Department of Physics and Astronomy, FI-20014, University of Turku, Finland
 \and 
 Department of Physics, University of Warwick, Coventry CV4 7AL, United Kingdom
 \and 
 National Astronomical Observatory of Japan, National Institutes of Natural Sciences, 2-21-1 Osawa, Mitaka, Tokyo 181-8588, Japan
 \and 
 School of Physics and Astronomy, Faculty of Science, Monash University, Clayton, Victoria 3800, Australia
 \and 
 European Southern Observatory, Alonso de C\'ordova 3107, Casilla 19, Santiago, Chile
 \and 
Department of Physics, Virginia Tech, Blacksburg, VA 24061, USA
\and
George P. and Cynthia Woods Mitchell Institute for Fundamental Physics and Astronomy, Department of Physics and Astronomy, Texas A\&M University, College Station, TX 77843, USA
\and 
INAF - Osservatorio Astronomico di Capodimonte, Salita Moiariello 16, 80131 Napoli, Italy
\and 
Department of Physics, Ariel University, Ariel, Israel
\and 
CENTRA-Centro de Astrofísica e Gravitacao and Departamento de Fisica, Instituto Superior Tecnico, Universidade de Lisboa, Avenida Rovisco Pais, 1049-001 Lisboa, Portugal
\and 
Astronomical Observatory, University of Warsaw, Al. Ujazdowskie 4, 00-478 Warszawa, Poland
\and 
Department of Physics and Astronomy, Michigan State University, East Lansing, MI 48824, USA
\and 
Astrophysics Research Centre, School of Mathematics and Physics, Queens University Belfast, Belfast BT7 1NN, UK
\and 
Department of Physics and Astronomy, Johns Hopkins University, Baltimore, MD 21218, USA
\and 
Space Telescope Science Institute, 3700 San Martin Drive, Baltimore, MD 21218, USA
}

\date{Received XX September, 2023 \ Accepted  XX XXXXX, XXXX.}
 
 \abstract{We present a comprehensive data set of supernova (SN) 2016adj located within the central dust lane of Centaurus~A. 
 SN~2016adj is significantly reddened and  after correcting the peak apparent $B$-band  magnitude ($m_B = 17.48\pm0.05$) for Milky Way reddening and our inferred host-galaxy reddening parameters  (i.e., $R_{V}^{host} = 5.7\pm0.7$ and $A_{V}^{host} = 6.3\pm0.2$), we estimate it reached a peak absolute magnitude of $M_B \sim  -18$. Detailed inspection of the optical/NIR spectroscopic time-series reveals  a carbon-rich SN~Ic and not a SN~Ib/IIb as  previously suggested in the literature. The NIR spectra shows prevalent  carbon-monoxide formation occurring already by $+$41 days past $B$-band maximum, which is $\approx 11$ days earlier  than previously reported in the literature for this object. Interestingly around two months past maximum, the NIR spectrum of SN~2016adj begins to exhibit H features, with a $+$97~d medium resolution spectrum revealing both  Paschen and Bracket lines  with absorption minima of $\sim 2000$~km~s$^{-1}$, full-width-half-maximum emission velocities of $\sim  1000$~km~s$^{-1}$, and emission line ratios consistent with a dense emission region.  We speculate these attributes are due to circumstellar interaction (CSI) between the rapidly expanding SN ejecta and a H-rich shell of material formed during the pre-SN phase. A bolometric light curve is constructed and a semi-analytical model fit suggests the supernova synthesized 0.5~$M_{\sun}$ of $^{56}$Ni and ejected 4.2~$M_{\sun}$ of material, though these values should be approached with caution given the large uncertainties associated with the adopted reddening parameters, possible CSI contamination, and known light echo emission.  Finally, inspection of Hubble Space Telescope archival data yielded no progenitor detection.}
 \keywords{supernovae: general, individual: SN~2016adj. galaxies: individual:  NGC~5128 (Centaurus~A).}
 
\maketitle

\section{Introduction}
\label{sec:intro}

Contemporary transient surveys are discovering thousands of supernovae (SNe) per year. The multitude of transients now being discovered early and in an unbiased manner is enabling the community to statistically characterize the observational properties  of many types of transients.  However, single objects displaying peculiar characteristics or occurring in nearby galaxies continue to play an important role in understanding late stages of stellar evolution and SN explosions. 

Nearby events favour detection close to the epoch of explosion, longer follow-up campaigns, the study of their environments, and in some instances, the direct detection of the progenitor stars in pre-explosion archival images. When located in close proximity even highly-reddened SNe can be detected and observed, further enabling the study of  circumstellar and interstellar dust properties of the  host galaxy. 

In this paper we present a comprehensive data set gathered from an assortment of ground- and space-based facilities of the stripped-envelope (SE) SN~2016adj, located in NGC~5128. As indicated by Fig.~\ref{fig:FC}, SN~2016adj is positioned in close proximity to a bright foreground star,  slightly  North-West of the center of the galaxy, and well within the iconic central dust lane of the parent galaxy, which hosts the active galactic nucleus known as Centaurus~A. 

Given the proximity to Earth, SN~2016adj offers an excellent opportunity to study a SE~SN in a significantly dusty environment and the conditions associated with the formation of carbon-monoxide (CO) molecules.  Before delving into our analysis, we visit SN~2016adj's  story of  discovery and the  community's earlier efforts in determining spectral classification.

\subsection{Discovery and a flurry of spectral classification}

\begin{figure}[th!]
 \centering
 \resizebox{\hsize}{!}
    {\includegraphics[]{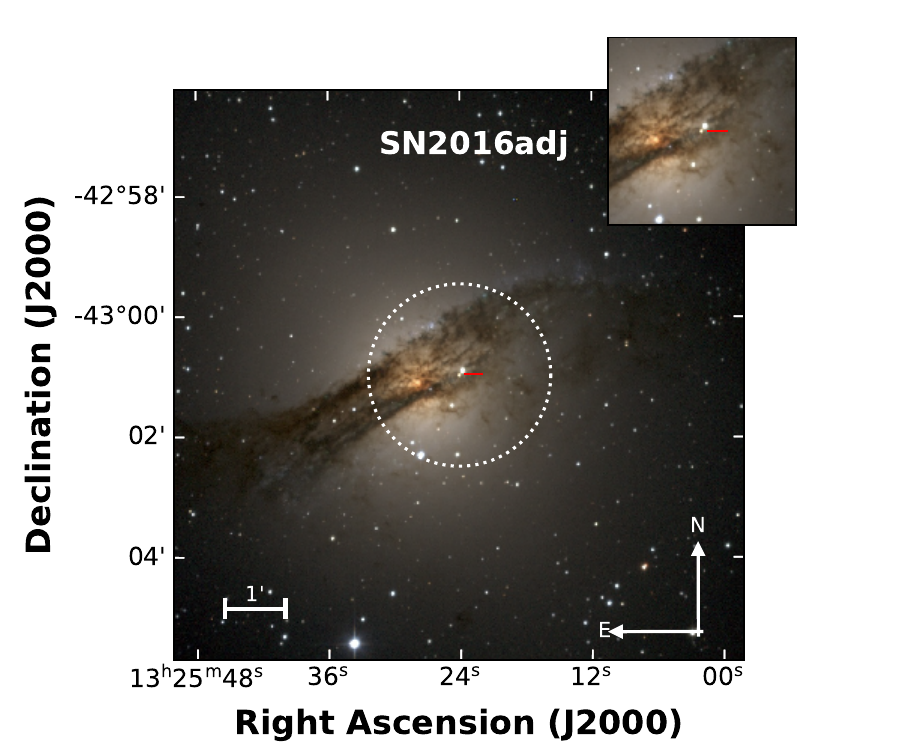}}
\caption{Colored image of Centaurus~A (NGC~5128) constructed from RGB images obtained with the Swope telescope located at the Las Campanas Observatory.  SN~2016adj occurred within the central dust lane,  just South-East of a very bright, and typically saturated foreground star.}
\label{fig:FC}
\end{figure}

SN~2016adj was discovered on 2016 February 08.56 UT (i.e., JD-2457427.06) by the Backyard Observatory Supernova Search (BOSS; \citealp{marples2016}) at an apparent $V$-band magnitude of 14.0~mag. As previously mentioned, SN~2016adj occurred in Centaurus~A with (J2000.0) coordinates of R.A.  $= 13^{h}25^{m}24^{s}.118$ and  
Decl. $= -43^\circ00'$57\farcs96 \citep{kelly2016}. Inspection of Fig.~\ref{fig:FC} reveals that SN~2016adj is located (on the sky)  in close proximity to  a bright foreground star.

 \citet{kiyota2016}  reported on 2016 February 08.69 UT  multi-color photometry of SN~2016adj  suggesting the light emitted by SN~2016adj suffered significant  reddening.
Inspection of a low signal-to-noise optical spectrum of SN~2016adj obtained with the Lijiang 2.4-meter telescope on 2016 February 08.9 UT, led \citet{yi2016} to initially report a tentative hydrogen-rich SN~II classification.
A spectrum taken several hours later (2016 February 09.2 UT) at the Las Campanas Observatory (LCO), led to the first indication of SN~2016adj of being a stripped envelope SN~Ib  \citep{stritzinger2016}.  
Soon after, \citet{hounsell2016}  and
\citet{thomas2016}  both  proposed  a SN~IIb classification.
From our detailed study of the optical and near-IR (NIR) spectra presented in Sect.~\ref{sec:spectrosocpy}, and following the standard SN classification taxonomy, we   will demonstrate that SN~2016adj is in fact a carbon-rich SN~Ic.

\subsection{Distance to Centaurus~A}
\label{sec:distance}

Centaurus~A  was discovered  by James Dunlop in the 1820s and is the fifth brightest galaxy in the night sky.  
To date,  Centaurus~A has served as a laboratory to study  black-hole accretion physics, determine AGN feedback effects, and understand spectacular X-ray and radio jets  \citep[see][for a review]{israel1998}.
The Hubble type of  Centaurus~A is widely debated with different camps of thought favoring either  an S0p or E0p Hubble type, however, as discussed by  \citet{Harris2010}, there are indications from its halo and individual stars it is more akin to an Ep galaxy.

To date Centaurus~A has hosted the 
peculiar  type~Ia SN~1986G, which was also located within the central dust lane, South-East from the center. 
Given the discovery of these two SNe within the central region of Centaurus~A strongly suggests it experienced  merger activity which generated a burst of star formation \citep[see, e.g.,][]{Dellavalle2003}.   The observed colors of SN~1986G  suggest significant host-galaxy reddening, which \citet{phillips1987} characterized with a host-galaxy color excess of  $E(B-V)_{host} \sim 0.7$ mag  
\citep[see also][]{ashall2016}, while polarimetry studies  suggest a total-to-selective absorption coefficient of   $R_V^{host} \sim 2.4\pm0.13$ \citep{hough1987,patat2015}.
As  shown in Sect.~\ref{sec:photmetry}, we infer host-galaxy reddening parameters  along the line-of-sight to  SN~2016adj of $R^{host}_V = 5.7\pm0.7$ and $A^{host}_{V} = 6.3\pm0.2$ mag. 

\begin{figure*}
\centering
\resizebox{0.75\hsize}{!}
 {\includegraphics[]{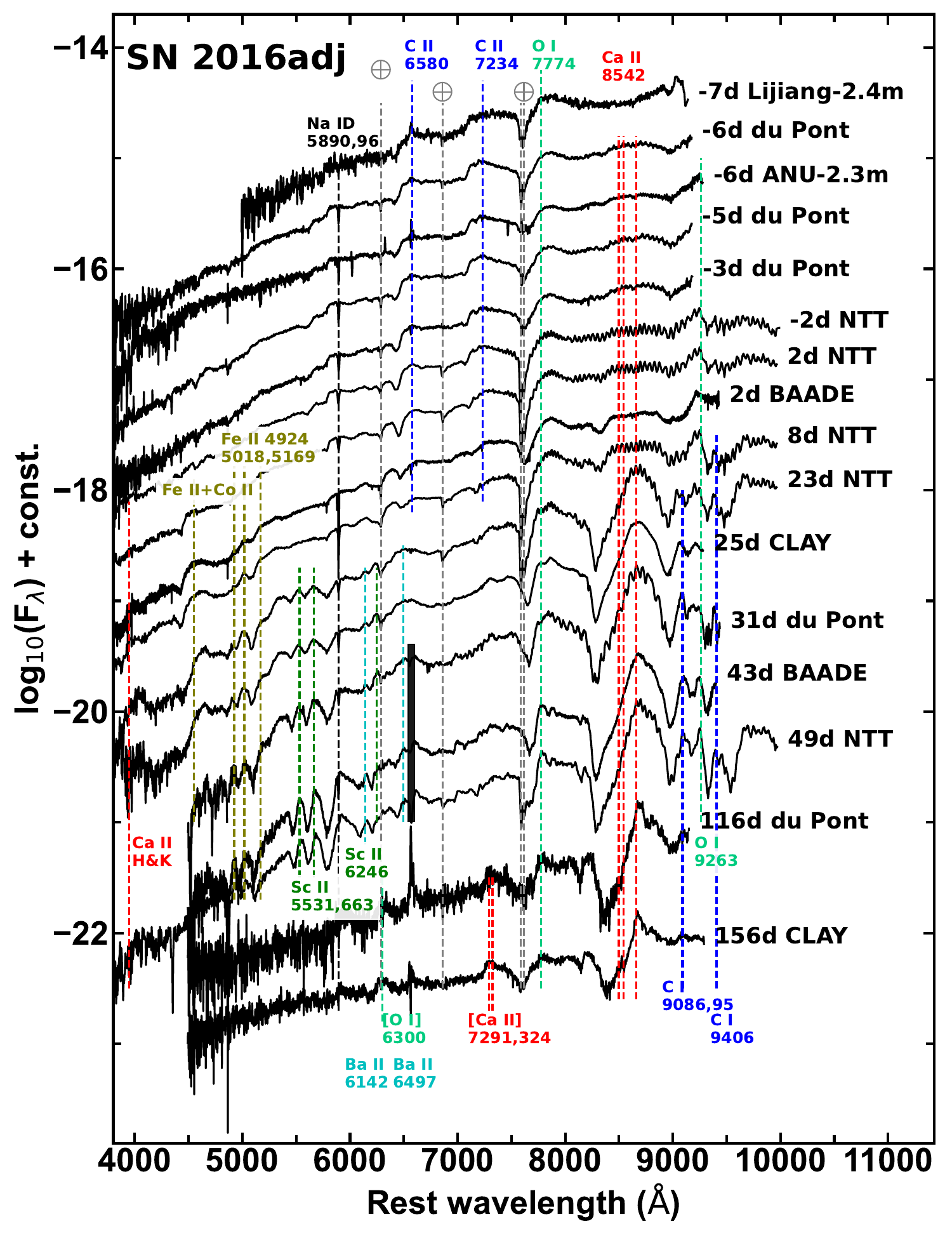}}
\caption{Optical spectra of SN~2016adj  from $-$7~d to $+$156~d. Phase and telescope facility are indicated on the right of each spectrum. Prevalent spectral features are marked with vertical dashed lines and labeled. Telluric absorption features are labeled with  Earth symbols, while for the +43~d to +49~d spectra  over-subtraction of host galaxy H$\alpha$ is masked by a black solid region.}
\label{fig:optspec}
\end{figure*}

The NASA Extragalactic Database (NED)\footnote{\href{http://ned.ipac.caltech.edu}{http://ned.ipac.caltech.edu}} provides a heliocentric red-shift  $z=0.00183\pm0.00002$ and a red-shift velocity of $547\pm5$~km~s$^{-1}$ \citep{Fouque1992}. 
NED provides  a number of direct distance measurements  included the Cepheid distance  $d = 3.42\pm0.18$ (random)  $\pm0.25$ (systematic) Mpc \citep{ferrarese07}, or a distance modulus $\mu = 27.67\pm0.12$ (random) $\pm0.16$ (systematic) mag, which is adopted throughout our analysis to set the absolute flux scale. 

The structure of this paper is as follows. Brief properties of our  data set are  presented in Sect.~\ref{sec:data}, followed by our analysis  of the spectroscopic and  photometric evolution in Sects.~\ref{sec:spectrosocpy} and \ref{sec:photmetry}, respectively. Rough estimates of key explosion parameters are  estimated in  Sect.~\ref{sec:explosionparameters}, while
 in Sect.~\ref{sec:progenitoranalysis}
 pre-explosion images are examined to place loose  constraints on the progenitor star.   
 Next in Sect.~\ref{sec:COmodeling} our results of modeling the CO first overtone feature are presented, followed by Sect.~\ref{sec:progenitorscenario} dedicated to the emergence of hydrogen features in the post maximum near-IR spectra of SN~2016adj as well as a discussion on the increasing incidences of hydrogen features appearing in other SNe~Ic which interact with H-rich circumstellar material.
The manuscript ends with our conclusions in  Sect.~\ref{sec:conclusion}, which is then followed  by four appendices including a  number of complementary figures based on the analysis of both our unpublished and published data.

\section{Data acquisition and reduction}
\label{sec:data}

We present an extensive ground-based set of observations of SN~2016adj, complemented with some public ultraviolet (UV) observations. 
More complete details on the various facilities, the data obtained, and the reduction techniques applied are provided in  Appendix~\ref{sec:appendixobs}. 

In short, the bulk of the ground-based optical and near-infrared (NIR) photometry and spectroscopy were taken by Carnegie Supernova Project-II \citep[][hereafter CSP-II]{Phillips2019} and the Public ESO Spectroscopic Survey of Transient Objects  \citep[][hereafter PESSTO]{smartt2015}. Finally, the earliest optical image was fortuitously obtained with the ESO-Paranal VST (VST Survey Telescope) $\sim 20$ days prior to maximum light.

Turning to space-based data, a measurement of the pre-maximum ultraviolet (UV) flux is obtained from  single channel images obtained with the UVOT camera on-board the \textit{Neil Gehrels Swift Observatory} \citep{gehrels04}.

\section{Spectroscopy}
\label{sec:spectrosocpy}

\begin{figure}[t]
\centering
\resizebox{0.95\hsize}{!}
{\includegraphics{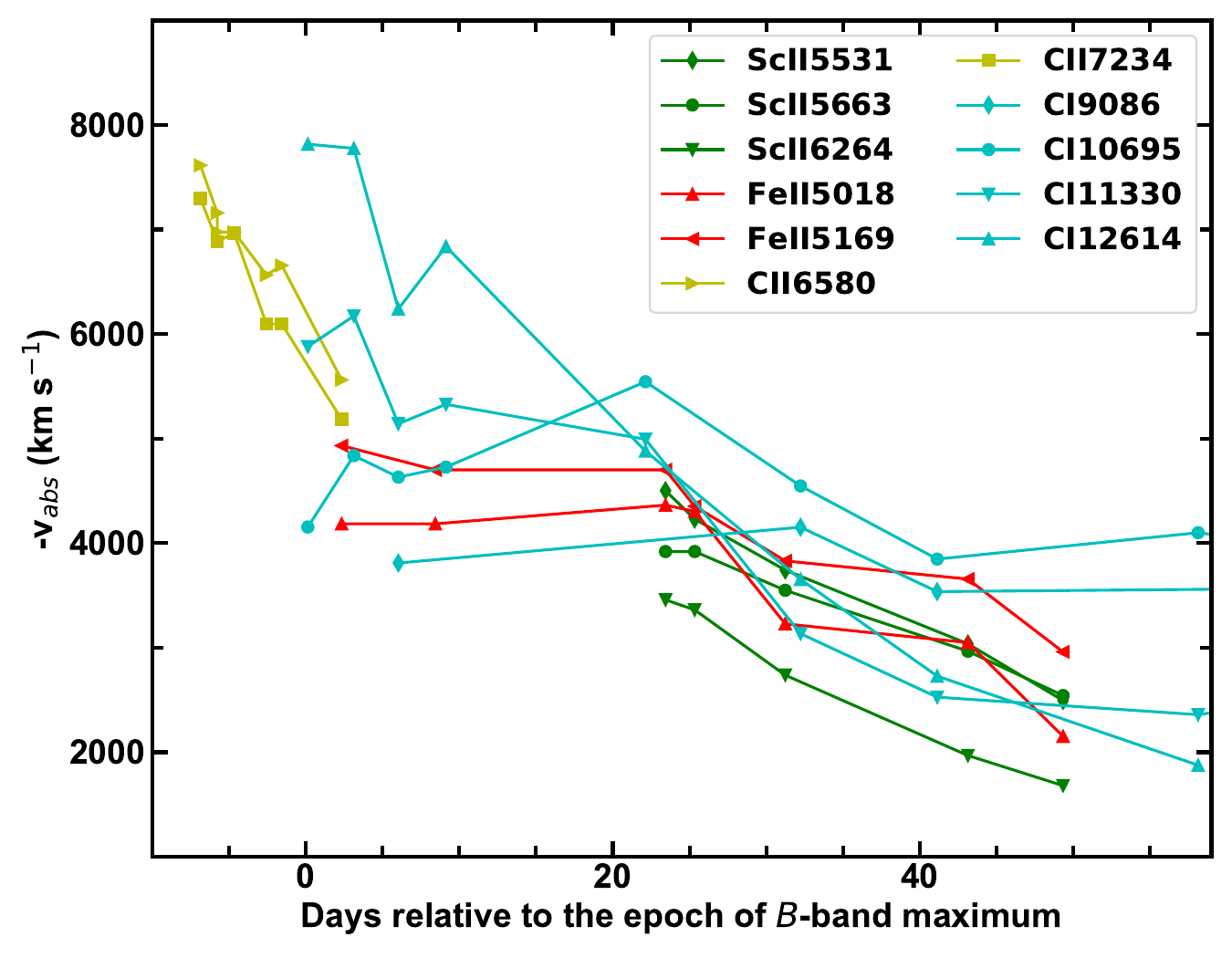}}
\caption{Measurements of $-v_{abs}$ for a handful of mostly prominent features (see legend) in the optical and NIR spectra of SN~2016adj, plotted vs. days relative to the epoch of $B$-band maximum.}
\label{fig:optvel}
\end{figure}

\subsection{Optical}
\label{sec:optspectra}

Sixteen optical  spectra of SN~2016adj  are plotted in Fig.~\ref{fig:optspec} covering 14 epochs  between $-$7~d to $+$156~d.\footnote{Rest-frame phases of observations are provided throughout, unless explicitly stated, with respect to the epoch of $B$-band maximum, which occurred on JD--2,457,433.5 (see Table~\ref{tab:peak}).} 
A journal of all spectroscopic observations of SN~2016adj is provided in Table~\ref{tab:specjor}.
As previously noted by the initial classification reports and clearly apparent in Fig.~\ref{fig:optspec}, the blue end of the optical spectra suffers prevalent suppression in flux due to significant host reddening. Additional signatures of high host reddening  take the forms of the conspicuous \ion{Na}{i}~D $\lambda\lambda$5890, 5896  doublet feature, and as well with the 5780~\AA\ diffuse interstellar band (DIB) line resolved in the $+2.4$~d spectrum obtained with the Magellan telescope equipped with the medium resolution spectrograph MagE. 
The \ion{Na}{i}~D features along with other prominent features in the spectra of SN~2016adj are marked with vertical lines and labeled in Fig.~\ref{fig:optspec}, while Table~\ref{tab:linelogopt} summarizes the ions  attributed to the various spectral features identified in the optical time-series.

The earliest optical spectra are relatively featureless exhibiting only a handful of notches, however by a week past maximum, a number of prominent P~Cygni features do emerge.
Specifically, the \ion{Ca}{ii} NIR triplet and the nearby \ion{O}{i} $\lambda$7774 lines are observed from the first epochs with ever-increasing pEW values. At the blue end of the spectra  extended below 4000~\AA, narrow \ion{Ca}{ii} H\&K  features are also identified around maximum light. 
Between $-7$~d to  $+$8~d  the spectra also exhibit a feature around 6500~\AA, which the classification reports mentioned in Sect.~\ref{sec:intro} largely attribute to H$\alpha$ and hence the initial SN~Ib/IIb classification of SN~2016adj. If this classification were correct then \ion{He}{i} features are expected to be present. 
 However the maximum phase optical spectra do  not exhibit any evidence of such  features. Moreover,  \ion{He}{i} features do not emerge in any of the post maximum spectra when such features are known to emerge in SNe~IIb/Ib \citep[see][and references therein]{Holmbo2023}.
As an alternative,   we suggest  the   6500~\AA\ feature is formed by  \ion{C}{ii} $\lambda$6580 \citep[e.g.,][]{valenti2008}. Furthermore, an absorption feature around $\sim$7100~\AA\ could be produced by  \ion{C}{ii} $\lambda$7234. 
In support of these \ion{C}{ii} associations are at least ten individual \ion{C}{i} features identified in the NIR spectra  of SN~2016adj (see below).
 
Scrutiny  of the last two spectra obtained on $+$116~d and $+$156~d of SN 2016adj in Fig.~\ref{fig:optspec}  reveals the emergence of forbidden   [\ion{O}{i}] $\lambda\lambda$6300, 6363 and  [\ion{Ca}{ii}] $\lambda\lambda$7291, 7324 lines, as well as narrow emission lines of H$\beta$ and  H$\alpha$. The Balmer lines are characterized by a full-width-half-maximum velocity ($v_{FWHM}$) of a few $\cdot$100~km~s$^{-1}$, and are attributed to an underlying \ion{H}{ii} region. 

\begin{figure*}
\centering
\resizebox{\hsize}{!}
{\includegraphics{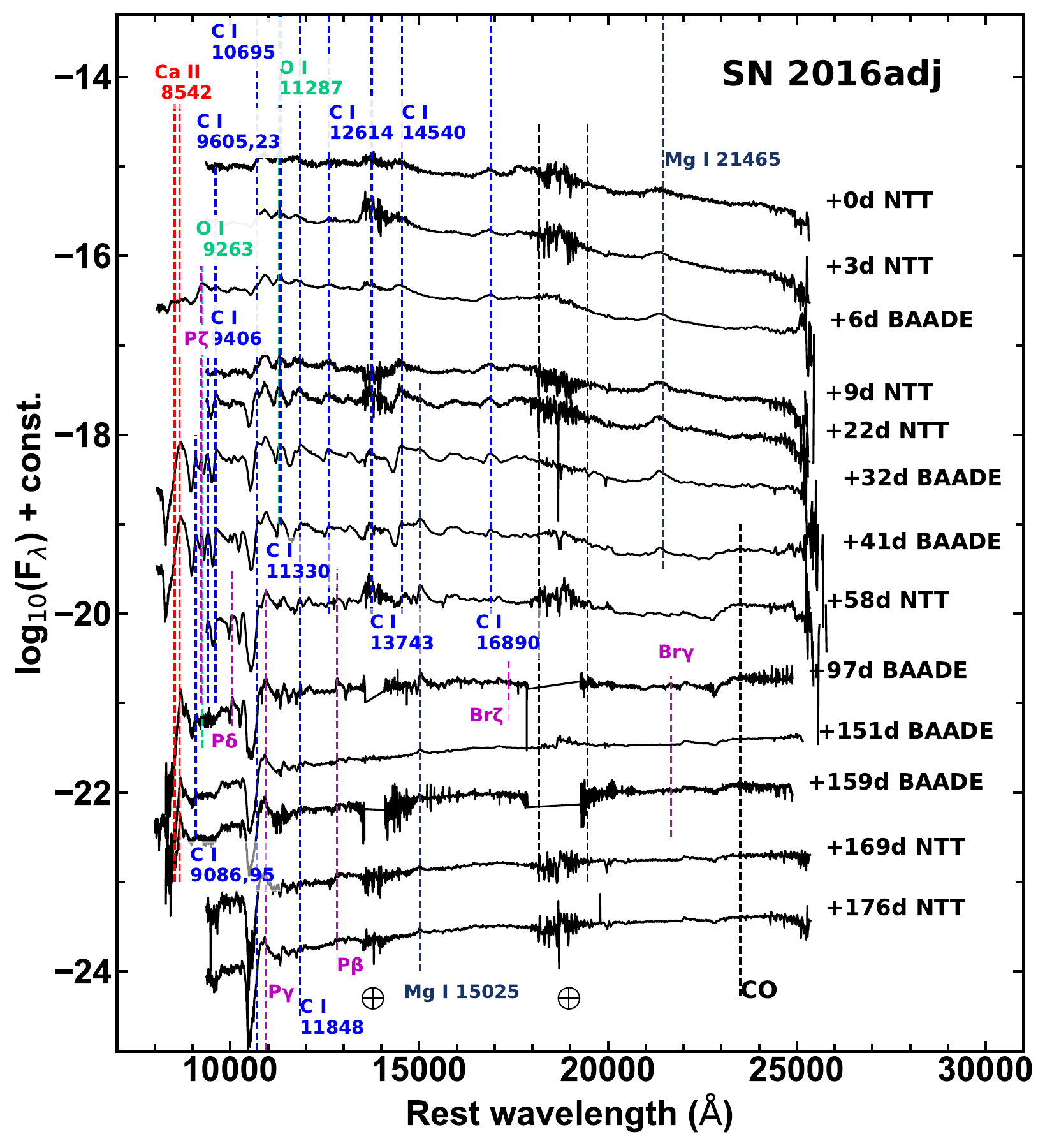}}
\caption{NIR spectral time-series of SN~2016adj covering 6 months of evolution beginning from the epoch of $B$-band maximum. Phase and telescope  are indicated to the right of the spectra. The locations of various spectral features are identified with vertical dashed lines and labeled. Lines indicate the spectra where that the features are identified. Telluric regions are labeled with Earth symbol.} 
\label{fig:nirspec}
\end{figure*}  

 Fig.~\ref{fig:optvel} shows the time evolution of the  Doppler velocity at maximum absorption ($-v_{abs}$) inferred from the early \ion{C}{ii} $\lambda$6580 and \ion{C}{ii} $\lambda$7234 features, the  \ion{Fe}{ii} $\lambda\lambda$5018, 5169 features, the \ion{Sc}{ii} $\lambda\lambda$5531, 5663, 6246 features, as well as the   \ion{C}{i}   $\lambda$9086, \ion{C}{i} $\lambda$10695, \ion{C}{i} $\lambda$11330 and   \ion{C}{i} $\lambda$12615 features.
 During the  covered epochs of spectra no  $-v_{abs}$ values  exceed 8000 km~s$^{-1}$.  Indeed, the  \ion{C}{ii} features are consistent,  evolving from   $\sim$7700~km~s$^{-1}$ to $\sim$5000 km~s$^{-1}$ between  $-$7~d  to  +2~d. Measurements of the \ion{Fe}{ii} lines in the post-maximum spectra indicate $-v_{abs}$  values of $\approx 4000-5000$ km~s$^{-1}$ on day $+$25~d down to  2000-3000~km~s$^{-1}$ by $+$49~d. Similarly, the \ion{Sc}{ii}  lines present between $+$23~d and $+$49~d  evolve from $-v_{abs} \sim 3500-4500$ km~s$^{-1}$ to $\sim1700-3000$ km~s$^{-1}$. 
 In comparison to  $-v_{abs}$ values inferred for literature samples of  SNe~Ic, SN~2016adj exhibits values on the low side of the distribution  \citep[e.g.][]{liu2016,fremling2018,Holmbo2023}. This hints at SN~2016adj having a low explosion energy, a high ejecta mass, or a combination of the two (see below).

\subsection{NIR Spectroscopy: early  C~I and post-maximum  H~I  features}
\label{sec:nirspectra}

Our NIR time-series offers a rare chance to view in detail the evolution of this wavelength region of a SN~Ic over a six month period. The  NIR spectral time-series of SN~2016adj plotted in Fig.~\ref{fig:nirspec} extends from $+0$~d to $+176$~d. Prominent features in the spectra are indicated with vertical lines and labeled. A summary of these features are provided in Table~\ref{tab:linelognir}, and include the \ion{Ca}{ii} triplet, ten conspicuous \ion{C}{i} features, two subtle  \ion{O}{i} $\lambda$9263 and \ion{O}{i} $\lambda$11287 features, and the \ion{Mg}{i} $\lambda$15025  and \ion{Mg}{i} $\lambda$21465 features. 
The $-v_{abs}$ values of four of these \ion{C}{i} features  are  plotted in Fig.~\ref{fig:optvel} along with the values inferred from the various optical lines.
The velocity of these \ion{C}{i} features are in good agreement with those inferred from the  \ion{C}{ii} features present in the early optical spectra.

 \citet{banerjee2018} presented a NIR time-series of SN~2016adj  from $-5$~d to $+$61~d, and in their analysis preferred a  SN~IIb classification. Contrary to  \citeauthor{banerjee2018}, we do not ``unambiguously identify" from the onset of observations hydrogen  features associated with Pa-$\delta$ (Pa-7) $\lambda$10049, 
 Pa-$\gamma$ $\lambda$10938, 
 Pa-$\beta$ $\lambda$12822, and 
 Br-$\gamma$ $\lambda$21661. 
Furthermore, we do not find features attributable to Pa-$\delta$ and Pa-$\beta$ $\lambda$12822. A  feature around $\sim$21500~\AA\ is more likely  \ion{Mg}{i} $\lambda$21465 rather than Br-$\gamma$ $\lambda$21661.
 There is a notch in the early spectra that could be Pa-$\gamma$ $\lambda$10938, but the identification is not conclusive. 
 \citeauthor{banerjee2018} also suggest a feature around   $\sim$9230~\AA\ that might be produced by a blend of  Pa-$\zeta$ (Pa-9) $\lambda$9232 and \ion{Mg}{i} $\lambda$9258 blend, however, based on the location and time evolution of the feature, we suggest a \ion{O}{i} $\lambda$9263 identification. 
  We note that the NIR time-series shows no evidence of Pa-$\alpha$ $\lambda$18751 at any time; however, Pa-$\alpha$ must be very strong to see the signal through the telluric haze \citep[see, e.g.,][]{Shahbandeh2021}.

\begin{figure*}[t]
\resizebox{\hsize}{!}
{\includegraphics{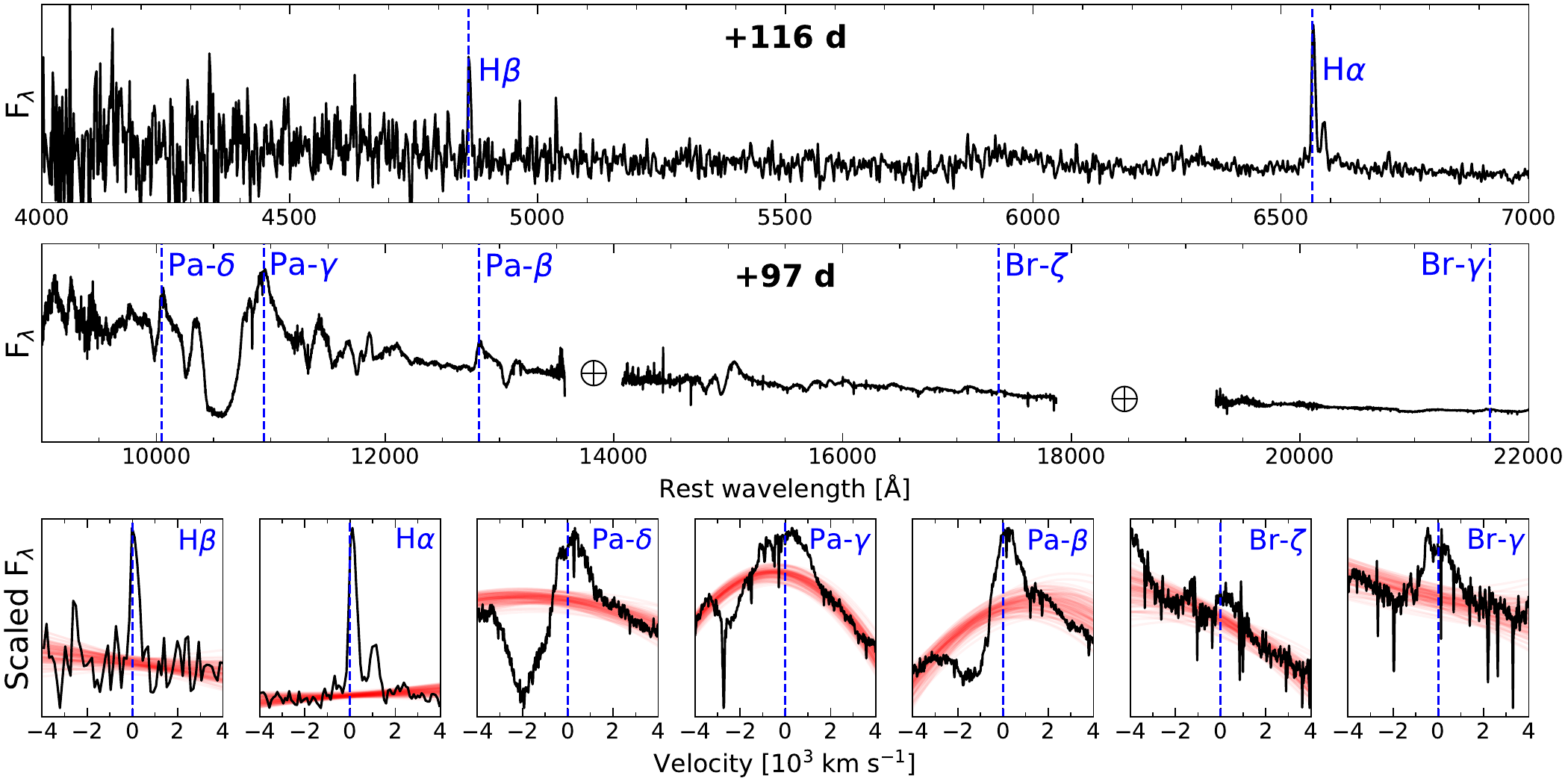}}
\caption{\textit{(top)} Low-resolution optical- $+$116~d   and \textit{(middle)} echelle NIR  $+$97~d spectra of SN~2016adj. The optical and NIR spectra have been color-corrected to match coeval broad-band photometry. Both spectra are   de-reddened using the reddening parameters estimated in Sect.~\ref{sec:reddening}. The positions of   conspicuous hydrogen features are marked with vertical lines and labeled. \textit{(bottom)} Zoom in velocity space of the Balmer, Paschen and Bracket series features.
The  Paschen features exhibit  P~Cygni profiles with  $-v_{abs}$ values $\sim 1500-3000$~km~s$^{-1}$ and $v_{FWHM} \sim 1000$ km~s$^{-1}$. 
The red lines correspond to pseudo-continuum fits determined by Monte Carlo simulations. Note that the flux of the nebular [\ion{N}{ii}]  $\lambda\lambda$6548,6584 lines on either side  of H$\alpha$ is not included in  its emission-line flux measurement.}
\label{fig:Hlines}
\end{figure*}

 Interestingly, by  $+$58~d  (corresponding to the last epoch spectrum presented by \citeauthor{banerjee2018}) we identify the emergence of a handful of  hydrogen features. Among our time-series these features are best revealed in the medium-resolution   echelle spectrum obtained on $+$97~d.
  Plotted in the top panel of Fig.~\ref{fig:Hlines} is the $+$116~d optical spectrum  with the  two narrow Balmer emission features marked and labeled, while the middle panel displays the $+$97~d spectrum containing  hydrogen features attributed to: Pa-$\delta$, Pa-$\gamma$, Pa-$\beta$, Br-$\epsilon$, and Br-$\gamma$. 
  The seven hydrogen features contained within the top two panels are
  also plotted in velocity space  within the bottom row of
  Fig.~\ref{fig:Hlines}. The optical spectrum reveals  narrow
  H$\beta$ and H$\alpha$ emission with $v_{FWHM} \sim$ a few
  $\times$100~km~s$^{-1}$. Inspection of the features (see also
  Fig.~\ref{fig:optspec}) seem to tentatively suggest a broader
  component for both features, however, this is speculative due to the
  poor signal-to-noise of the data. In the optical data  the narrow
  component is likely due to nebular emission lines associated with an underlying \ion{H}{ii} region (see Appendix~\ref{sec:appendixMUSE}).
   Turning to the NIR Paschen lines, they exhibit P~Cygni profiles with  $-v_{abs} \sim 1500-2500$ km~s$^{-1}$.  The wavelength regions containing the Bracket lines are of lower signal-to-noise, but they also exhibit emission features with $v_{FWHM} \sim 1000$ km~s$^{-1}$ with some line structure. 
  
 We next measure the emission and absorption  line flux values associated with  the H features in the two spectra shown in Fig.~\ref{fig:Hlines}. To do so, a blue and red edge for the features is first defined by applying an iterative second order polynomial fit, which  enables  the  pseudo-continuum associated with the features to be defined. Emission and absorption flux values are then estimated  by integration of the flux contained within the area defined by the emission/absorption components and the pseudo-continuum. This process is applied following a Monte Carlo  approach consisting of 100 realizations, which provide an estimate on the uncertainty of the estimated flux values. The MC realizations are over-plotted with the features shown in the bottom panels of  Fig.~\ref{fig:Hlines}, and the emission and absorption flux values are summarized  in Table~\ref{tab:H1fluxes}.  In Sect.~\ref{sec:progenitorscenario} the emission-line flux values are used as a diagnostic to estimate the density of the line forming region.
 
 \begin{figure}[t]
 \resizebox{\hsize}{!}
{\includegraphics{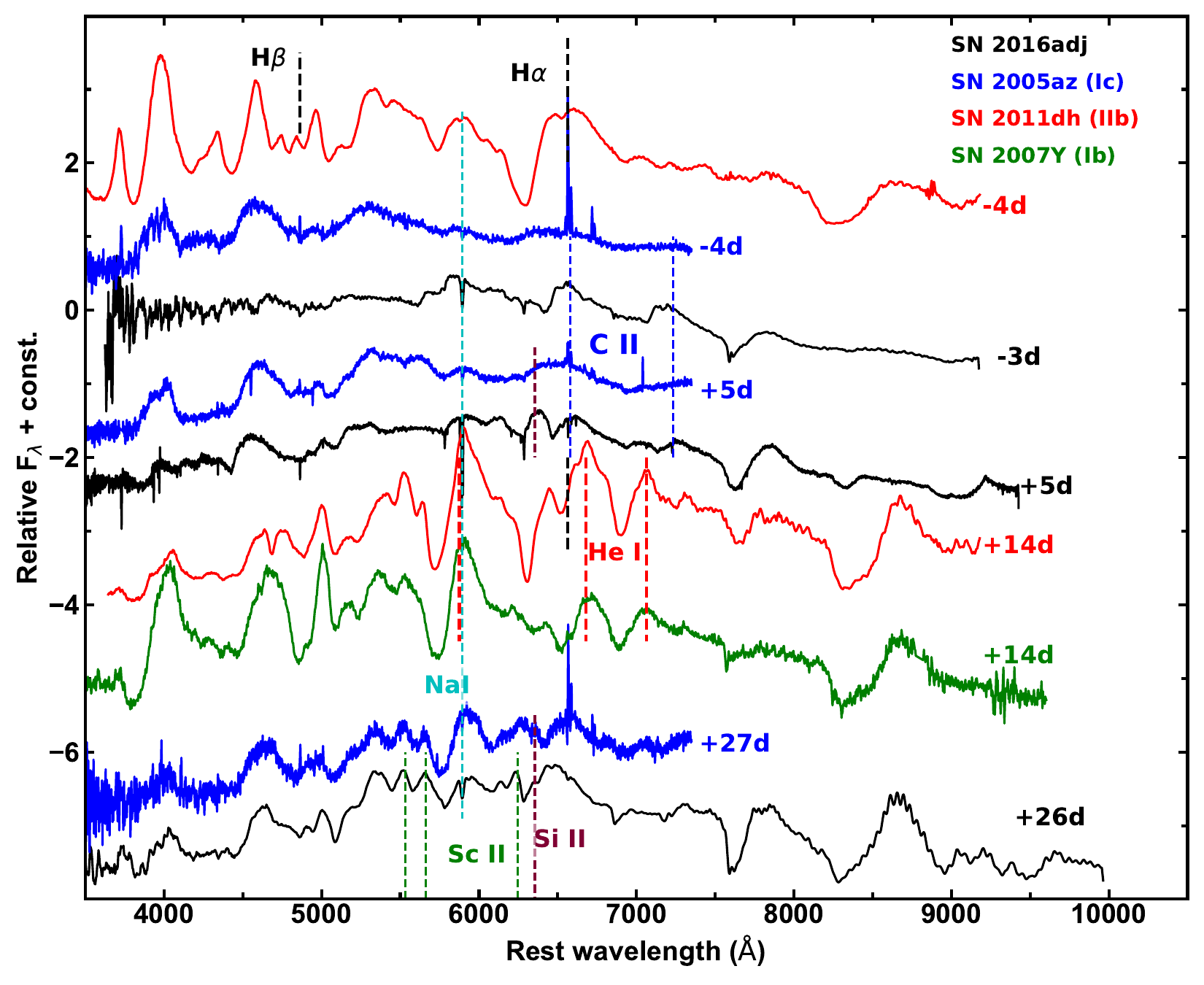}}
\caption{A selection of  optical spectra of SN~2016adj de-reddened and  compared with similar  phase spectra of the type~IIb SN~2011dh  \citep{ergon14}, the type~Ib SN~2007Y \citep{stritzinger09}, and the type~Ic  SN~2005az \citep{bianco2014,modjaz_spectra_2014}. The epochs labeled next to each spectra are relative to the epoch of  $r$-band maximum. Similar to the type~Ic SN~2005az, the spectra of SN~2016adj lack  Balmer lines (dashed magenta) and \ion{He}{i} features (red dash lines), both of which  develop past maximum in the  SNe~2007Y and 2011dh.  \ion{C}{ii} lines   are indicated with blue dashed lines.}
\label{fig:comp_spec_opt}
\end{figure}

\begin{figure}[t]
\centering
\resizebox{\hsize}{!}
{\includegraphics{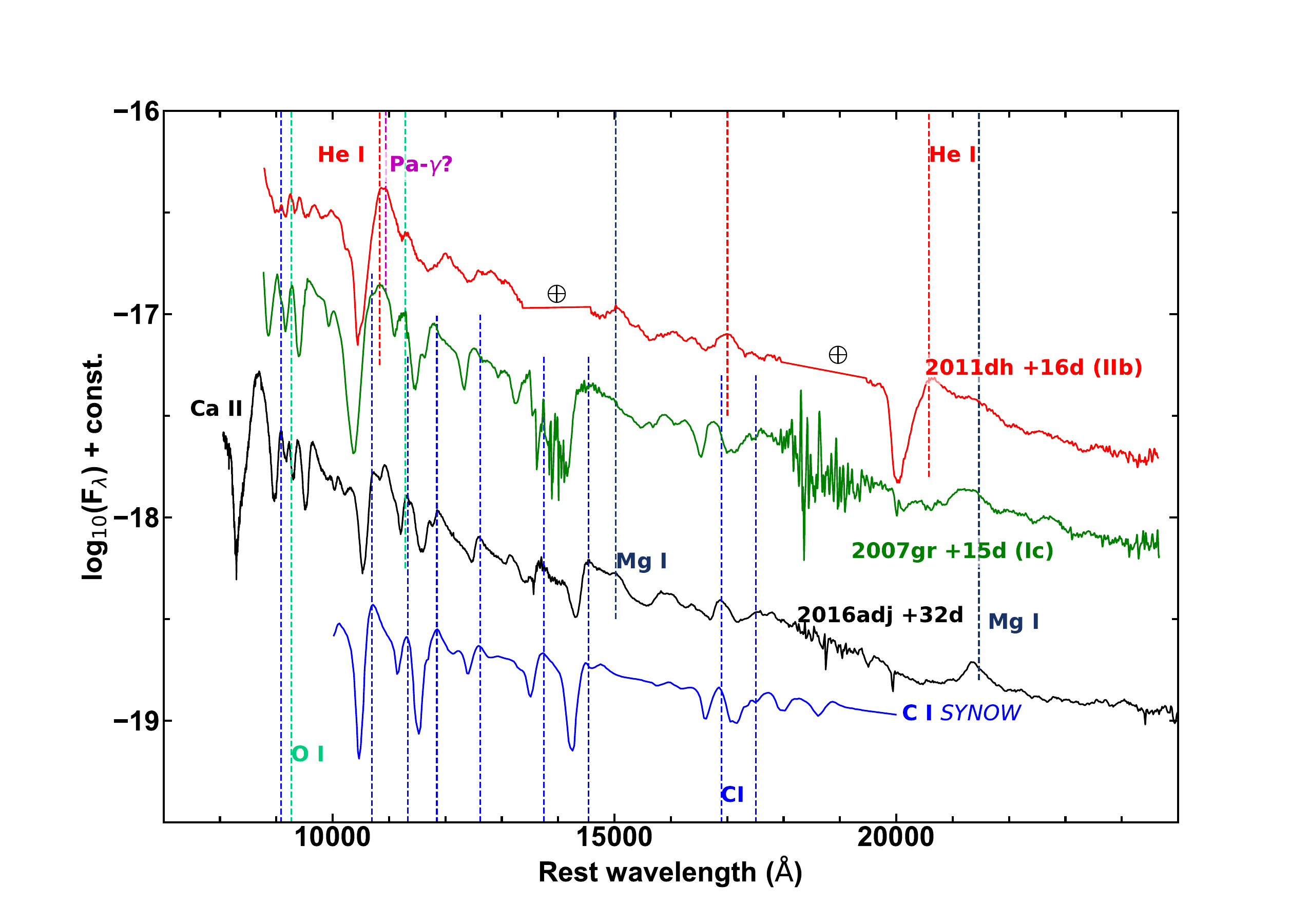}}
\caption{The de-reddened  NIR spectrum of SN~2016adj  taken at +32~d compared with a $+$15~d spectrum of the type~Ic SN~2007gr \citep{valenti2008},  a $+16$~d spectrum of the type~IIb SN~2011dh \citep{ergon14}, and the \texttt{SYNOW} synthetic spectrum computed  for \ion{C}{i} \citep{millard99}. Prevalent \ion{C}{i} lines, as for example the strong feature around 11000~\AA,  are present in the NIR spectrum of SN 2016adj. As in the optical, no  traces of \ion{He}{i} are identified in the longer wavelength spectrum of SN~2016adj. This is contrary to the  \ion{He}{i} features  in the spectrum of SN~2011dh, particularly  the prevalent P~Cygni feature at 20581~\AA.}
\protect\label{fig:comp_spec_nir}
\end{figure}

\subsection{SN~2016adj is a type Ic supernova}
\label{sec:classification}

\citeauthor{banerjee2018} alluded  to the possibility that SN~2016adj could be a hydrogen- and helium-deficient SNe~Ic,  but due to a lack of optical spectra and  interpretation of the NIR spectra that is contrary to our own, they  preferred a SNe~IIb classification. Based on the detailed comparisons to other SE SNe, we find SN~2016adj is a SN~Ic. Here the optical and NIR spectra of SN~2016adj are compared to similar epoch observations of the various SE SN  sub-types in order to support our reclassification of SN~2016adj. 
 
First, Fig.~\ref{fig:comp_spec_opt} compares the $-$3~d, $+$5~d and $+$26~d optical spectra of SN~2016adj to similar phase spectra of the type~IIb SN~2011dh \citep{ergon14}, the type~Ib SN~2007Y \citep{stritzinger09}, and the type~Ic SN~2005az \citep{bianco2014}. These objects were chosen as they are excellent representatives of their SE SN subtype and have similar phase spectra as SN~2016adj. The spectra of SN~2005az are found to be very similar to those of SN~2016adj. 
Both objects clearly lack the Balmer  features that are quite prevalent in SN~2011dh. Moreover, SN~2005az and SN~2016adj do not exhibit over their evolution \ion{He}{i} lines  which are so apparent  in the post-maximum spectra of  SN~2007Y and SN~2011dh. From this comparison alone, we find that SN~2016adj is fully consistent with being a SN~Ic. 

Turning  to longer wavelengths,  Fig.~\ref{fig:comp_spec_nir} compares the $+$32~d NIR spectrum of SN~2016adj with  the $\approx$ +15~d spectrum of the type~Ic SN~2007gr \citep{valenti2008}, the $+$16~d spectrum of  SN~2011dh \citep{ergon14}, and a \texttt{SYNOW} synthetic spectrum  for  solely \ion{C}{i} \citep{millard99}. 
  The spectra of SN~2007gr and SN~2016adj are nearly identical, exhibiting the same features including, most notably, the \ion{C}{i} lines. As compared to SN~2011dh, both objects lack the prominent \ion{He}{i} $\lambda$20581 line. Within this context, the strong 10500~\AA\ feature in SN~2007gr and SN~2016adj is attributed to \ion{C}{i} $\lambda$10695 as is supported by the comparison with the \texttt{SYNOW} spectrum. In the case of SN~2011dh, its 10500~\AA\ feature is then attributed to \ion{He}{i} $\lambda$10831 with a possible contribution from \ion{C}{i} $\lambda$10695. 
  Based on these findings, the NIR spectra of SN~2016adj are in full agreement with it  being a SN~Ic.

\subsection{Detection of the first CO overtone}
\label{sec:CO}
 
CO emission was first identified in SN~2016adj by \citet{banerjee2016},  who noted the presence of the first CO overtone (2-0) band head ($2.25-2.45$~$\mu m$)  in a spectrum obtained at $+$52.6~d. The same authors  confirmed  this initial assessment with a spectrum  obtained  at   $+$58.7~d  \citep{banerjee2018}.
Turning to our NIR spectroscopic time-series extending through $+$176~d (see Fig.~\ref{fig:nirspec}),  the first CO overtone is found to have emerged already by $+$41~d. This is  $\approx10$~d earlier than previously reported, and  to our knowledge, among the earliest  CO  signature detected  in the wake of a  supernova \citep{banerjee2018,Ravi2023}.

In Sect.~\ref{sec:COmodeling} we turn to  CO models to estimate  key physical parameters characterizing the underlying CO emission region. 

\begin{figure*}[t]
\centering
 \resizebox{0.95\hsize}{!}
{\includegraphics{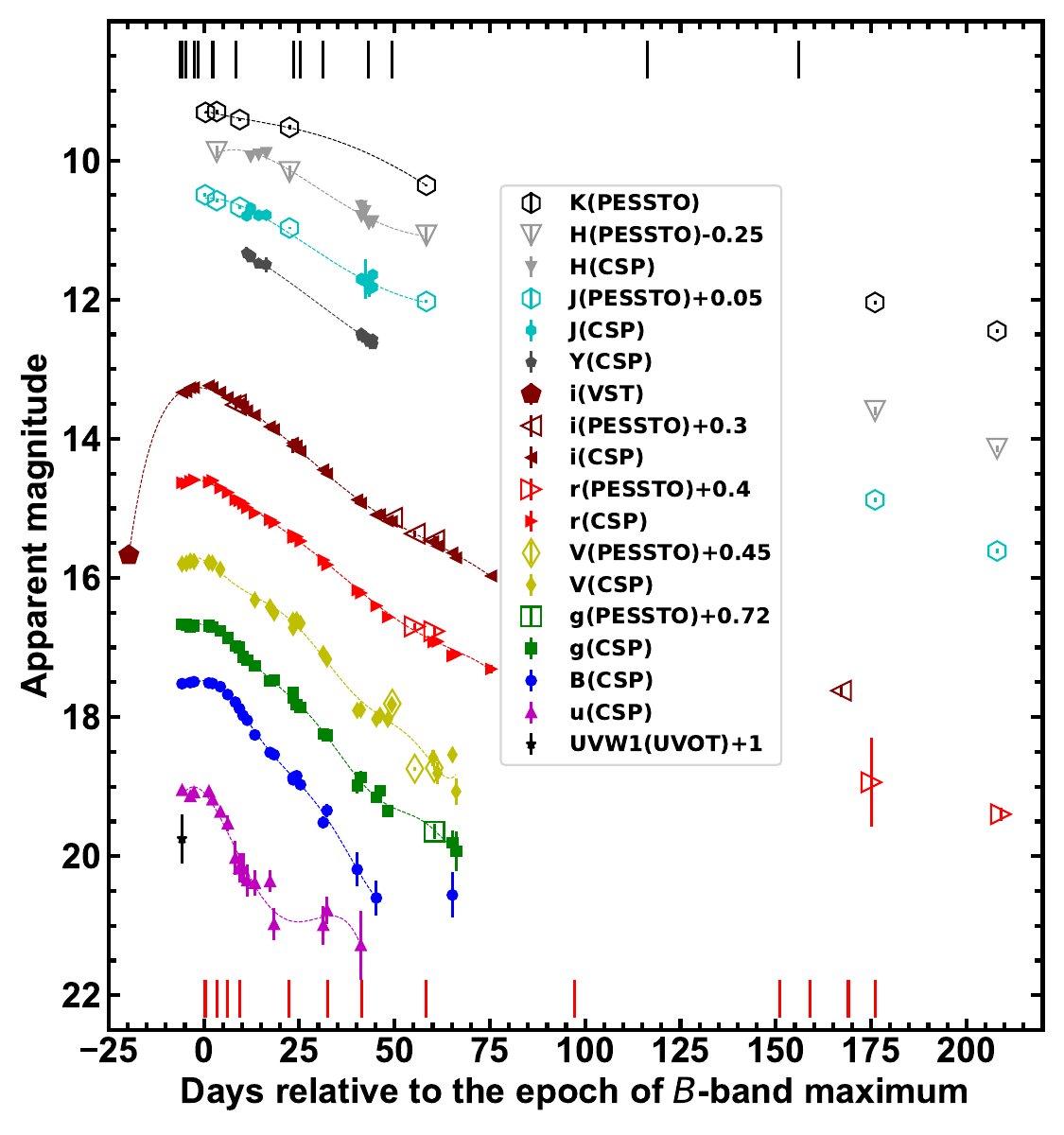}}
\caption{Optical and NIR photometry of SN~2016adj plotted relative to the epoch of $B$-band maximum from observations obtained by CSP-II, PESSTO, with the ESO-Paranal VLT Survey Telescope (VST) equipped with OmegaCAM, and with the UVOT camera on-board  \textit{Swift}. To facilitate the comparison  of photometry obtained with different instruments, in some cases offsets have been applied as indicated in the legend. A low-order polynomial function is overplotted on each light curve and used to infer the time and value of peak. Epochs of spectroscopic observations are indicated by black (visual) and red (NIR) segments.} 
\protect\label{fig:lcs}
\end{figure*}

\section{Photometry}
\label{sec:photmetry}

\subsection{Light curve parameters}

Our UV/optical ($UVW1,u,B,g,V,r,i$) and NIR ($Y,J,H,K$) light curves of SN~2016adj are shown in  Fig.~\ref{fig:lcs}. The bulk of the photometry follows the flux evolution beginning a few days prior to the epoch of  $B$-band maximum through  $\approx+$70~d in the optical and $\approx+$210~d in the NIR. In addition, a very early ($-$19.7~d) $i$-band photometric measurement was computed from a serendipitous image obtained by the VST.

The light curves of SN~2016adj in the bluer bands are much fainter than the light curves of the redder bands. 
This is a result of the significant host-galaxy reddening affecting the light of SN~2016adj. Indeed, the $K$-band photometry is  $\approx$6 magnitudes brighter than the $V$-band photometry, suggesting a host-galaxy visual extinction of the same order.     
An indication of significant extinction is also evident from the comparison between the observed colors of SN~2016adj with those of SE SNe intrinsic color-curve templates (see below).
 
Comparison among the light curves indicates that the bluer bands decline faster than the redder bands. This is typical of  the evolution documented in  samples of SE SNe studied in  the literature \citep[e.g.,][]{taddia15sdss,taddia18csp}.
 The epoch and apparent magnitude at peak for each light curve was determined through the use of low order polynomial fits, as well as the light-curve decline rate parameter $\Delta$m$_{15}$ \citep{phillips1993}. The values of these light curve parameters and the absolute peak magnitudes (see below) are summarized in  Table~\ref{tab:peak}.

\begin{figure*}[t]
\centering
 \resizebox{0.9\hsize}{!}
{\includegraphics{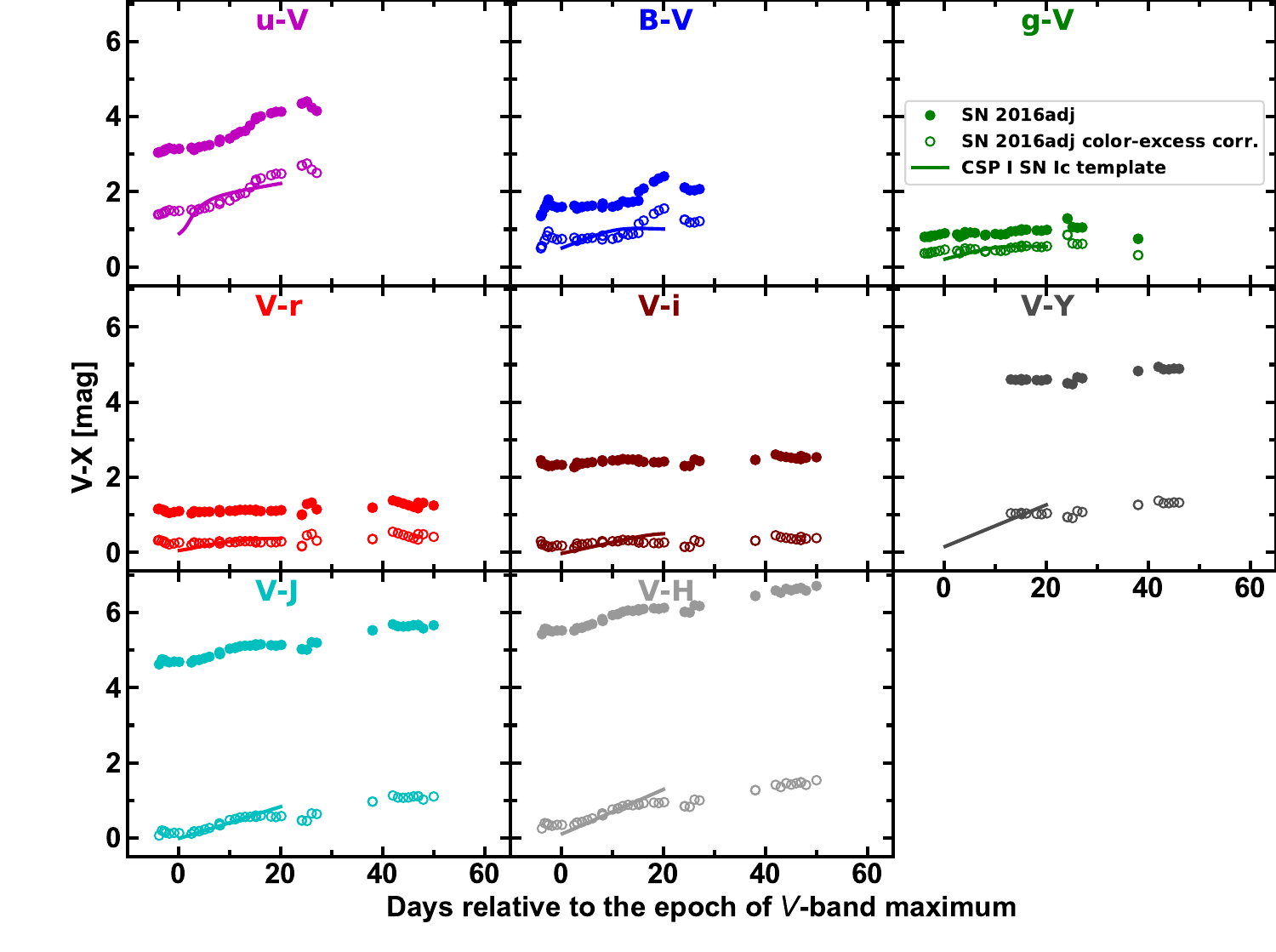}}
\caption{Optical and optical/NIR colors of SN~2016adj corrected for Milky Way reddening (filled symbols) vs.  days relative to the time of $V$-band maximum.  The  solid lines correspond to the SN~Ic intrinsic color-curve templates presented by \citet{stritzinger2018b}, while the unfilled symbols correspond to the fit of the templates to the filled symbols. The  $E(V-H)_{host}$ color excess suggests an  $A_V^{host} \gtrsim 5$ mag.}
\protect\label{fig:colors}
\end{figure*}

\subsection{Reddening of SN~2016adj}
\label{sec:reddening}

\begin{figure}[t]
\resizebox{\hsize}{!}
{\includegraphics{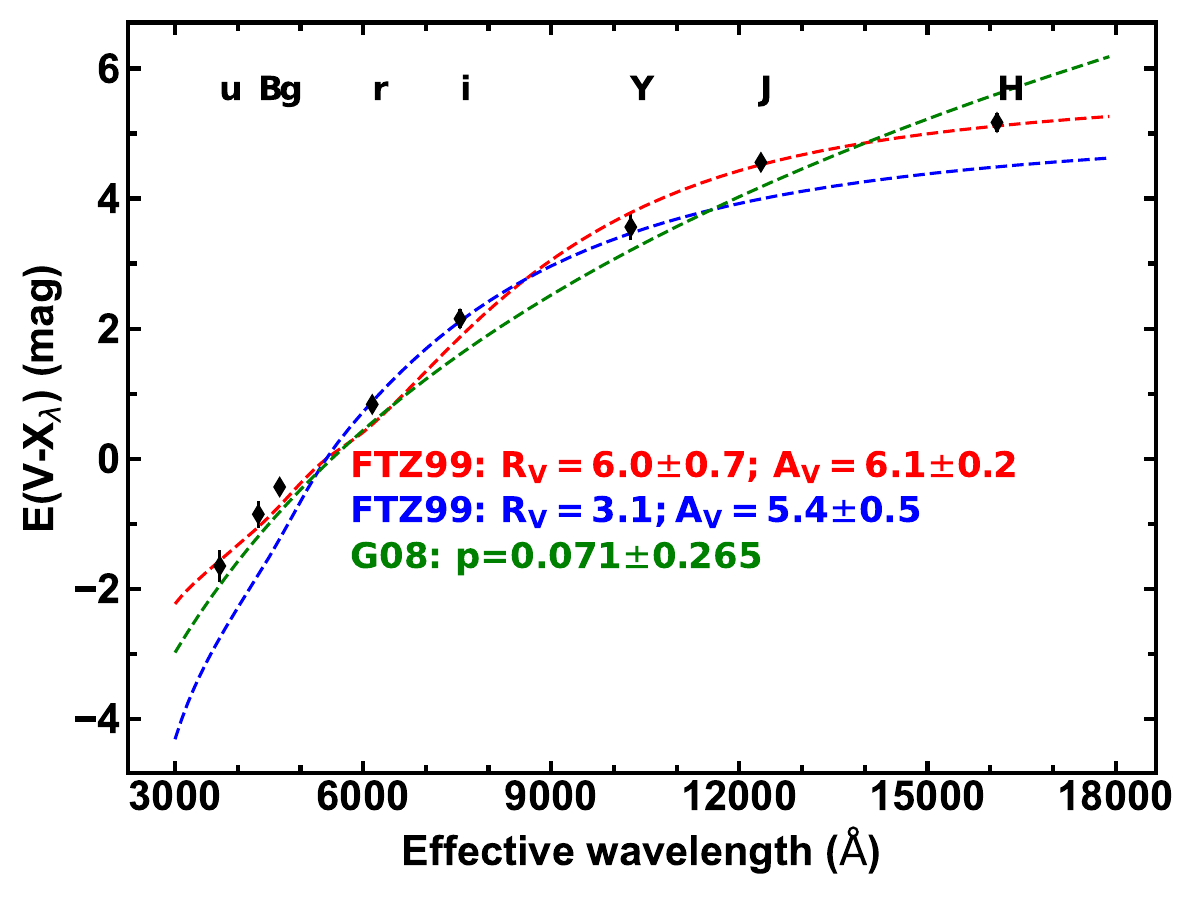}}
\caption{Optical and NIR $E(V-X)_{host}$  (where $X = {u, B, g, r, i, Y, J, H}$) color excess measurements of SN~2016adj, plotted as a function of the effective wavelength of passband $X$. Each color excess measurement represents the difference between  the observed color (corrected for $E(B-V)_{MW}$) and the  intrinsic color-curve template for SN~Ic  shown in Fig.~\ref{fig:colors}. The red dashed line corresponds to the best \citet{fitzpatrick99} reddening law  model fit characterized by a ratio of total-to-selective absorption $R_V^{host} = 5.7\pm0.7$ and a visual extinction $A_V^{host} = 6.3\pm0.2$ mag. Also shown are the best  \citet{fitzpatrick99} reddening law model fits  with $R_V^{host} = 3.1$ (blue dashed line), 
and the best-fit \citet{goobar08}  power-law
model (green dashed line).}
\protect\label{fig:colorexcess}
\end{figure}

According to \citet{schlafly11} and reported by NED, the Milky Way reddening  in the direction of Centaurus~A is non-negligible with a color excess of $E(B-V)_{MW} = 0.1$ mag, which upon adopting the standard Milky Way total-to-selective absorption coefficient, $R_V = 3.1$, corresponds to the Milky Way visual extinction value $A_{V}^{MW} = 0.31$ mag. 
Located within the central dust lane of Centaurus~A, it is not surprising that  SN~2016adj suffers significant host-galaxy reddening. This is manifested in the significant suppression of flux on the blue end of the optical spectra (see Fig.~\ref{fig:optspec}) and the presence of conspicuous and saturated \ion{Na}{i}~D features. 

Determining an accurate estimate of the host reddening of SN~2016adj is a challenging task. As a first step the equivalent width (EW) of the  DIB 5780~\AA\ feature measured from our MagE spectrum implies, following  Equation~(6) of \citet{phillips2013}, an $A_V^{host} = 3.0^{+1.9}_{-1.2}$ mag.
Assuming an $R_V \sim$ 3.1 this corresponds to  $E(B-V)_{host} \sim 1.0$ mag. 

Turning to pre-explosion (February 2015) integrated field spectroscopy observations obtained with the VLT (+MUSE; see Appendix~\ref{sec:appendixMUSE}) and assuming a canonical H$\alpha$/H$\beta$ intrinsic ratio of 2.86,  the Balmer decrement  implies a gas-phase color excess at the location of SN~2016adj  of  $E(B-V)_{gas} = 0.92\pm0.37$ mag.

We next estimated the host reddening by comparing the  observed colors of SN~2016adj to the intrinsic SE SN color-curve templates  presented by  \citet{stritzinger2018b}. Plotted in Fig.~\ref{fig:colors} are the   ($X-V$, where $X = {u, B, g, r, i, Y, J, H}$ colors) of SN~2016adj corrected for Milky Way reddening and the SN~Ic color-curve templates. Taking the mean difference between  the Milky Way reddening-corrected colors and the color-curve  templates  for each color combination, yields  the  $E(V-X)_{host}$ values plotted as a function of the effective filter wavelength in Fig.~\ref{fig:colorexcess}. Overplotted to these  values are three different  best-fit reddening laws representing the
 \citet[][hereafter FTZ99]{fitzpatrick99} reddening law with 
 $R_{V}^{host} = 3.1$ and  $A_V^{host} = 5.7\pm0.5$ mag (reduced $\chi^2= 20.7$),  the   \citet[][G08]{goobar08} power-law model (reduced $\chi^2 = 11.5$), and the FTZ99 law with $R_{V}^{host}$ set as a free parameter (reduced $\chi^2 = 5.6$).

Given the reduced $\chi^2$ values  we adopt in our analysis the host reddening parameters of $R_V^{host} = 5.7\pm0.7$ and  $A_V^{host} = 6.3\pm0.2$ mag (i.e.,  $E(B-V)_{host} = 1.1$ mag), which is  consistent with the reddening estimated from Balmer decrement measurements computed using a MUSE spectrum.

\begin{figure}[t]
\resizebox{\hsize}{!}
{\includegraphics{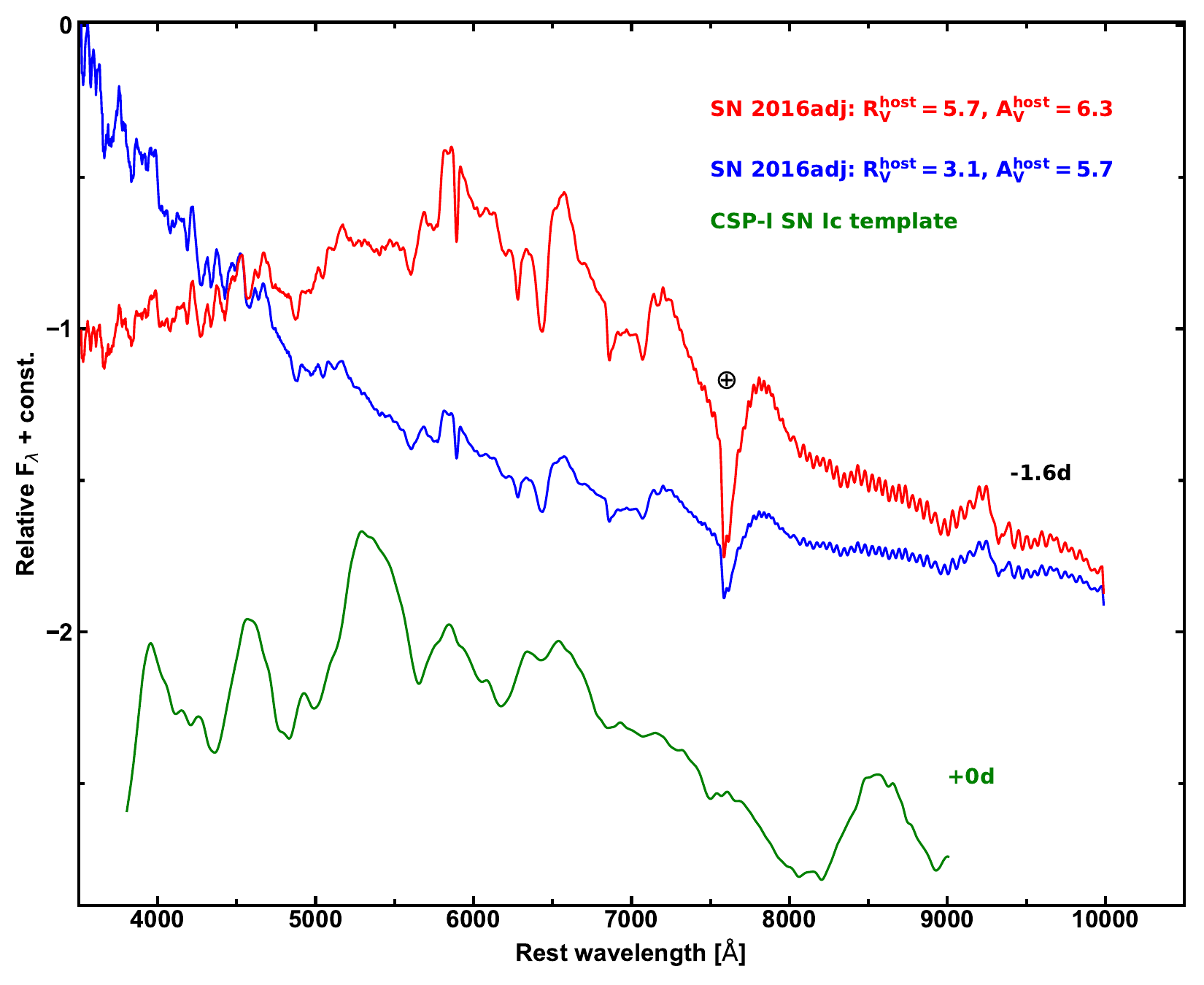}}
\caption{Comparison of the $-1.6$~d spectrum of SN~2016adj de-reddened for Milky Way reddening and also assuming the two different sets of reddening parameters discussed in Sect.~\ref{sec:reddening}  and indicated in the plot. Also plotted in green is the CSP-I SN~Ic spectral template at $+$0~d \citep{Holmbo2023}.}
\label{fig:spectral_comparison_for_reddening}
\end{figure}

To assess the validity of the FTZ99 model characterized by a high $R_{V}^{host}$ value, we  turn to Fig.~\ref{fig:spectral_comparison_for_reddening} which presents a comparison between an intrinsic SN~Ic template spectrum at maximum \citep{Holmbo2023} and SN~2016adj at $-1.6$~d.
The spectrum of SN~2016adj is shown corrected for the two  sets of  reddening parameters, i.e.,   $R_V^{host} = 3.1$ and $A_V^{host} = 5.7$ mag, and $R_V^{host} = 5.7$ and $A_V^{host} = 6.3$ mag. Clearly the spectrum of SN~2016adj corrected for the higher reddening parameters  provides a much better match to the shape of the template spectrum. This gives an additional measure of confidence that the reddening values inferred with a higher $R_V^{host}$ value more  accurately describe the reddening of this system.

 A high  $R_V^{host}$ value as inferred for SN~2016adj is  not without precedent. For example, $R_V$ values on the level of 4--6 have been inferred from observations of the Ophiuchus and  Taurus molecular clouds \citep[e.g.,][]{mathis1990}.  Moreover, \citet{stritzinger2018b} demonstrated that SNe~Ic  are more likely to occur in environments characterized by larger $R_V^{host}$ values  compared to SNe~IIb/Ib, while high values of $R_V$ are expected as   SNe~Ic are preferentially  associated with regions of on-going star formation \citep{anderson2015,Galbany2017,Sextl2023} .

\begin{figure*}
\centering
 \resizebox{0.85\hsize}{!}
{\includegraphics{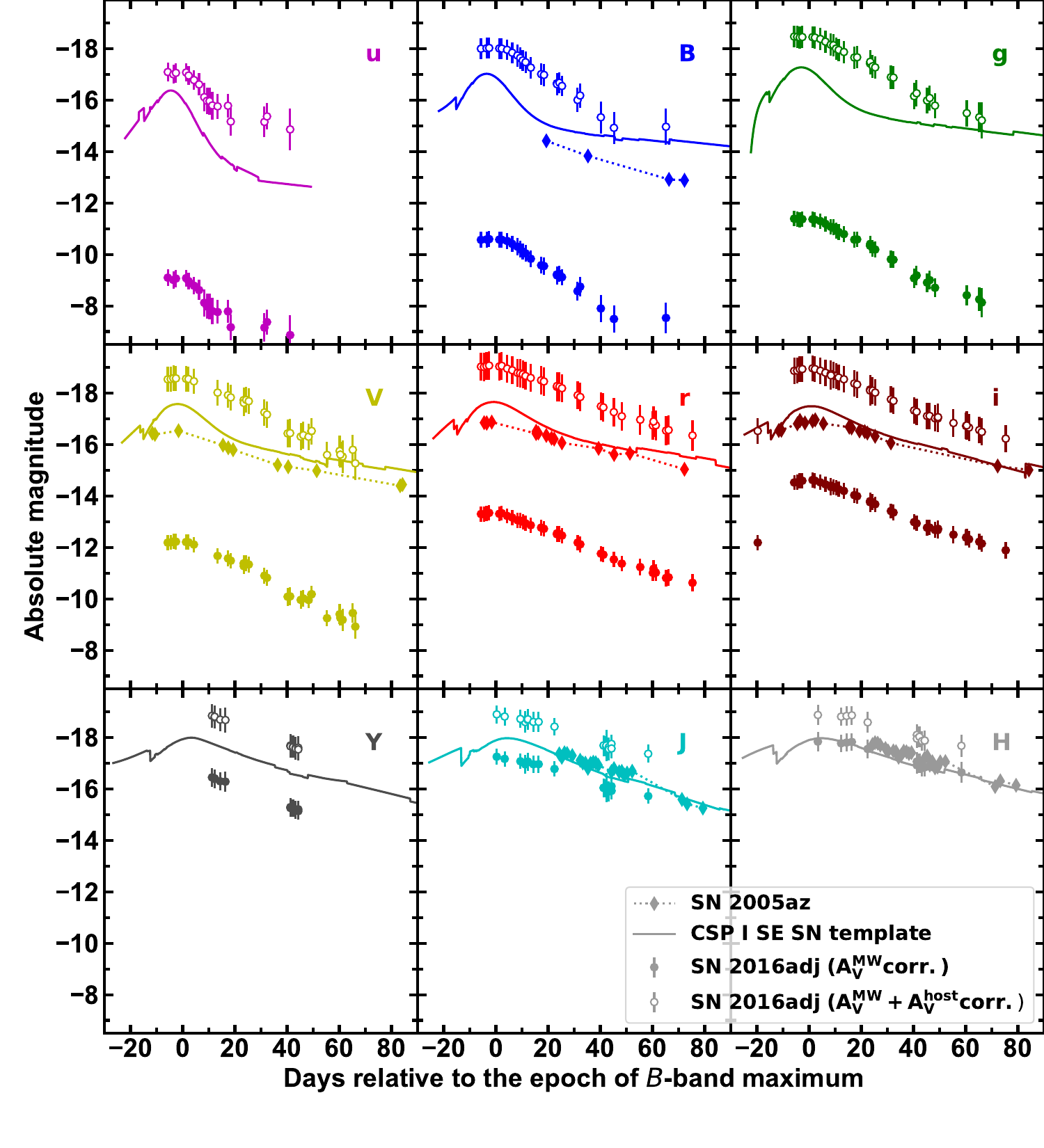}}
\caption{Absolute magnitude light curves of SN 2016adj compared to the CSP-I SE SN template light curve scaled in time and to the average peak absolute magnitude determined from the CSP-I SN~Ic sample. The light curves of SN~2016adj are plotted with and without host-extinction correction as indicated in the legend. The prevalent reddening is visible in the bluer bands with the Milky Way reddening corrected light curves of SN~2016adj being much fainter compared with the average  SN~Ic light curves. Also shown for comparison are the light curves of the type~Ic SN~2005az \citep{bianco2014} which, as  demonstrated in Fig.~\ref{fig:comp_spec_opt}, is spectroscopically similar to SN~2016adj. The error bars accompanying the magnitudes of SN~2016adj account for the  uncertainties in  the distance and the reddening parameters  $R_V^{host}$ and $A_V^{host}$.}
\label{fig:abs_lc}
\end{figure*}

\subsection{Absolute magnitude light curves of SN~2016adj}

With reddening parameters and distance in hand, the absolute magnitude light curves of SN~2016adj are readily computed as shown in Fig.~\ref{fig:abs_lc}. To demonstrate the effects of host-reddening, absolute magnitudes are plotted with (empty circles) and without (filled circles) host-reddening correction. Also shown in the figure are the canonical SE SN light curves from \citet{taddia18csp}, scaled to match the average SN~Ic  peak absolute magnitudes and shifted to match the average time differences between the epochs of maximum in the different filters.
Clearly,  the absolute light curves of SN~2016adj  corrected for host reddening provide a much better match to the template light curves.  
Also plotted  in Fig.~\ref{fig:abs_lc} are  the light curves of the  type~Ic SN~2005az \citep{bianco2014}, which is spectroscopically similar to SN~2016adj.
  
The extinction-corrected light curves of SN 2016adj suggest it reached an absolute peak $B$-band magnitude $M_B \sim -18.0\pm0.1$, which is approximately a magnitude more luminous than the ``average'' SN~Ic. The peak absolute magnitudes for the entire sequence of light curves computed with and without host-extinction corrections are listed in Table~\ref{tab:peak}. Inspection of the other bands red-ward of the $B$ band reveals peak values  between $\sim -18.5$ to $-19.1$ mag.

\begin{figure}[t]
 \resizebox{\hsize}{!}
{\includegraphics{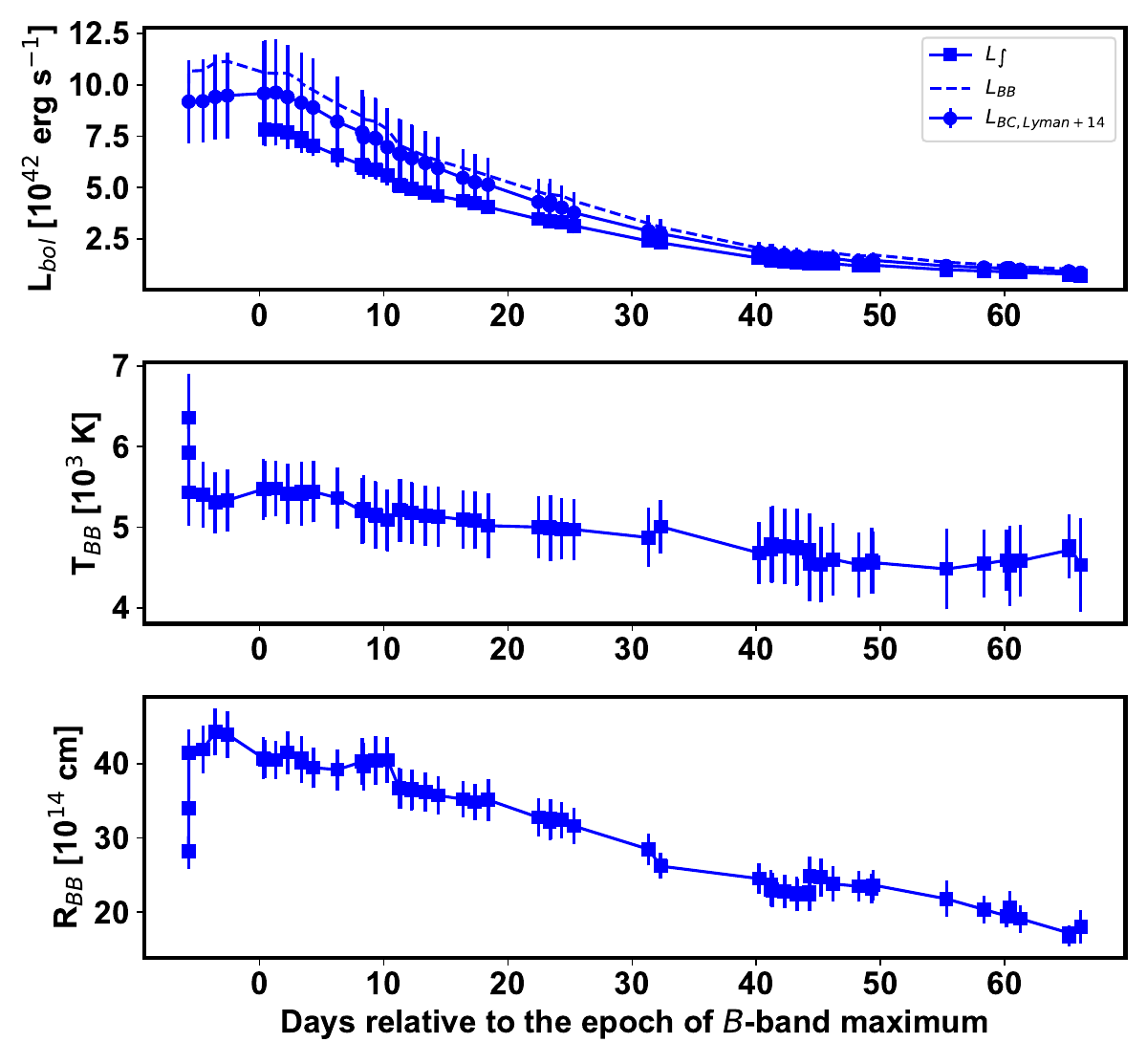}}
\caption{Bolometric light curves \textit{(top)}, black-body temperature profile \textit{(middle)} and black-body radius profile \textit{(bottom)} of SN~2016adj. Using broadband photometry corrected for both Milky Way and host-galaxy reddening, bolometric light curves were computed by: (i) integrating the best-fit Planck function to the SEDs (dashed line denoted by  $L_{BB}$), (ii) trapezoidal integration of the SEDs (filled squares denoted by $L_{\int}$) and (iii) through the combination of the $g$- and $i$-band photometry combined with the bolometric corrections presented by 
\citet[][filled circles denoted L$_{BC, Lyman}$]{Lyman2014}. We adopted our best host-reddening estimate, i.e., a FTZ99 reddening law with $R^{host}_V = 5.7\pm0.7$ and $A^{host}_{V} = 6.3\pm0.2$ mag. The uncertainties reported for $R_{BB}$ and $T_{BB}$ correspond to the BB fit errors, while those  of L$_{BC, Lyman}$ account for the uncertainties of the adopted reddening and distance to Centaurus~A.}
\label{fig:bolo}
\end{figure}

\begin{figure}
 \resizebox{\hsize}{!}
{\includegraphics{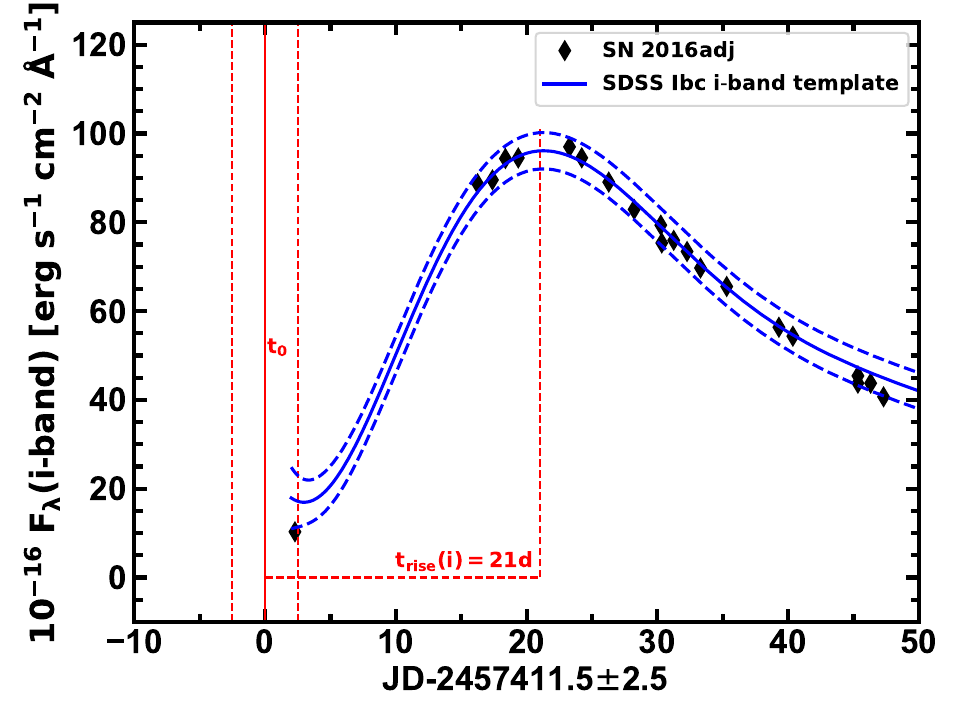}}
\caption{Observed $i$-band photometry of SN~2016adj in flux units (black symbols) vs. days relative to the estimated explosion epoch (JD--2,457,411.5$\pm$2.5).
Overplotted as a blue solid line is the SN~Ib/c $i$-band template light curve presented by \citet{taddia15sdss}, scaled and shifted to match the light curve of SN~2016adj. The blue dashed lines corresponds to a  1-$\sigma$ uncertainty error snake. The epoch of the inferred explosion time, $t_0$, is indicated by the vertical red solid line and assumes an  $i$-band rise time of 21 days  \citep[see][]{taddia15sdss}. The time difference between the inferred explosion date and the first $i$-band epoch is 2.5~d, which  corresponds to the uncertainty of our estimated  rise time.}
\label{fig:plfit}
\end{figure}

\section{Bolometric properties and explosion parameters}
\label{sec:explosionparameters}

\subsection{Pseudo-bolometric light curves and peak luminosity}

Here we describe the techniques used to construct the pseudo-bolometric light curves of SN~2016adj which are used to estimate  key explosion parameters based on semi-analytical models  appropriate for SE~SNe \citep[e.g.,][]{arnett82,kathami2019}.

First, the broad-band optical and NIR  photometry of SN~2016adj  was linearly interpolated in time and  corrected for reddening.  The resulting magnitudes were converted to their respective AB magnitude values using the terms provided in Table~16 of  \citet{krisciunas2017} and then converted to  monochromatic fluxes. These flux points ranging over optical to NIR wavelengths  enable us to construct spectral energy distributions (SEDs) of SN~2016adj for each  epoch of observations and are  corrected for the dust reddening using the FTZ99 reddening law characterized by $R_V^{host}$=5.7  $A_V^{host}$=6.3 mag (see Sect.~\ref{sec:reddening}).

With SEDs extending from the $u$ to the $K_s$ bands in hand, black body (BB) functions were fit to each SED enabling estimates of the BB temperature ($T_{BB}$) and the BB radius ($R_{BB}$) of the underlying emission region. The results of this procedure are plotted in Fig.~\ref{fig:bolo} with the $T_{BB}$ and $R_{BB}$ profiles plotted in the middle and bottom panels, respectively. 
The inferred BB parameters reveal a maximum $R_{BB}$ of about 4$\times$10$^{15}$ cm and $T_{BB}$ values decreasing over time from 6000 to 5000 K.\footnote{\citet{banerjee2018} estimated $T_{BB} = 3680$K, from a BB fit to a combined optical/NIR spectrum. The discrepancy is largely attributed to their  adopted lower  host-galaxy color excess of  $E(B-V)_{host}=0.60$ mag.}

 The pseudo-bolometric luminosity of SN~2016adj was computed following three methods. In the first method each SED was summed over wavelength using trapezoidal integration  to obtain the UltaViolet-Optical-nIR (UVOIR) flux ($F_{UVOIR}$), which was multiplied with  4$\pi$D$^2$ to produce the UVOIR luminosity ($L_{UVOIR}$). In the second method, the best-fit BB Planck functions were  integrated and multiplied  by 4$\pi$D$^2$ to obtain $L_{BB}$, and in the third approach the  $g$- and $i$-band photometry was combined with the bolometric corrections presented by \citet[][]{Lyman2014}. 
 The resulting pseudo-bolometric light curves constructed following these methods are plotted in the top panel of Fig.~\ref{fig:bolo}, while the middle and bottom panel display the temporal evolution of $T_{BB}$ and $R_{BB}$, respectively. 
 
  The pseudo-bolometric light curve computed using the \citeauthor{Lyman2014} corrections indicates the SN reached a peak luminosity $\sim 9.5\pm1.5\times10^{42}$ erg~s$^{-1}$. 
  Here the uncertainty accounts for the error in the adopted distance translating to about 12\% uncertainty in luminosity. An additional 16\% systematic error should be added due to the reddening uncertainty.
  The bolometric light curve computed with the simple integration of the SEDs starts at around maximum when all the bands are covered, and it is obviously fainter than the other two bolometric light curves which include extrapolation corrections for  wavelengths not covered by the observed passbands. The bolometric light curve computed via BB integration is slightly brighter than the one obtained with the \citeauthor{Lyman2014} bolometric corrections, as it tends to overestimate the flux in the UV which does not follow the emission of a BB function.

\begin{figure}[t]
 \resizebox{\hsize}{!}
{\includegraphics[width=15cm]{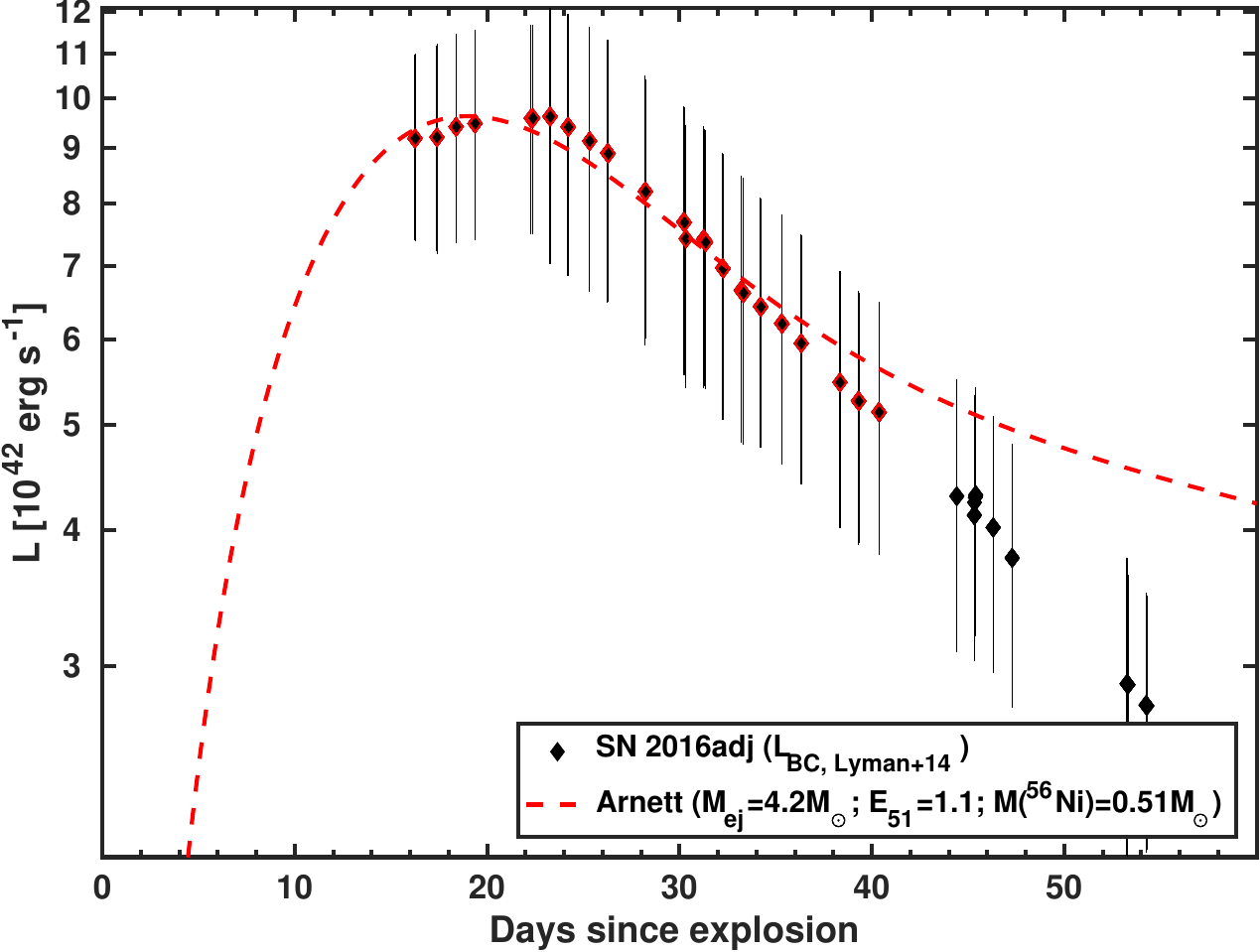}}
\caption{Bolometric light curve of SN 2016adj constructed with the bolometric corrections from \citet{Lyman2014} along with the best-fit \citet{arnett82} model. The model is fit to the light curve points up to $+$40~days past the explosion epoch (black/red diamonds). 
The corresponding model fit yields the explosion parameters reported in the legend.}
\label{fig:bolofit}
\end{figure}

\subsection{Explosion epoch}

An estimate of the explosion epoch of SN~2016adj is needed before we can accurately fit its bolometric light curve  with semi-analytical models. In  Fig.~\ref{fig:plfit} the $i$-band light curve  of SN~2016adj (in flux units) is plotted along  with that of the  SDSS-II SN~Ibc  template $i$-band  light curve  \citep{taddia15sdss}; scaled and shifted to match the peak of SN~2016adj. 
The template reproduces the light curve shape quite well without the need to use a light-curve stretch parameter.

If we assume that the template peak 
corresponds to the peak of the observed
 $i$-band light-curve of SN~2016adj on JD-2,457,432.5$\pm2.5$,  we infer an explosion epoch of JD-2,457,411.5$\pm2.5$.
 The quoted  uncertainty corresponds to the difference between the derived explosion epoch and the first $i$-band detection of SN~2016adj.
The difference between our inferred explosion epoch and the epoch of $i$-band maximum indicates a rise time of $21.0\pm2.5$ days, which is consistent with the average values inferred from the SDSS-II SNe~Ibc sample, i.e., $\sim 20$ days in  the $r$ band and $\sim 21$ days in the $i$ band.

\begin{figure*}[t]
\centering
 \resizebox{0.9\hsize}{!}
{\includegraphics{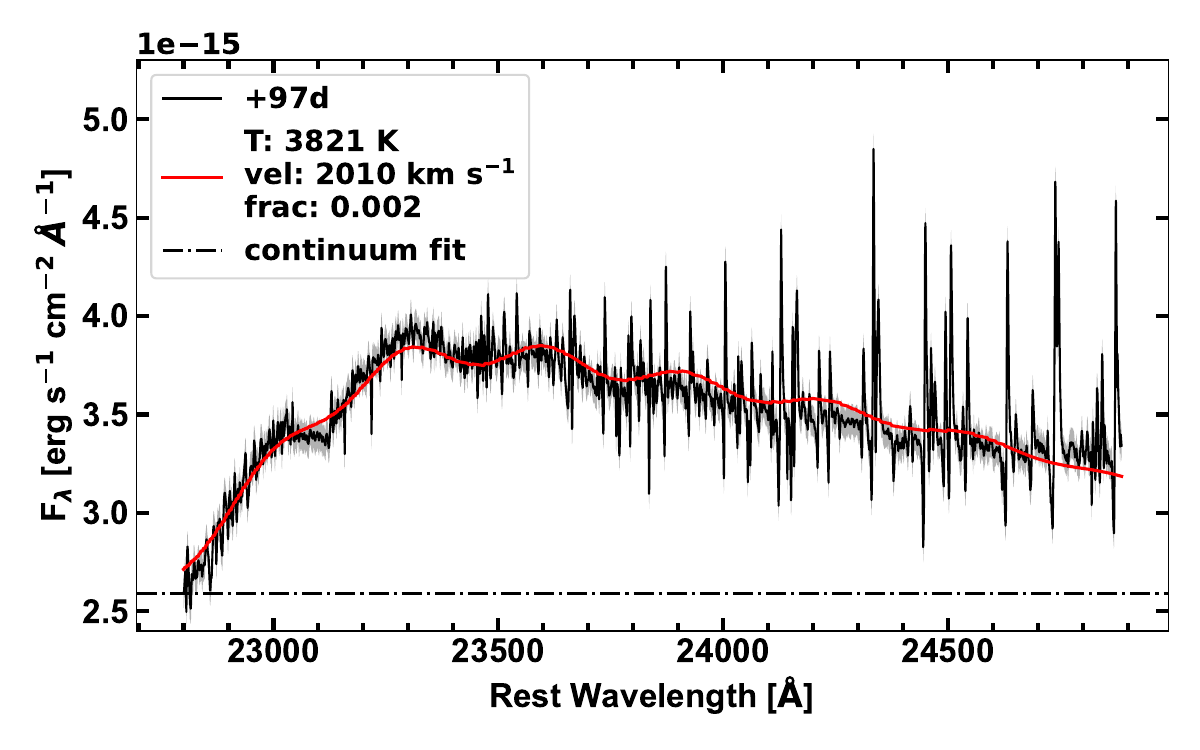}}
\caption{The first CO  overtone feature  in the $+$97~d NIR medium-resolution spectrum of SN~2016adj (black solid line) compared with the best-fit model (red solid line) characterized by the parameters listed in the legend. The black dot-dashed line corresponds to the underlying continuum flux.} 
\label{fig:CO}
\end{figure*}

\subsection{Explosion parameters via Arnett's model}

With the epoch of explosion in hand, the  pseudo-bolometric light curve constructed using the \citeauthor{Lyman2014} bolometric correction is plotted in Fig.~\ref{fig:bolofit}, along with our best-fit  \citeauthor{arnett82} model as determined using the data extending to $+$40~d post maximum.
The  model  assumes the ejecta has a constant density and a constant opacity of 0.07~cm$^2$~g$^{-1}$ \citep[see][]{taddia18csp}, and that the velocity of the bulk of the ejecta at maximum is $6500\pm2500$ km~s$^{-1}$, as inferred from the optical \ion{C}{ii}  features (see Fig.~\ref{fig:optvel}). 

The best-fit Arnett model plotted in Fig.~\ref{fig:bolofit} corresponds to  an  ejecta mass $M_{ej}~\sim~4.2^{+1.6}_{-1.6}$~$M_{\odot}$, an explosion kinetic energy $E{_K}~\sim~1.1^{+0.4}_{-0.4}\times10^{51}$~erg~s$^{-1}$, and a $^{56}$Ni mass $\sim~0.51^{+0.03}_{-0.03}~M_{\odot}$.
Here the  uncertainties account for the errors on the  ejecta velocity and the inferred explosion epoch. The ejecta velocity  error dominates the uncertainties on $E_K$ and $M_{ej}$, while the explosion epoch error mainly contributes to the $^{56}$Ni estimate. 
In addition to these uncertainties, we add an error of  $\pm0.1$~$M_{\odot}$  to $M_{ej}$  to account for the fitting error, which  is negligible to the  $E_K$ and $^{56}$Ni estimates. Furthermore,   related to the uncertainties in the adopted distance (12\%), the reddening parameters (16\%),  and possible (10\%) contamination from an underlying light echo \citep[see][]{Stritzinger2022},  we tack onto the $^{56}$Ni uncertainty  value  $\pm0.06$~$M_{\odot}$, $\pm0.08$~$M_{\odot}$, and $\pm0.05$~$M_{\odot}$, respectively.  

With caveats related to the high uncertainty in the peak bolometric luminosity of SN~2016adj and that the basis of Arnett's Rule  relies on a number of  assumptions, we  briefly compare our results for SN~2016adj to those in the literature.
 \citet[][see their Table~7]{Barbarino2021} compares   average explosion parameters estimates of the iPTF SN~Ic sample to those found by other authors who consider various size SN~Ic samples \citep{Drout2011,Lyman2016,Prentice2016,taddia18csp,Prentice2019}.
In general, the explosion parameters computed for SN~2016adj are not radically different than the average values listed by \citet{Barbarino2021}. The KE and ejecta mass estimates are within the range of the average values found by most of the other works, while the $^{56}$Ni mass of 0.5 $M_{\sun}$ is a factor of two more than the typically average value of 0.25 $M_{\sun}$. Though it is worth mentioning the SN~Ic sample of  \citet{Barbarino2021} contained a number of objects with $^{56}$Ni masses of $\sim 0.5$ $M_{\odot}$.

When considering all of the uncertainties, the minimum of the $^{56}$Ni mass confidence interval of  $\approx 0.3 M_{\odot}$  is relatively  high compared to the typical sample median values (see Table~9 in \citealt{taddia18csp}).  However, $^{56}$Ni masses in excess of 0.3~$M_{\odot}$ for SNe~Ic are not without precedent.   \citet{anderson2019} presented a meta-analysis regarding $^{56}$Ni mass estimates for both  hydrogen-rich and  SE SNe, finding a median SN~Ic  $^{56}$Ni mass of $0.16^{+0.84}_{-0.03}~M_{\odot}$, with  a handful of objects,  for example,  PTF~12gzk \citep{Prentice2016}, iPTF15dtg \citep{Taddia2016}, and SN~2011bm \citep{Valenti2012,Lyman2016,Prentice2016} with values exceeding 0.3~$M_{\sun}$. 
We conclude here by pointing out that within standard   CC SNe  simulations  it is difficult to produce $\gtrsim 0.2$ $M_{\odot}$ of $^{56}$Ni \citep[see, e.g.,][and references therein]{Mueller2016},  implying values inferred from  Arnett's model are overestimated and/or objects with high  $^{56}$Ni estimates have an additional energy source contributing to their bolometric emission.

\section{Progenitor analysis using pre-explosion imaging}
\label{sec:progenitoranalysis}

We searched the ESO and HST archives for pre-explosion images of Centaurus~A containing the position of SN~2016adj. A full summary of the analysis is presented in Appendix~\ref{sec:progenitoranalysis-appendix}. In short, no source is detected at the position of SN~2016adj in a series of HST (+WFC3; Wide Field Camera 3) and VLT (+NACO; Nasmyth Adaptive Optics System (NAOS) Near-Infrared Imager and Spectrograph (CONICA)) pre-explosion images. Nevertheless, two of the HST images were of high enough quality that limiting magnitudes could be determined, though these do suffer from the uncertainties related to the estimated dust reddening properties.  As described in Sect.~\ref{sec:upperlimits}, we find limiting apparent magnitudes  of $m_{F814W} > 26.4$ mag and $m_{F545M} = 25.9$ mag.  Similarly, the NACO images allow us to place limits of $J>21.6$, $H>20.9$ and $K_s>21.1$~mag on the progenitor. Unfortunately neither the optical limits from HST not the IR limits from NACO allow us to place any meaningful constraint on the progenitor luminosity. The vast majority of known WR stars are fainter than $-5.7$ in $F814W$ \citep{Eldridge13}, while the dusty WR stars that are bright in the IR \citep[e.g.][]{Rate20} would go similarly undetected in our data ($K>-7.5$).

\section{CO emission}
\label{sec:COmodeling}

\subsection{The CO first overtone}

The first CO  overtone feature in the $+$41~d NIR spectrum  of SN~2016adj is among the earliest detection yet documented in a SN. For comparison, the feature was detected in the type~Ic SN~2021krf by $+$43~d, in the type~Ic SN~2013ge by $+$48~d  \citep{drout2016}, while the type~Ic SN~2007gr showed signatures of CO  by $+$70~d \citep{hunter09}. In addition, in the $\sim 14$  mostly SNe~II with  CO emission, the CO signature  typically emerged months later  \citep[see][their Table~4]{banerjee2018}.

\subsection{CO model fitting}

We now turn to comparison of a grid of CO emission models with the first CO overtone feature in the +97~d and +159~d  medium-resolution NIR spectra of SN~2016adj. This enables the estimate of key physical parameters of the CO gas including: the temperature ($T_{CO}$), the velocity ($v_{CO}$), the ratio of CO$^{+}$ to CO (aka frac), and a lower limit on the CO mass. First, a grid of CO emission models was computed using a recently developed module contained within  the non-LTE HYDrodynamical RAdiation code \texttt{HYDRA} \citep[see][and references therein]{hoeflich2003,hoeflich2009,Hristov2021,Hoeflich2021}, which enables the determination of the  vibrational  transition opacities for CO and SiO gas over a range of parameter space. 
 
To determine the best-fit model(s) appropriate for the CO emission in SN~2016adj,  a six parameter function was created and used to determine the best-fit. Fitting was performed following a Markov chain Monte Carlo (MCMC) calculation making use of the Python \text{emcee} package. The model parameters consist of: (i) an amplitude ($A$) parameter proportional to the CO mass ($M_{CO}$), (ii) an underlying continuum fit parameter ($b$) extending between 22,850~\AA\ to 25,000~\AA, (iii) $T_{CO}$, (iv)  $v_{CO}$,  (v) frac, and (vi) a velocity parameter ($z$). Parameter $b$  takes the functional form of $\lambda^{-2}$ and stems from  the continuum radiation being formed by  free-free emission \citep{Rybicki1979}. The $z$ parameter accounts for an arbitrary shift between $v_{CO}$ and the grid of models, although the origin of such velocity offset is unclear. Possibilities include the offset of the progenitors orbital velocity to its host's systemic velocity, a peculiar velocity of the progenitor within its host, or if the progenitor belongs to a double star system, the binary orbital velocity. 

 Figure~\ref{fig:CO} displays the first CO overtone feature in the color-corrected and de-reddened $+97$~d spectrum of SN~2016adj. Over-plotted is the best-fit MCMC model and the underlying continuum flux extrapolation.
 For completeness, the  MCMC corner plot  containing the posterior probability distributions of the model-fit  parameters is presented in Fig.~\ref{fig:cornerplot}. 
 Overall the model agrees well with the data and corresponds to model-fit parameters of $A= 1.43 \pm 0.12$, $b = 2.59 \pm 0.11$ $T_{CO} = 3830\pm170$~K, $v_{CO} = 2010\pm270$ km~s$^{-1}$,  frac $= 0.002$, $z = 0.00145\pm0.00091$.
The low inferred value of frac suggests the emission region lacks gamma-rays, non-thermal electrons, and/or  He$^{+}$. The underlying CO mass is estimated following:   For an optically thin emission region the emissivity per unit mass is given by    $\eta_\lambda =  \kappa_\lambda(\rho,v,T)*B_\lambda(T)$. 
Thus the total flux observed will be given by 
$F_{\lambda} = M_{CO} \eta_{\lambda} / 4 \pi D^2$.  
Using the fitting results, we find $M_{CO} \approx 2.5 \times 10^{-3} M_\odot$. This value is an upper limit  as radiative transfer effects will reduce the emitted luminosity by a factor $< 1$, however, a non-isothermal emission region could increase the emitted flux.


As previously mentioned, \citet{banerjee2018} reported on CO emission in SN~2016adj present in their  NIR spectral time series. Using a model developed by \citet{das2009} with no details on how their fitting was accomplished, they estimated from a $+$64~d spectrum the model parameters: $T_{CO} = 4600\pm400$ K, $v_{CO} = 3400\pm150$ km~s$^{-1}$, and $M_{CO}= 2.1 \pm 0.4 \times 10^{-4} M_{\sun}$. 

\section{Late-time Hydrogen Features}
\label{sec:progenitorscenario}

\begin{figure*}
\centering
\resizebox{0.95\hsize}{!}
{\includegraphics{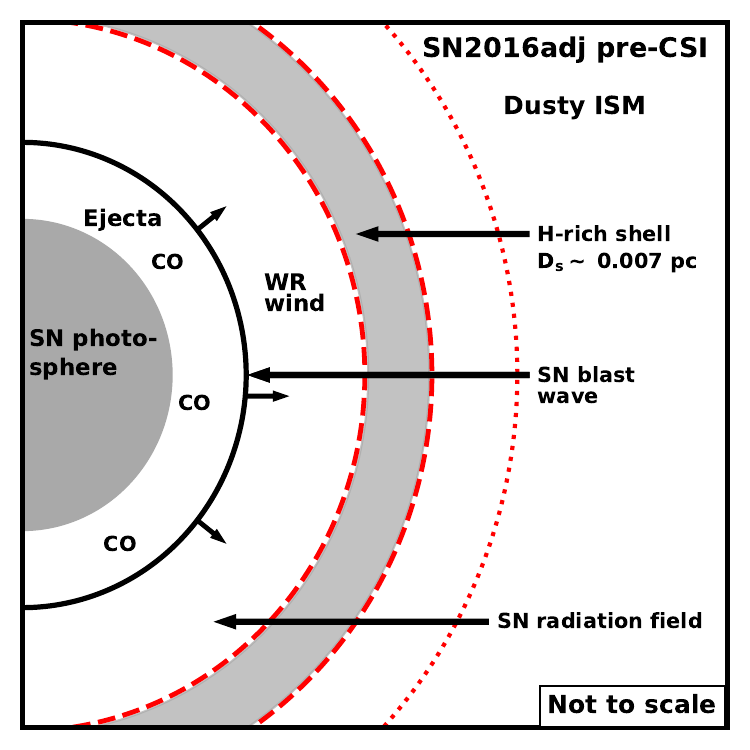}
\includegraphics{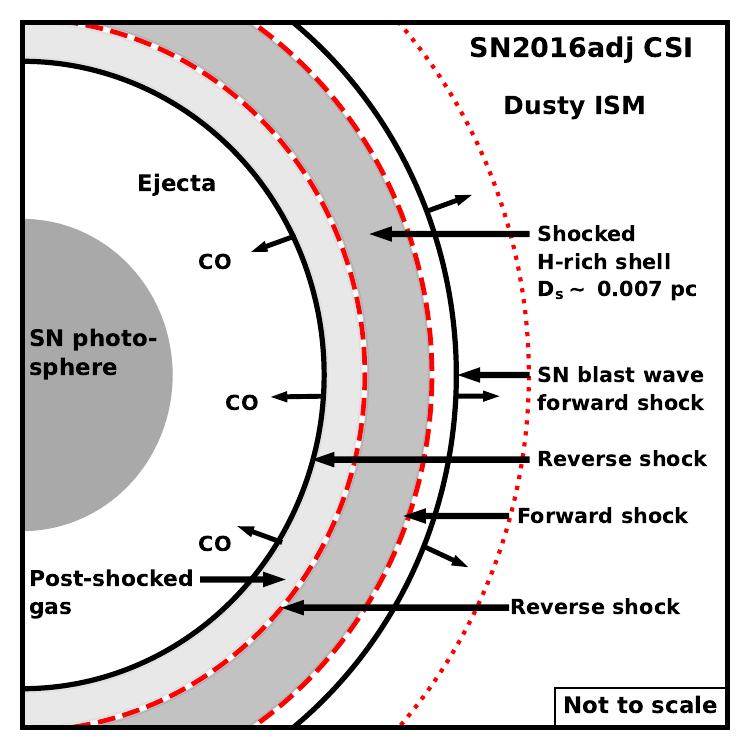}}
\caption{Cartoon  of SN~2016adj experiencing  circumstellar interaction (CSI). 
\textit{(Left)} A H-rich shell located within $D_{s} \sim 0.007$ pc of an infant SN delimited with a cavity formed by its progenitor WR wind.  The expanding SN ejecta is indicated, along with  the CO  freshly synthesized within the wake of the SN ejecta.  
\textit{Right:} Schematic after the shell has been shocked by the expanding SN ejecta. CSI produces both forward and reverse shocks, and the ionization of hydrogen.}
\label{fig:cartoon}
\end{figure*}

\subsection{Hydrogen features from circumstellar interaction?}

The  H features in the post maximum NIR spectra of SN~2016adj  with  $-v_{abs} \approx 1500-3000$ km~s$^{-1}$ and $v_{FWHM} \sim 1000$ km~s$^{-1}$, coupled with the time scale of their appearance suggests they arise from a narrow decoupling region. The density of the emitting gas can be estimated through the use of line ratios in the classical emission-line nebular case \citep{osterbrock2006}. Using the emission-line flux values  determined from the post-maximum spectra presented in Fig.~\ref{fig:Hlines} and listed in  Table~\ref{tab:H1fluxes}, H emission-line ratios are computed relative to 
Pa-$\beta$. The resulting line ratios are listed in Table~\ref{tab:lineratios}. 
The inferred line ratios  are inconsistent with 
pure recombination values (i.e., $P_\gamma/P_\beta \sim 0.6$) for a gas with a reasonable range of temperature and  in pure local thermal equilibrium (LTE). In fact both the Paschen and Bracket line ratios indicate these features are associated with an optically thick gas in non-LTE (NLTE) with electron densities  of  $n_e \gtrsim 10^{10}-10^{11}$~cm$^{-3}$ \citep[e.g.,][]{Lynch2000}.

Turning to temporal emergence of the H features in the NIR spectra of SN~2016adj, adopting the phase of the spectrum plotted  in Fig.~\ref{fig:Hlines}, photoionization of the shell by the SN shock breakout radiation field, implies a distance (hereafter $D_s$) between the progenitor and a circumstellar  (CS) shell of $\sim 0.08$ pc. However, if the shell was located at this  $D_s$ then the H features should have appeared earlier.  We therefore assume in the following that the ionization of the shell is caused by CS interaction (hereafter CSI) with the SN ejecta \citep[e.g.,][]{Simon1966,Chevalier1994}. 
Adopting a velocity  for the fastest portion of the SN ejecta to be  25,000~km~s$^{-1}$,  CSI implies  $D_s \sim 0.007$ pc. Adopting  a Wolf-Rayet (WR) wind velocity of 1,000~km~s$^{-1}$ indicates that the purported  shell formed around a decade prior to the supernova.  

A confined, optically thick emission region could be associated with a shell located within the CS environment of a WR star \citep[see][for a review]{Crowther2007}. 
Often located at or near the center of a complex ring nebula \citep[e.g.,][]{smith1967}, WR stars  exhibit significant diversity \citep[e.g.,][]{Grosdidier1998,Marchenko2010} linked to their robust line-driven winds and other potential mass-loss mechanisms \citep[see][]{Puls2008}.

The morphology and sub-structures of known WR associated nebula led to the establishment of a classification system \citep[see][and references therein]{Chu1981}. Firstly, R-type nebula  are photoionized \ion{H}{ii} regions in the nearby vicinity of an WR star.    
Bona fide associations between WR stars and nebulae   include wind-driven  W-type WR stars \citep[e.g.,][]{Johnson1965,Avedisova1972,Cappa2005} 
and post common envelope evolution  E-type WR stars \citep[e.g.,][]{Podsiadlowski2010,JP2020,Schroder2020}. 
W-type WR stars exhibit a variety of CS shell(s) and/or wind blown CS bubbles, which can form by a variety of processes.

 CS shells may form via wind-wind interaction \citep[e.g.,][]{Bransford1999}, for example, putative luminous blue variable (LBV)-like mass ejections  
\citep[see][]{Smith2014,Vink2017}, or binary interaction with a companion \citep[e.g.,][]{Yoon2017}.
On the other hand, W-type WR stars associated with colliding-wind binary systems (e.g., WR~140) and pin wheel dust nebulae form a shell from wind-wind interaction between two massive stars  during  their periastron  passage  \citep[e.g.,][]{Tuthill2006,Williams2021}.
In the case of E-type WR stars such as WR~124 located within the nebula M1-67, its complex 3-D structure consisting of bipolar outflows, wind-blown bubbles, and a toroidal structure  is aligned with expectations of a binary system that experienced post common envelope evolution \citep[e.g.,][]{Chu1981,Zavala2022}.  

 The $D_s$ values inferred for a CS shell surrounding SN~2016adj  are inconsistent with the order of magnitude higher values reported in the literature for resolved  shells produced by wind-wind interaction around WR stars that  exhibit  $D_s$ values of $\gtrsim 10$ pc \citep[see][their Tables 1 and 2]{Bransford1999}. 
 In other words, as the early SN spectra of SN~2016adj are devoid of H features and  the appearance time scale of H in the post maximum spectra  could be explained by  shell  formation from binary interaction. This could take the form of  either Roche Lobe overflow,   or an LBV-like eruption from a companion star  \citep[see, e.g.,][]{Kuncarayakti2018}.

Figure~\ref{fig:cartoon} contains a cartoon schematic of a SN~Ic interacting with a CS shell located within $D_s \sim 0.007$ pc from the SN progenitor. The left panel is a snapshot post SN explosion and includes the formation of CO within the wake of the SN ejecta, the WR wind of the progenitor sweeping up CS material, and the H-rich shell. The right panel  is a  snapshot after the CS shell is shocked by the expanding SN ejecta that generates X-rays which then photoionize the shell. The shell  consists of a forward shock, a  reverse shock, and post shocked gas. 

If the origins of the CSM are from a companion star then one would naturally not expect to see any He features. Alternatively, if the CSM originated from the progenitor He could remain hidden due to its high-ionization potential.

\subsection{Incidences of H signatures in SNe~Ic}

 The nearby and well-observed type~IIb SN~1993J was the first SE SN to exhibit strong evidence of CSI at late phases in the form of Balmer emission features \citep{Filippenko1994,Patat1995,Fransson1996,Houck1996,Chevalier1997,Matheson2000}. Over the past decade a growing number of SE SNe  have been recognized to exhibit (typically late phase) signatures of CSI.
In  cases of SNe~IIb/Ib, these signatures take the form of either CSI emission features, high-ionization coronal features, broadband emission excesses, and/or X-ray emission
\citep[e.g.,][]{Ben-Ami2014,Morales-Garoffolo2014,Maeda2015,Milisavljevic2015,Margutti2017,Mauerhan2018,Bostroem2020,Chandra2020,Kilpatrick2021,Zenati2022,Maeda2023}.  

Focusing on SNe~Ic with claimed CSI produced H signatures, \citet{Roy2016} attributed an excess of flux at peak  and the presence of a secondary  post-maximum light-curve peak in SN~2012aa to  CSI between SN ejecta and a massive H-shell, despite only a tentative detection of H$\alpha$. 
On the other hand, SN~2017dio  was the first SN~Ic to exhibit prominent hydrogen and helium emission features at early times suggested  to be produced from CSI \citep[see][]{Kuncarayakti2018}. 
CSI driven H emission features and other coronal lines  have also been observed in the late phase optical/NIR spectrum of the superluminous type~Ic SN~2017ens  \citep{Chen2018}, and in the late phase optical spectrum of the bright type~Ic-BL SN~2018ijp \citep{Tartaglia2021}. 

Recently,  the late phase observations of the SN~2021ocs were shown to exhibit features associated with intermediate-mass elements typically not seen in SE SNe \citep{Kuncarayakti2022}. \citeauthor{Kuncarayakti2022} attributed these features to allowed and forbidden transitions of O and Mg which get illuminated by CSI \citep{Kuncarayakti2022}. 
Most recently, the type~Ic-BL SN~2022xxf containing a double hump light curve with each hump reaching the same brightness has also been suggested to experience significant CSI \citep{Kuncarayakti2023}.
Finally, \citet{Ravi2023} just published observations on the type~Ic SN~2021krf  exhibiting an excess of late-phase emission relative to expectations of energy deposition being solely due to $^{56}$Co decay.  However, due to a lack of CSI spectral features of H, He and/or coronal lines,  \citeauthor{Ravi2023} considered the possibilities of an IR echo associated with either pre-existing and/or newly  synthesized  dust, or emission linked to magnetic dipole
radiation from a newly formed   neutron star.

 Clearly, the time scales and strength of the CSI signatures observed to date in SE SN progenitors  suggests a diversity in pre-SN mass-loss histories and underlying progenitor systems. If CSM originates from erupted mass loss of the progenitor star in the lead-up  of going core collapse \citep{Dessart2010,Owocki2019}, \citet{Tsuna2023} suggest fallback could be  suppressed by the star's radiation pressure. Depending on the amount of ejected material and the radiation pressure of the progenitor star, a cavity devoid of CSM will form around the star. This could account for CSI to naturally occur weeks to months after explosion.  A similar effect would also be expected if CSM originated from a companion.

\section{Conclusion}
\label{sec:conclusion}

We have presented a detailed analysis of the carbon-rich type~Ic SN~2016adj located in the iconic dust lane of the famous early type galaxy Centaurus~A. 
Unsurprisingly, SN~2016adj is found to suffer significant host reddening, preventing an accurate estimate of its peak luminosity and explosion parameters. 
However, our unique post maximum NIR spectroscopic time-series reveals two interesting aspects.  
The  CO first overtone feature appears by $+$41~d making this the earliest detection in a SN~Ic. 
Modeling of the CO bandhead  as captured by two medium-resolution NIR spectra provides an upper limit on the CO mass of $M_{CO} \sim 10^{-3} M_{\sun}$.
Secondly, the NIR spectra document the emergence of a handful of P~Cygni spectral features that we attribute to H Paschen and Bracket line transitions. These features could arise from CSI between rapidly expanding SN ejecta and a H-rich shell.  Such a shell could originate from mass loss experience by the progenitor star in the decades prior to undergoing core-collapse, or alternatively from a companion star which experienced an episode of mass loss.
Whatever the origin, the post-maximum H features  present in the NIR spectra of SN~2016adj place it among a growing group of  SNe~Ic, SNe~Ic-BL and SLSN-Ic   in the literature displaying  signatures of CSI involving H-rich material.

A key take away from this study  is that medium-resolution NIR spectroscopy offers significant potential  to further unravel the pre-SN mass loss history of SE SNe progenitor systems. 
We therefore recommend future SE SN observational campaigns seek to obtain such data out to late phases. Such observations could serve as a forensic tool to  map out the  heterogeneous nature of SE SNe mass-loss histories, CSI, and perhaps, a means to disentangle binary versus single star progenitor systems. Fortunately new facilities such as  SOXS (son of x-shooter) and NTE (NOT transient explorer) will soon come online, making such data commonplace. 

\begin{acknowledgements}
We appreciate constructive discussions with Subhash Bose and N.~B.~Suntzeff.
The work of the CSP-II has been generously supported by the National Science Foundation under grants AST-1008343, AST-1613426, AST-1613455, and AST-1613472, and by the Danish Research Foundation via a Sapere Aude Fellowship.
 M.D.S.  is supported by grants from the Independent Research Fund Denmark (IRFD; 8021-00170B and 10.46540/2032-00022B). L.G. acknowledges financial support from the Spanish Ministerio de Ciencia e Innovaci\'on (MCIN), the Agencia Estatal de Investigaci\'on (AEI) 10.13039/501100011033, and the European Social Fund (ESF) ``Investing in your future" under the 2019 Ram\'on y Cajal program RYC2019-027683-I and the PID2020-115253GA-I00 HOSTFLOWS project, from Centro Superior de Investigaciones Cient\'ificas (CSIC) under the PIE project 20215AT016, and the program Unidad de Excelencia Mar\'ia de Maeztu CEX2020-001058-M.
This work was funded in part by ANID, Millennium Science Initiative, ICN12\_009.
MN is supported by the European Research Council (ERC) under the European Union’s Horizon 2020 research and innovation programme (grant agreement No.~948381) and by UK Space Agency Grant No.~ST/Y000692/1.
 This research has made use of the NASA/IPAC Extragalactic Database (NED), which is operated by the Jet Propulsion Laboratory, California Institute of Technology, under contract with the National Aeronautics and Space Administration.

\textit{Facilities.} Based on observations made with facilities at the Las Campanas Observatory including the Swope telescope, the du Pont telescope,  and the Magellan telescopes. Data was also obtained   with the ESO-La Silla NTT observatory (ESO programme IDs 191.D-0935 and 197.D-1075), the ESO-Paranal VLT  (Programme IDs 094.B-0298 and 098.D-0540), and the ESO-Paranal VST (Program ID 60.A-9800). Finally, some data presented were obtained with the   \textit{Swift} Space Telescope.

\textit{Software.} This research made use of \texttt{Astropy}  \citep{price2018} and \texttt{Photutils} \citep{bradley2020}.  Photometry of SN~2016adj was computing using  the Aarhus-Barcelona FLOWS projects automated photometry pipeline available for download on
\href{https://github.com/SNflows}{github}. \texttt{Hydra} has been used to calculate the CO model spectra, and the gamma-ray transport \citep{h90,hoeflich2003,hoeflich2009,Hristov2021}.

\end{acknowledgements}

\bibliographystyle{aa}
\bibliography{v1.bib}

\begin{thebibliography}{150}
\expandafter\ifx\csname natexlab\endcsname\relax\def\natexlab#1{#1}\fi

\bibitem[{{Anderson}(2019)}]{anderson2019}
{Anderson}, J.~P. 2019, \aap, 628, A7

\bibitem[{{Anderson} {et~al.}(2015){Anderson}, {James}, {Habergham}, {Galbany},
  \& {Kuncarayakti}}]{anderson2015}
{Anderson}, J.~P., {James}, P.~A., {Habergham}, S.~M., {Galbany}, L., \&
  {Kuncarayakti}, H. 2015, \pasa, 32, e019

\bibitem[{{Arnaboldi} {et~al.}(1998){Arnaboldi}, {Capaccioli}, {Mancini},
  {Rafanelli}, {Scaramella}, {Sedmak}, \& {Vettolani}}]{arnaboldi1998}
{Arnaboldi}, M., {Capaccioli}, M., {Mancini}, D., {et~al.} 1998, The Messenger,
  93, 30

\bibitem[{{Arnett}(1982)}]{arnett82}
{Arnett}, W.~D. 1982, \apj, 253, 785

\bibitem[{{Ashall} {et~al.}(2016){Ashall}, {Mazzali}, {Pian}, \&
  {James}}]{ashall2016}
{Ashall}, C., {Mazzali}, P.~A., {Pian}, E., \& {James}, P.~A. 2016, \mnras,
  463, 1891

\bibitem[{{Asplund} {et~al.}(2009){Asplund}, {Grevesse}, {Sauval}, \&
  {Scott}}]{asplund2009}
{Asplund}, M., {Grevesse}, N., {Sauval}, A.~J., \& {Scott}, P. 2009, \araa, 47,
  481

\bibitem[{{Avedisova}(1972)}]{Avedisova1972}
{Avedisova}, V.~S. 1972, \sovast, 15, 708

\bibitem[{{Bacon} {et~al.}(2014){Bacon}, {Vernet}, {Borisova}, {Bouch{\'e}},
  {Brinchmann}, {Carollo}, {Carton}, {Caruana}, {Cerda}, {Contini}, {Franx},
  {Girard}, {Guerou}, {Haddad}, {Hau}, {Herenz}, {Herrera}, {Husemann},
  {Husser}, {Jarno}, {Kamann}, {Krajnovic}, {Lilly}, {Mainieri}, {Martinsson},
  {Palsa}, {Patricio}, {P{\'e}contal}, {Pello}, {Piqueras}, {Richard},
  {Sandin}, {Schroetter}, {Selman}, {Shirazi}, {Smette}, {Soto}, {Streicher},
  {Urrutia}, {Weilbacher}, {Wisotzki}, \& {Zins}}]{bacon2014}
{Bacon}, R., {Vernet}, J., {Borisova}, E., {et~al.} 2014, The Messenger, 157,
  13

\bibitem[{{Banerjee} {et~al.}(2016){Banerjee}, {Connelley}, Tokunaga, {Gehrz},
  {Geballe}, {Evans}, {Spyromilio}, {Joshi}, {Ashok}, \&
  {Srivastava}}]{banerjee2016}
{Banerjee}, D.~P.~K., {Connelley}, M.~S., Tokunaga, A.~T., {et~al.} 2016, The
  Astronomer's Telegram, 8976, 1

\bibitem[{{Banerjee} {et~al.}(2018){Banerjee}, {Joshi}, {Evans}, {Srivastava},
  {Ashok}, {Gehrz}, {Connelley}, {Geballe}, {Spyromilio}, {Rho}, \&
  {Roy}}]{banerjee2018}
{Banerjee}, D.~P.~K., {Joshi}, V., {Evans}, A., {et~al.} 2018, \mnras, 481, 806

\bibitem[{{Barbarino} {et~al.}(2021){Barbarino}, {Sollerman}, {Taddia},
  {Fremling}, {Karamehmetoglu}, {Arcavi}, {Gal-Yam}, {Laher}, {Schulze},
  {Wozniak}, \& {Yan}}]{Barbarino2021}
{Barbarino}, C., {Sollerman}, J., {Taddia}, F., {et~al.} 2021, \aap, 651, A81

\bibitem[{{Ben-Ami} {et~al.}(2014){Ben-Ami}, {Gal-Yam}, {Mazzali}, {Gnat},
  {Modjaz}, {Rabinak}, {Sullivan}, {Bildsten}, {Poznanski}, {Yaron}, {Arcavi},
  {Bloom}, {Horesh}, {Kasliwal}, {Kulkarni}, {Nugent}, {Ofek}, {Perley},
  {Quimby}, \& {Xu}}]{Ben-Ami2014}
{Ben-Ami}, S., {Gal-Yam}, A., {Mazzali}, P.~A., {et~al.} 2014, \apj, 785, 37

\bibitem[{{Bianco} {et~al.}(2014){Bianco}, {Modjaz}, {Hicken}, {Friedman},
  {Kirshner}, {Bloom}, {Challis}, {Marion}, {Wood-Vasey}, \&
  {Rest}}]{bianco2014}
{Bianco}, F.~B., {Modjaz}, M., {Hicken}, M., {et~al.} 2014, \apjs, 213, 19

\bibitem[{{Bostroem} {et~al.}(2020){Bostroem}, {Valenti}, {Sand}, {Andrews},
  {Van Dyk}, {Galbany}, {Pooley}, {Amaro}, {Smith}, {Yang}, {Anupama},
  {Arcavi}, {Baron}, {Brown}, {Burke}, {Cartier}, {Hiramatsu}, {Dastidar},
  {DerKacy}, {Dong}, {Egami}, {Ertel}, {Filippenko}, {Fox}, {Haislip},
  {Hosseinzadeh}, {Howell}, {Gangopadhyay}, {Jha}, {Kouprianov}, {Kumar},
  {Lundquist}, {Milisavljevic}, {McCully}, {Milne}, {Misra}, {Reichart},
  {Sahu}, {Sai}, {Singh}, {Smith}, {Vinko}, {Wang}, {Wang}, {Wheeler},
  {Williams}, {Wyatt}, {Zhang}, \& {Zhang}}]{Bostroem2020}
{Bostroem}, K.~A., {Valenti}, S., {Sand}, D.~J., {et~al.} 2020, \apj, 895, 31

\bibitem[{{Bradley} {et~al.}(2020){Bradley}, {Sip{\H{o}}cz}, {Robitaille},
  {Tollerud}, {Vin{\'\i}cius}, {Deil}, {Barbary}, {Wilson}, {Busko},
  {G{\"u}nther}, {Cara}, {Conseil}, {Bostroem}, {Droettboom}, {Bray}, {Andersen
  Bratholm}, {Lim}, {Barentsen}, {Craig}, {Pascual}, {Perren}, {Greco},
  {Donath}, {De Val-Borro}, {Kerzendorf}, {Bach}, {Weaver}, {D'Eugenio},
  {Souchereau}, \& {Ferreira}}]{bradley2020}
{Bradley}, L., {Sip{\H{o}}cz}, B., {Robitaille}, T., {et~al.} 2020,
  {astropy/photutils: 1.0.0}, Zenodo

\bibitem[{{Bransford} {et~al.}(1999){Bransford}, {Thilker}, {Walterbos}, \&
  {King}}]{Bransford1999}
{Bransford}, M.~A., {Thilker}, D.~A., {Walterbos}, R.~A.~M., \& {King}, N.~L.
  1999, \aj, 118, 1635

\bibitem[{{Breeveld} {et~al.}(2011){Breeveld}, {Landsman}, {Holland}, {Roming},
  {Kuin}, \& {Page}}]{breeveld2011}
{Breeveld}, A.~A., {Landsman}, W., {Holland}, S.~T., {et~al.} 2011, in American
  Institute of Physics Conference Series, Vol. 1358, Gamma Ray Bursts 2010, ed.
  J.~E. {McEnery}, J.~L. {Racusin}, \& N.~{Gehrels}, 373--376

\bibitem[{{Brown} {et~al.}(2015){Brown}, {Baron}, {Milne}, {Roming}, \&
  {Wang}}]{Brown2015}
{Brown}, P.~J., {Baron}, E., {Milne}, P., {Roming}, P. W.~A., \& {Wang}, L.
  2015, \apj, 809, 37

\bibitem[{{Brown} {et~al.}(2014){Brown}, {Breeveld}, {Holland}, {Kuin}, \&
  {Pritchard}}]{brown2014}
{Brown}, P.~J., {Breeveld}, A.~A., {Holland}, S., {Kuin}, P., \& {Pritchard},
  T. 2014, \apss, 354, 89

\bibitem[{{Buzzoni} {et~al.}(1984){Buzzoni}, {Delabre}, {Dekker}, {Dodorico},
  {Enard}, {Focardi}, {Gustafsson}, {Nees}, {Paureau}, \&
  {Reiss}}]{buzzoni1984}
{Buzzoni}, B., {Delabre}, B., {Dekker}, H., {et~al.} 1984, The Messenger, 38, 9

\bibitem[{{Cappa} {et~al.}(2005){Cappa}, {Niemela}, {Mart{\'\i}n}, \&
  {McClure-Griffiths}}]{Cappa2005}
{Cappa}, C., {Niemela}, V.~S., {Mart{\'\i}n}, M.~C., \& {McClure-Griffiths},
  N.~M. 2005, \aap, 436, 155

\bibitem[{{Chandra} {et~al.}(2020){Chandra}, {Chevalier}, {Chugai},
  {Milisavljevic}, \& {Fransson}}]{Chandra2020}
{Chandra}, P., {Chevalier}, R.~A., {Chugai}, N., {Milisavljevic}, D., \&
  {Fransson}, C. 2020, \apj, 902, 55

\bibitem[{{Chen} {et~al.}(2018){Chen}, {Inserra}, {Fraser}, {Moriya}, {Schady},
  {Schweyer}, {Filippenko}, {Perley}, {Ruiter}, {Seitenzahl}, {Sollerman},
  {Taddia}, {Anderson}, {Foley}, {Jerkstrand}, {Ngeow}, {Pan}, {Pastorello},
  {Points}, {Smartt}, {Smith}, {Taubenberger}, {Wiseman}, {Young}, {Benetti},
  {Berton}, {Bufano}, {Clark}, {Della Valle}, {Galbany}, {Gal-Yam},
  {Gromadzki}, {Guti{\'e}rrez}, {Heinze}, {Kankare}, {Kilpatrick},
  {Kuncarayakti}, {Leloudas}, {Lin}, {Maguire}, {Mazzali}, {McBrien},
  {Prentice}, {Rau}, {Rest}, {Siebert}, {Stalder}, {Tonry}, \& {Yu}}]{Chen2018}
{Chen}, T.~W., {Inserra}, C., {Fraser}, M., {et~al.} 2018, \apjl, 867, L31

\bibitem[{{Chevalier}(1997)}]{Chevalier1997}
{Chevalier}, R.~A. 1997, Science, 276, 1374

\bibitem[{{Chevalier} \& {Fransson}(1994)}]{Chevalier1994}
{Chevalier}, R.~A. \& {Fransson}, C. 1994, \apj, 420, 268

\bibitem[{{Chu}(1981)}]{Chu1981}
{Chu}, Y.~H. 1981, \apj, 249, 195

\bibitem[{{Cid Fernandes} {et~al.}(2005){Cid Fernandes}, {Mateus}, {Sodr{\'e}},
  {Stasi{\'n}ska}, \& {Gomes}}]{2005MNRAS.358..363C}
{Cid Fernandes}, R., {Mateus}, A., {Sodr{\'e}}, L., {Stasi{\'n}ska}, G., \&
  {Gomes}, J.~M. 2005, \mnras, 358, 363

\bibitem[{{Crowther}(2007)}]{Crowther2007}
{Crowther}, P.~A. 2007, \araa, 45, 177

\bibitem[{{Das} {et~al.}(2009){Das}, {Banerjee}, \& {Ashok}}]{das2009}
{Das}, R.~K., {Banerjee}, D.~P.~K., \& {Ashok}, N.~M. 2009, \mnras, 398, 375

\bibitem[{{Della Valle} \& {Panagia}(2003)}]{Dellavalle2003}
{Della Valle}, M. \& {Panagia}, N. 2003, \apjl, 587, L71

\bibitem[{{Della Valle} {et~al.}(2005){Della Valle}, {Panagia}, {Padovani},
  {Cappellaro}, {Mannucci}, \& {Turatto}}]{Dellavalle2005}
{Della Valle}, M., {Panagia}, N., {Padovani}, P., {et~al.} 2005, \apj, 629, 750

\bibitem[{{Dessart} {et~al.}(2010){Dessart}, {Livne}, \&
  {Waldman}}]{Dessart2010}
{Dessart}, L., {Livne}, E., \& {Waldman}, R. 2010, \mnras, 405, 2113

\bibitem[{{Dopita} {et~al.}(2016){Dopita}, {Kewley}, {Sutherland}, \&
  {Nicholls}}]{dopita2016}
{Dopita}, M.~A., {Kewley}, L.~J., {Sutherland}, R.~S., \& {Nicholls}, D.~C.
  2016, \apss, 361, 61

\bibitem[{{Dressler} {et~al.}(2011){Dressler}, {Bigelow}, {Hare}, {Sutin},
  {Thompson}, {Burley}, {Epps}, {Oemler}, {Bagish}, {Birk}, {Clardy},
  {Gunnels}, {Kelson}, {Shectman}, \& {Osip}}]{dressler2011}
{Dressler}, A., {Bigelow}, B., {Hare}, T., {et~al.} 2011, \pasp, 123, 288

\bibitem[{{Drout} {et~al.}(2016){Drout}, {Milisavljevic}, {Parrent},
  {Margutti}, {Kamble}, {Soderberg}, {Challis}, {Chornock}, {Fong}, {Frank},
  {Gehrels}, {Graham}, {Hsiao}, {Itagaki}, {Kasliwal}, {Kirshner}, {Macomb},
  {Marion}, {Norris}, \& {Phillips}}]{drout2016}
{Drout}, M.~R., {Milisavljevic}, D., {Parrent}, J., {et~al.} 2016, \apj, 821,
  57

\bibitem[{{Drout} {et~al.}(2011){Drout}, {Soderberg}, {Gal-Yam}, {Cenko},
  {Fox}, {Leonard}, {Sand}, {Moon}, {Arcavi}, \& {Green}}]{Drout2011}
{Drout}, M.~R., {Soderberg}, A.~M., {Gal-Yam}, A., {et~al.} 2011, \apj, 741, 97

\bibitem[{{Eldridge} {et~al.}(2013){Eldridge}, {Fraser}, {Smartt}, {Maund}, \&
  {Crockett}}]{Eldridge13}
{Eldridge}, J.~J., {Fraser}, M., {Smartt}, S.~J., {Maund}, J.~R., \&
  {Crockett}, R.~M. 2013, \mnras, 436, 774

\bibitem[{{Ergon} {et~al.}(2014){Ergon}, {Sollerman}, {Fraser}, {Pastorello},
  {Taubenberger}, {Elias-Rosa}, {Bersten}, {Jerkstrand}, {Benetti},
  {Botticella}, {Fransson}, {Harutyunyan}, {Kotak}, {Smartt}, {Valenti},
  {Bufano}, {Cappellaro}, {Fiaschi}, {Howell}, {Kankare}, {Magill}, {Mattila},
  {Maund}, {Naves}, {Ochner}, {Ruiz}, {Smith}, {Tomasella}, \&
  {Turatto}}]{ergon14}
{Ergon}, M., {Sollerman}, J., {Fraser}, M., {et~al.} 2014, \aap, 562, A17

\bibitem[{{Ferrarese} {et~al.}(2007){Ferrarese}, {Mould}, {Stetson}, {Tonry},
  {Blakeslee}, \& {Ajhar}}]{ferrarese07}
{Ferrarese}, L., {Mould}, J.~R., {Stetson}, P.~B., {et~al.} 2007, \apj, 654,
  186

\bibitem[{{Filippenko} {et~al.}(1994){Filippenko}, {Matheson}, \&
  {Barth}}]{Filippenko1994}
{Filippenko}, A.~V., {Matheson}, T., \& {Barth}, A.~J. 1994, \aj, 108, 2220

\bibitem[{{Fitzpatrick}(1999)}]{fitzpatrick99}
{Fitzpatrick}, E.~L. 1999, \pasp, 111, 63

\bibitem[{{Fouque} {et~al.}(1992){Fouque}, {Gourgoulhon}, {Chamaraux}, \&
  {Paturel}}]{Fouque1992}
{Fouque}, P., {Gourgoulhon}, E., {Chamaraux}, P., \& {Paturel}, G. 1992, \aaps,
  93, 211

\bibitem[{{Fransson} {et~al.}(1996){Fransson}, {Lundqvist}, \&
  {Chevalier}}]{Fransson1996}
{Fransson}, C., {Lundqvist}, P., \& {Chevalier}, R.~A. 1996, \apj, 461, 993

\bibitem[{{Fremling} {et~al.}(2018){Fremling}, {Sollerman}, {Kasliwal},
  {Kulkarni}, {Barbarino}, {Ergon}, {Karamehmetoglu}, {Taddia}, {Arcavi},
  {Cenko}, {Clubb}, {De Cia}, {Duggan}, {Filippenko}, {Gal-Yam}, {Graham},
  {Horesh}, {Hosseinzadeh}, {Howell}, {Kuesters}, {Lunnan}, {Matheson},
  {Nugent}, {Perley}, {Quimby}, \& {Saunders}}]{fremling2018}
{Fremling}, C., {Sollerman}, J., {Kasliwal}, M.~M., {et~al.} 2018, \aap, 618,
  A37

\bibitem[{{Galbany} {et~al.}(2018){Galbany}, {Anderson}, {S{\'a}nchez},
  {Kuncarayakti}, {Pedraz}, {Gonz{\'a}lez-Gait{\'a}n}, {Stanishev},
  {Dom{\'\i}nguez}, {Moreno-Raya}, {Wood-Vasey}, {Mour{\~a}o}, {Ponder},
  {Badenes}, {Moll{\'a}}, {L{\'o}pez-S{\'a}nchez}, {Rosales-Ortega},
  {V{\'\i}lchez}, {Garc{\'\i}a-Benito}, \& {Marino}}]{galbany2018}
{Galbany}, L., {Anderson}, J.~P., {S{\'a}nchez}, S.~F., {et~al.} 2018, \apj,
  855, 107

\bibitem[{{Galbany} {et~al.}(2017){Galbany}, {Mora}, {Gonz{\'a}lez-Gait{\'a}n},
  {Bolatto}, {Dannerbauer}, {L{\'o}pez-S{\'a}nchez}, {Maeda}, {P{\'e}rez},
  {P{\'e}rez-Torres}, {S{\'a}nchez}, {Wong}, {Badenes}, {Blitz}, {Marino},
  {Utomo}, \& {Van de Ven}}]{Galbany2017}
{Galbany}, L., {Mora}, L., {Gonz{\'a}lez-Gait{\'a}n}, S., {et~al.} 2017,
  \mnras, 468, 628

\bibitem[{{Gehrels} {et~al.}(2004){Gehrels}, {Chincarini}, {Giommi}, {Mason},
  {Nousek}, {Wells}, {White}, {Barthelmy}, {Burrows}, {Cominsky}, {Hurley},
  {Marshall}, {M{\'e}sz{\'a}ros}, {Roming}, {Angelini}, {Barbier}, {Belloni},
  {Campana}, {Caraveo}, {Chester}, {Citterio}, {Cline}, {Cropper}, {Cummings},
  {Dean}, {Feigelson}, {Fenimore}, {Frail}, {Fruchter}, {Garmire}, {Gendreau},
  {Ghisellini}, {Greiner}, {Hill}, {Hunsberger}, {Krimm}, {Kulkarni}, {Kumar},
  {Lebrun}, {Lloyd-Ronning}, {Markwardt}, {Mattson}, {Mushotzky}, {Norris},
  {Osborne}, {Paczynski}, {Palmer}, {Park}, {Parsons}, {Paul}, {Rees},
  {Reynolds}, {Rhoads}, {Sasseen}, {Schaefer}, {Short}, {Smale}, {Smith},
  {Stella}, {Tagliaferri}, {Takahashi}, {Tashiro}, {Townsley}, {Tueller},
  {Turner}, {Vietri}, {Voges}, {Ward}, {Willingale}, {Zerbi}, \&
  {Zhang}}]{gehrels04}
{Gehrels}, N., {Chincarini}, G., {Giommi}, P., {et~al.} 2004, \apj, 611, 1005

\bibitem[{{Goobar}(2008)}]{goobar08}
{Goobar}, A. 2008, \apjl, 686, L103

\bibitem[{{Grosdidier} {et~al.}(1998){Grosdidier}, {Moffat}, {Joncas}, \&
  {Acker}}]{Grosdidier1998}
{Grosdidier}, Y., {Moffat}, A. F.~J., {Joncas}, G., \& {Acker}, A. 1998, \apjl,
  506, L127

\bibitem[{{Hamuy} {et~al.}(2006){Hamuy}, {Folatelli}, {Morrell}, {Phillips},
  {Suntzeff}, {Persson}, {Roth}, {Gonzalez}, {Krzeminski}, {Contreras},
  {Freedman}, {Murphy}, {Madore}, {Wyatt}, {Maza}, {Filippenko}, {Li}, \&
  {Pinto}}]{hamuy2006}
{Hamuy}, M., {Folatelli}, G., {Morrell}, N.~I., {et~al.} 2006, \pasp, 118, 2

\bibitem[{{Harris}(2010)}]{Harris2010}
{Harris}, G. L.~H. 2010, \pasa, 27, 475

\bibitem[{{Hoeflich} {et~al.}(2021){Hoeflich}, {Ashall}, {Bose}, {Baron},
  {Stritzinger}, {Davis}, {Shahbandeh}, {Anand}, {Baade}, {Burns}, {Collins},
  {Diamond}, {Fisher}, {Galbany}, {Hristov}, {Hsiao}, {Phillips}, {Shappee},
  {Suntzeff}, \& {Tucker}}]{Hoeflich2021}
{Hoeflich}, P., {Ashall}, C., {Bose}, S., {et~al.} 2021, \apj, 922, 186

\bibitem[{{H{\"o}flich}(1990)}]{h90}
{H{\"o}flich}, P. 1990, PhD thesis, -

\bibitem[{{H{\"o}flich}(2003)}]{hoeflich2003}
{H{\"o}flich}, P. 2003, in Astronomical Society of the Pacific Conference
  Series, Vol. 288, Stellar Atmosphere Modeling, ed. I.~{Hubeny}, D.~{Mihalas},
  \& K.~{Werner}, 371

\bibitem[{{H{\"o}flich}(2009)}]{hoeflich2009}
{H{\"o}flich}, P. 2009, in American Institute of Physics Conference Series,
  Vol. 1171, Recent Directions in Astrophysical Quantitative Spectroscopy and
  Radiation Hydrodynamics, ed. I.~{Hubeny}, J.~M. {Stone}, K.~{MacGregor}, \&
  K.~{Werner}, 161--172

\bibitem[{{Holmbo} {et~al.}(2023){Holmbo}, {Stritzinger}, {Karamehmetoglu},
  {Burns}, {Morrell}, {Ashall}, {Hsiao}, {Galbany}, {Folatelli}, {Phillips},
  {Baron}, {Guti{\'e}rrez}, {Leloudas}, {M{\"u}ller-Bravo}, {Hoeflich},
  {Taddia}, \& {Suntzeff}}]{Holmbo2023}
{Holmbo}, S., {Stritzinger}, M.~D., {Karamehmetoglu}, E., {et~al.} 2023, \aap,
  675, A83

\bibitem[{{Houck} \& {Fransson}(1996)}]{Houck1996}
{Houck}, J.~C. \& {Fransson}, C. 1996, \apj, 456, 811

\bibitem[{{Hough} {et~al.}(1987){Hough}, {Bailey}, {Rouse}, \&
  {Whittet}}]{hough1987}
{Hough}, J.~H., {Bailey}, J.~A., {Rouse}, M.~F., \& {Whittet}, D.~C.~B. 1987,
  \mnras, 227, 1P

\bibitem[{{Hounsell} {et~al.}(2016){Hounsell}, {Miller}, {Pan}, {Foley},
  {Rest}, {Jha}, {Scolnic}, {Smith}, {Wright}, {Smartt}, {Huber}, {Chambers},
  {Flewelling}, {Willman}, {Primak}, {Schultz}, {Gibson}, {Magnier}, {Waters},
  {Tonry}, \& {Wainscoat}}]{hounsell2016}
{Hounsell}, R.~A., {Miller}, J.~A., {Pan}, Y.~C., {et~al.} 2016, The
  Astronomer's Telegram, 8663, 1

\bibitem[{{Hristov} {et~al.}(2021){Hristov}, {Hoeflich}, \&
  {Collins}}]{Hristov2021}
{Hristov}, B., {Hoeflich}, P., \& {Collins}, D.~C. 2021, \apj, 923, 210

\bibitem[{{Hsiao} {et~al.}(2019){Hsiao}, {Phillips}, {Marion}, {Kirshner},
  {Morrell}, {Sand}, {Burns}, {Contreras}, {Hoeflich}, {Stritzinger},
  {Valenti}, {Anderson}, {Ashall}, {Baltay}, {Baron}, {Banerjee}, {Davis},
  {Diamond}, {Folatelli}, {Freedman}, {F{\"o}rster}, {Galbany}, {Gall},
  {Gonz{\'a}lez-Gait{\'a}n}, {Goobar}, {Hamuy}, {Holmbo}, {Kasliwal},
  {Krisciunas}, {Kumar}, {Lidman}, {Lu}, {Nugent}, {Perlmutter}, {Persson},
  {Piro}, {Rabinowitz}, {Roth}, {Ryder}, {Schmidt}, {Shahbandeh}, {Suntzeff},
  {Taddia}, {Uddin}, \& {Wang}}]{hsiao2019}
{Hsiao}, E.~Y., {Phillips}, M.~M., {Marion}, G.~H., {et~al.} 2019, \pasp, 131,
  014002

\bibitem[{{Hunter} {et~al.}(2009){Hunter}, {Valenti}, {Kotak}, {Meikle},
  {Taubenberger}, {Pastorello}, {Benetti}, {Stanishev}, {Smartt}, {Trundle},
  {Arkharov}, {Bufano}, {Cappellaro}, {Di Carlo}, {Dolci}, {Elias-Rosa},
  {Frandsen}, {Fynbo}, {Hopp}, {Larionov}, {Laursen}, {Mazzali}, {Navasardyan},
  {Ries}, {Riffeser}, {Rizzi}, {Tsvetkov}, {Turatto}, \& {Wilke}}]{hunter09}
{Hunter}, D.~J., {Valenti}, S., {Kotak}, R., {et~al.} 2009, \aap, 508, 371

\bibitem[{{Israel}(1998)}]{israel1998}
{Israel}, F.~P. 1998, \aapr, 8, 237

\bibitem[{{Jim{\'e}nez-Hern{\'a}ndez}
  {et~al.}(2020){Jim{\'e}nez-Hern{\'a}ndez}, {Arthur}, \& {Toal{\'a}}}]{JP2020}
{Jim{\'e}nez-Hern{\'a}ndez}, P., {Arthur}, S.~J., \& {Toal{\'a}}, J.~A. 2020,
  \mnras, 497, 4128

\bibitem[{{Johnson} \& {Hogg}(1965)}]{Johnson1965}
{Johnson}, H.~M. \& {Hogg}, D.~E. 1965, \apj, 142, 1033

\bibitem[{{Kelly} {et~al.}(2016){Kelly}, {Ciardi}, {Beichman}, {Filippenko},
  {Fox}, {Patel}, \& {Sinukoff}}]{kelly2016}
{Kelly}, P.~L., {Ciardi}, D.~R., {Beichman}, C.~A., {et~al.} 2016, The
  Astronomer's Telegram, 8720, 1

\bibitem[{{Kennicutt}(1998)}]{1998ApJ...498..541K}
{Kennicutt}, Robert~C., J. 1998, \apj, 498, 541

\bibitem[{{Khatami} \& {Kasen}(2019)}]{kathami2019}
{Khatami}, D.~K. \& {Kasen}, D.~N. 2019, \apj, 878, 56

\bibitem[{{Kilpatrick} {et~al.}(2021){Kilpatrick}, {Drout}, {Auchettl},
  {Dimitriadis}, {Foley}, {Jones}, {DeMarchi}, {French}, {Gall}, {Hjorth},
  {Jacobson-Gal{\'a}n}, {Margutti}, {Piro}, {Ramirez-Ruiz}, {Rest}, \&
  {Rojas-Bravo}}]{Kilpatrick2021}
{Kilpatrick}, C.~D., {Drout}, M.~R., {Auchettl}, K., {et~al.} 2021, \mnras,
  504, 2073

\bibitem[{{Kiyota} {et~al.}(2016){Kiyota}, {Shappee}, {Stanek}, \&
  {Dong}}]{kiyota2016}
{Kiyota}, S., {Shappee}, B.~J., {Stanek}, K.~Z., \& {Dong}, S. 2016, The
  Astronomer's Telegram, 8654, 1

\bibitem[{{Krisciunas} {et~al.}(2017){Krisciunas}, {Contreras}, {Burns},
  {Phillips}, {Stritzinger}, {Morrell}, {Hamuy}, {Anais}, {Boldt}, {Busta},
  {Campillay}, {Castell{\'o}n}, {Folatelli}, {Freedman}, {Gonz{\'a}lez},
  {Hsiao}, {Krzeminski}, {Persson}, {Roth}, {Salgado}, {Ser{\'o}n}, {Suntzeff},
  {Torres}, {Filippenko}, {Li}, {Madore}, {DePoy}, {Marshall}, {Rheault}, \&
  {Villanueva}}]{krisciunas2017}
{Krisciunas}, K., {Contreras}, C., {Burns}, C.~R., {et~al.} 2017, \aj, 154, 211

\bibitem[{{Kr{\"u}hler} {et~al.}(2017){Kr{\"u}hler}, {Kuncarayakti}, {Schady},
  {Anderson}, {Galbany}, \& {Gensior}}]{kruehler2017}
{Kr{\"u}hler}, T., {Kuncarayakti}, H., {Schady}, P., {et~al.} 2017, \aap, 602,
  A85

\bibitem[{{Kuijken}(2011)}]{kuijken2011}
{Kuijken}, K. 2011, The Messenger, 146, 8

\bibitem[{{Kuijken} {et~al.}(2002){Kuijken}, {Bender}, {Cappellaro},
  {Muschielok}, {Baruffolo}, {Cascone}, {Iwert}, {Mitsch}, {Nicklas},
  {Valentijn}, {Baade}, {Begeman}, {Bortolussi}, {Boxhoorn}, {Christen},
  {Deul}, {Geimer}, {Greggio}, {Harke}, {H{\"a}fner}, {Hess}, {Hess}, {Hopp},
  {Ilijevski}, {Klink}, {Kravcar}, {Lizon}, {Magagna}, {M{\"u}ller},
  {Niemeczek}, {de Pizzol}, {Poschmann}, {Reif}, {Rengelink}, {Reyes},
  {Silber}, \& {Wellem}}]{kuijken2002}
{Kuijken}, K., {Bender}, R., {Cappellaro}, E., {et~al.} 2002, The Messenger,
  110, 15

\bibitem[{{Kuncarayakti} {et~al.}(2018){Kuncarayakti}, {Maeda}, {Ashall},
  {Prentice}, {Mattila}, {Kankare}, {Fransson}, {Lundqvist}, {Pastorello},
  {Leloudas}, {Anderson}, {Benetti}, {Bersten}, {Cappellaro}, {Cartier},
  {Denneau}, {Della Valle}, {Elias-Rosa}, {Folatelli}, {Fraser}, {Galbany},
  {Gall}, {Gal-Yam}, {Guti{\'e}rrez}, {Hamanowicz}, {Heinze}, {Inserra},
  {Kangas}, {Mazzali}, {Melandri}, {Pignata}, {Rest}, {Reynolds}, {Roy},
  {Smartt}, {Smith}, {Sollerman}, {Somero}, {Stalder}, {Stritzinger}, {Taddia},
  {Tomasella}, {Tonry}, {Weiland}, \& {Young}}]{Kuncarayakti2018}
{Kuncarayakti}, H., {Maeda}, K., {Ashall}, C.~J., {et~al.} 2018, \apjl, 854,
  L14

\bibitem[{{Kuncarayakti} {et~al.}(2022){Kuncarayakti}, {Maeda}, {Dessart},
  {Nagao}, {Fulton}, {Guti{\'e}rrez}, {Huber}, {Young}, {Kotak}, {Mattila},
  {Anderson}, {Ferrari}, {Folatelli}, {Gao}, {Magnier}, {Smith}, \&
  {Srivastav}}]{Kuncarayakti2022}
{Kuncarayakti}, H., {Maeda}, K., {Dessart}, L., {et~al.} 2022, \apjl, 941, L32

\bibitem[{{Kuncarayakti} {et~al.}(2023){Kuncarayakti}, {Sollerman}, {Izzo},
  {Maeda}, {Yang}, {Schulze}, {Angus}, {Aubert}, {Auchettl}, {Della Valle},
  {Dessart}, {Hinds}, {Kankare}, {Kawabata}, {Lundqvist}, {Nakaoka}, {Perley},
  {Raimundo}, {Strotjohann}, {Taguchi}, {Cai}, {Charalampopoulos}, {Fang},
  {Fraser}, {Gutierrez}, {Imazawa}, {Kangas}, {Kawabata}, {Kotak}, {Kravtsov},
  {Matilainen}, {Mattila}, {Moran}, {Murata}, {Salmaso}, {Anderson}, {Ashall},
  {Bellm}, {Benetti}, {Chambers}, {Chen}, {Coughlin}, {De Colle}, {Fremling},
  {Galbany}, {Gal-Yam}, {Gromadzki}, {Groom}, {Hajela}, {Inserra}, {Kasliwal},
  {Mahabal}, {Martin-Carrillo}, {Moore}, {Muller-Bravo}, {Nicholl}, {Ragosta},
  {Riddle}, {Sharma}, {Srivastav}, {Stritzinger}, {Wold}, \&
  {Young}}]{Kuncarayakti2023}
{Kuncarayakti}, H., {Sollerman}, J., {Izzo}, L., {et~al.} 2023, arXiv e-prints,
  arXiv:2303.16925

\bibitem[{{Landolt}(1992)}]{landolt1992}
{Landolt}, A.~U. 1992, \aj, 104, 340

\bibitem[{{Liu} {et~al.}(2016){Liu}, {Modjaz}, {Bianco}, \& {Graur}}]{liu2016}
{Liu}, Y.-Q., {Modjaz}, M., {Bianco}, F.~B., \& {Graur}, O. 2016, \apj, 827, 90

\bibitem[{{Lyman} {et~al.}(2014){Lyman}, {Bersier}, \& {James}}]{Lyman2014}
{Lyman}, J.~D., {Bersier}, D., \& {James}, P.~A. 2014, \mnras, 437, 3848

\bibitem[{{Lyman} {et~al.}(2016){Lyman}, {Bersier}, {James}, {Mazzali},
  {Eldridge}, {Fraser}, \& {Pian}}]{Lyman2016}
{Lyman}, J.~D., {Bersier}, D., {James}, P.~A., {et~al.} 2016, \mnras, 457, 328

\bibitem[{{Lyman} {et~al.}(2018){Lyman}, {Taddia}, {Stritzinger}, {Galbany},
  {Leloudas}, {Anderson}, {Eldridge}, {James}, {Kr{\"u}hler}, {Levan},
  {Pignata}, \& {Stanway}}]{lyman2018}
{Lyman}, J.~D., {Taddia}, F., {Stritzinger}, M.~D., {et~al.} 2018, \mnras, 473,
  1359

\bibitem[{{Lynch} {et~al.}(2000){Lynch}, {Rudy}, {Mazuk}, \&
  {Puetter}}]{Lynch2000}
{Lynch}, D.~K., {Rudy}, R.~J., {Mazuk}, S., \& {Puetter}, R.~C. 2000, \apj,
  541, 791

\bibitem[{{Maeda} {et~al.}(2023){Maeda}, {Chandra}, {Moriya}, {Reguitti},
  {Ryder}, {Matsuoka}, {Michiyama}, {Pignata}, {Hiramatsu}, {Bostroem},
  {Kundu}, {Kuncarayakti}, {Bersten}, {Pooley}, {Lee}, {Patnaude},
  {Rodr{\'\i}guez}, \& {Folatelli}}]{Maeda2023}
{Maeda}, K., {Chandra}, P., {Moriya}, T.~J., {et~al.} 2023, \apj, 942, 17

\bibitem[{{Maeda} {et~al.}(2015){Maeda}, {Hattori}, {Milisavljevic},
  {Folatelli}, {Drout}, {Kuncarayakti}, {Margutti}, {Kamble}, {Soderberg},
  {Tanaka}, {Kawabata}, {Kawabata}, {Yamanaka}, {Nomoto}, {Kim}, {Simon},
  {Phillips}, {Parrent}, {Nakaoka}, {Moriya}, {Suzuki}, {Takaki}, {Ishigaki},
  {Sakon}, {Tajitsu}, \& {Iye}}]{Maeda2015}
{Maeda}, K., {Hattori}, T., {Milisavljevic}, D., {et~al.} 2015, \apj, 807, 35

\bibitem[{{Marchenko} {et~al.}(2010){Marchenko}, {Moffat}, \&
  {Crowther}}]{Marchenko2010}
{Marchenko}, S.~V., {Moffat}, A.~F.~J., \& {Crowther}, P.~A. 2010, \apjl, 724,
  L90

\bibitem[{{Margutti} {et~al.}(2017){Margutti}, {Kamble}, {Milisavljevic},
  {Zapartas}, {de Mink}, {Drout}, {Chornock}, {Risaliti}, {Zauderer},
  {Bietenholz}, {Cantiello}, {Chakraborti}, {Chomiuk}, {Fong}, {Grefenstette},
  {Guidorzi}, {Kirshner}, {Parrent}, {Patnaude}, {Soderberg}, {Gehrels}, \&
  {Harrison}}]{Margutti2017}
{Margutti}, R., {Kamble}, A., {Milisavljevic}, D., {et~al.} 2017, \apj, 835,
  140

\bibitem[{{Marino} {et~al.}(2013){Marino}, {Rosales-Ortega}, {S{\'a}nchez},
  {Gil de Paz}, {V{\'\i}lchez}, {Miralles-Caballero}, {Kehrig},
  {P{\'e}rez-Montero}, {Stanishev}, {Iglesias-P{\'a}ramo}, {D{\'\i}az},
  {Castillo-Morales}, {Kennicutt}, {L{\'o}pez-S{\'a}nchez}, {Galbany},
  {Garc{\'\i}a-Benito}, {Mast}, {Mendez-Abreu}, {Monreal-Ibero}, {Husemann},
  {Walcher}, {Garc{\'\i}a-Lorenzo}, {Masegosa}, {Del Olmo Orozco},
  {Mour{\~a}o}, {Ziegler}, {Moll{\'a}}, {Papaderos},
  {S{\'a}nchez-Bl{\'a}zquez}, {Gonz{\'a}lez Delgado}, {Falc{\'o}n-Barroso},
  {Roth}, {van de Ven}, \& {Califa Team}}]{marino2013}
{Marino}, R.~A., {Rosales-Ortega}, F.~F., {S{\'a}nchez}, S.~F., {et~al.} 2013,
  \aap, 559, A114

\bibitem[{{Marples} {et~al.}(2016){Marples}, {Bock}, \& {Parker}}]{marples2016}
{Marples}, P., {Bock}, G., \& {Parker}, S. 2016, The Astronomer's Telegram,
  8651, 1

\bibitem[{{Massey}(2002)}]{Massey02}
{Massey}, P. 2002, \apjs, 141, 81

\bibitem[{{Matheson} {et~al.}(2000){Matheson}, {Filippenko}, {Ho}, {Barth}, \&
  {Leonard}}]{Matheson2000}
{Matheson}, T., {Filippenko}, A.~V., {Ho}, L.~C., {Barth}, A.~J., \& {Leonard},
  D.~C. 2000, \aj, 120, 1499

\bibitem[{{Mathis}(1990)}]{mathis1990}
{Mathis}, J.~S. 1990, \araa, 28, 37

\bibitem[{{Mauerhan} {et~al.}(2018){Mauerhan}, {Filippenko}, {Zheng}, {Brink},
  {Graham}, {Shivvers}, \& {Clubb}}]{Mauerhan2018}
{Mauerhan}, J.~C., {Filippenko}, A.~V., {Zheng}, W., {et~al.} 2018, \mnras,
  478, 5050

\bibitem[{{Maund} {et~al.}(2015){Maund}, {Fraser}, {Reilly}, {Ergon}, \&
  {Mattila}}]{Maund15}
{Maund}, J.~R., {Fraser}, M., {Reilly}, E., {Ergon}, M., \& {Mattila}, S. 2015,
  \mnras, 447, 3207

\bibitem[{{Milisavljevic} {et~al.}(2015){Milisavljevic}, {Margutti}, {Kamble},
  {Patnaude}, {Raymond}, {Eldridge}, {Fong}, {Bietenholz}, {Challis},
  {Chornock}, {Drout}, {Fransson}, {Fesen}, {Grindlay}, {Kirshner}, {Lunnan},
  {Mackey}, {Miller}, {Parrent}, {Sanders}, {Soderberg}, \&
  {Zauderer}}]{Milisavljevic2015}
{Milisavljevic}, D., {Margutti}, R., {Kamble}, A., {et~al.} 2015, \apj, 815,
  120

\bibitem[{{Millard} {et~al.}(1999){Millard}, {Branch}, {Baron}, {Hatano},
  {Fisher}, {Filippenko}, {Kirshner}, {Challis}, {Fransson}, {Panagia},
  {Phillips}, {Sonneborn}, {Suntzeff}, {Wagoner}, \& {Wheeler}}]{millard99}
{Millard}, J., {Branch}, D., {Baron}, E., {et~al.} 1999, \apj, 527, 746

\bibitem[{{Modjaz} {et~al.}(2014){Modjaz}, {Blondin}, {Kirshner}, {Matheson},
  {Berlind}, {Bianco}, {Calkins}, {Challis}, {Garnavich}, {Hicken}, {Jha},
  {Liu}, \& {Marion}}]{modjaz_spectra_2014}
{Modjaz}, M., {Blondin}, S., {Kirshner}, R.~P., {et~al.} 2014, \aj, 147, 99

\bibitem[{{Moorwood} {et~al.}(1998){Moorwood}, {Cuby}, \&
  {Lidman}}]{moorwood1998}
{Moorwood}, A., {Cuby}, J.~G., \& {Lidman}, C. 1998, The Messenger, 91, 9

\bibitem[{{Morales-Garoffolo} {et~al.}(2014){Morales-Garoffolo}, {Elias-Rosa},
  {Benetti}, {Taubenberger}, {Cappellaro}, {Pastorello}, {Klauser}, {Valenti},
  {Howerton}, {Ochner}, {Schramm}, {Siviero}, {Tartaglia}, \&
  {Tomasella}}]{Morales-Garoffolo2014}
{Morales-Garoffolo}, A., {Elias-Rosa}, N., {Benetti}, S., {et~al.} 2014,
  \mnras, 445, 1647

\bibitem[{{M{\"u}ller}(2016)}]{Mueller2016}
{M{\"u}ller}, B. 2016, \pasa, 33, e048

\bibitem[{{Osterbrock} \& {Ferland}(2006)}]{osterbrock2006}
{Osterbrock}, D.~E. \& {Ferland}, G.~J. 2006, {Astrophysics of gaseous nebulae
  and active galactic nuclei} (Mill Valley, CA: University Science Books)

\bibitem[{{Owocki} {et~al.}(2019){Owocki}, {Hirai}, {Podsiadlowski}, \&
  {Schneider}}]{Owocki2019}
{Owocki}, S.~P., {Hirai}, R., {Podsiadlowski}, P., \& {Schneider}, F. R.~N.
  2019, \mnras, 485, 988

\bibitem[{{Patat} {et~al.}(1995){Patat}, {Chugai}, \& {Mazzali}}]{Patat1995}
{Patat}, F., {Chugai}, N., \& {Mazzali}, P.~A. 1995, \aap, 299, 715

\bibitem[{{Patat} {et~al.}(2015){Patat}, {Taubenberger}, {Cox}, {Baade},
  {Clocchiatti}, {H{\"o}flich}, {Maund}, {Reilly}, {Spyromilio}, {Wang},
  {Wheeler}, \& {Zelaya}}]{patat2015}
{Patat}, F., {Taubenberger}, S., {Cox}, N.~L.~J., {et~al.} 2015, \aap, 577, A53

\bibitem[{{Persson} {et~al.}(1998){Persson}, {Murphy}, {Krzeminski}, {Roth}, \&
  {Rieke}}]{persson1998}
{Persson}, S.~E., {Murphy}, D.~C., {Krzeminski}, W., {Roth}, M., \& {Rieke},
  M.~J. 1998, \aj, 116, 2475

\bibitem[{{Phillips}(1993)}]{phillips1993}
{Phillips}, M.~M. 1993, \apjl, 413, L105

\bibitem[{{Phillips} {et~al.}(2019){Phillips}, {Contreras}, {Hsiao}, {Morrell},
  {Burns}, {Stritzinger}, {Ashall}, {Freedman}, {Hoeflich}, {Persson}, {Piro},
  {Suntzeff}, {Uddin}, {Anais}, {Baron}, {Busta}, {Campillay}, {Castell{\'o}n},
  {Corco}, {Diamond}, {Gall}, {Gonzalez}, {Holmbo}, {Krisciunas}, {Roth},
  {Ser{\'o}n}, {Taddia}, {Torres}, {Anderson}, {Baltay}, {Folatelli},
  {Galbany}, {Goobar}, {Hadjiyska}, {Hamuy}, {Kasliwal}, {Lidman}, {Nugent},
  {Perlmutter}, {Rabinowitz}, {Ryder}, {Schmidt}, {Shappee}, \&
  {Walker}}]{Phillips2019}
{Phillips}, M.~M., {Contreras}, C., {Hsiao}, E.~Y., {et~al.} 2019, \pasp, 131,
  014001

\bibitem[{{Phillips} {et~al.}(1987){Phillips}, {Phillips}, {Heathcote},
  {Blanco}, {Geisler}, {Hamilton}, {Suntzeff}, {Jablonski}, {Steiner},
  {Cowley}, {Schmidtke}, {Wyckoff}, {Hutchings}, {Tonry}, {Strauss},
  {Thorstensen}, {Honey}, {Maza}, {Ruiz}, {Landolt}, {Uomoto}, {Rich},
  {Grindlay}, {Cohn}, {Smith}, {Lutz}, {Lavery}, \& {Saha}}]{phillips1987}
{Phillips}, M.~M., {Phillips}, A.~C., {Heathcote}, S.~R., {et~al.} 1987, \pasp,
  99, 592

\bibitem[{{Phillips} {et~al.}(2013){Phillips}, {Simon}, {Morrell}, {Burns},
  {Cox}, {Foley}, {Karakas}, {Patat}, {Sternberg}, {Williams}, {Gal-Yam},
  {Hsiao}, {Leonard}, {Persson}, {Stritzinger}, {Thompson}, {Campillay},
  {Contreras}, {Folatelli}, {Freedman}, {Hamuy}, {Roth}, {Shields}, {Suntzeff},
  {Chomiuk}, {Ivans}, {Madore}, {Penprase}, {Perley}, {Pignata}, {Preston}, \&
  {Soderberg}}]{phillips2013}
{Phillips}, M.~M., {Simon}, J.~D., {Morrell}, N., {et~al.} 2013, \apj, 779, 38

\bibitem[{{Podsiadlowski} {et~al.}(2010){Podsiadlowski}, {Ivanova}, {Justham},
  \& {Rappaport}}]{Podsiadlowski2010}
{Podsiadlowski}, P., {Ivanova}, N., {Justham}, S., \& {Rappaport}, S. 2010,
  \mnras, 406, 840

\bibitem[{{Prentice} {et~al.}(2019){Prentice}, {Ashall}, {James}, {Short},
  {Mazzali}, {Bersier}, {Crowther}, {Barbarino}, {Chen}, {Copperwheat},
  {Darnley}, {Denneau}, {Elias-Rosa}, {Fraser}, {Galbany}, {Gal-Yam},
  {Harmanen}, {Howell}, {Hosseinzadeh}, {Inserra}, {Kankare}, {Karamehmetoglu},
  {Lamb}, {Limongi}, {Maguire}, {McCully}, {Olivares E}, {Piascik}, {Pignata},
  {Reichart}, {Rest}, {Reynolds}, {Rodr{\'\i}guez}, {Saario}, {Schulze},
  {Smartt}, {Smith}, {Sollerman}, {Stalder}, {Sullivan}, {Taddia}, {Valenti},
  {Vergani}, {Williams}, \& {Young}}]{Prentice2019}
{Prentice}, S.~J., {Ashall}, C., {James}, P.~A., {et~al.} 2019, \mnras, 485,
  1559

\bibitem[{{Prentice} {et~al.}(2016){Prentice}, {Mazzali}, {Pian}, {Gal-Yam},
  {Kulkarni}, {Rubin}, {Corsi}, {Fremling}, {Sollerman}, {Yaron}, {Arcavi},
  {Zheng}, {Kasliwal}, {Filippenko}, {Cenko}, {Cao}, \&
  {Nugent}}]{Prentice2016}
{Prentice}, S.~J., {Mazzali}, P.~A., {Pian}, E., {et~al.} 2016, \mnras, 458,
  2973

\bibitem[{{Price-Whelan} {et~al.}(2018){Price-Whelan}, {Crawford}, {Sipocz},
  {de Val-Borro}, {G{\"u}nther}, {Ginsburg}, {Lim}, {Robitaille}, {Tollerud},
  {Conseil}, {Sladen}, {Barmby}, {Vanderplas}, {lgbouma}, {Copin}, {Homeier},
  {Dencheva}, {Buddelmeijer}, {Jenness}, {Streicher}, {mdmueller}, {Shupe},
  {P{\'e}rez-Su{\'a}rez}, {Weaver}, {Cruz}, {Dietrich}, {Cano Rodr{\'\i}guez},
  {Kovacs}, {Muna}, \& {Bakanov}}]{price2018}
{Price-Whelan}, A., {Crawford}, S., {Sipocz}, B., {et~al.} 2018

\bibitem[{{Puls} {et~al.}(2008){Puls}, {Vink}, \& {Najarro}}]{Puls2008}
{Puls}, J., {Vink}, J.~S., \& {Najarro}, F. 2008, \aapr, 16, 209

\bibitem[{{Rate} \& {Crowther}(2020)}]{Rate20}
{Rate}, G. \& {Crowther}, P.~A. 2020, \mnras, 493, 1512

\bibitem[{{Ravi} {et~al.}(2023){Ravi}, {Rho}, {Park}, {Park}, {Yoon},
  {Geballe}, {Vink{\'o}}, {Tinyanont}, {Bostroem}, {Burke}, {Hiramatsu},
  {Howell}, {McCully}, {Newsome}, {Padilla Gonzalez}, {Pellegrino}, {Cartier},
  {Pritchard}, {Andersen}, {Blinnikov}, {Dong}, {Blanchard}, {Kilpatrick},
  {Hoeflich}, {Valenti}, {Filippenko}, {Suntzeff}, {Seok},
  {K{\"o}nyves-T{\'o}th}, {Foley}, {Siebert}, \& {Jones}}]{Ravi2023}
{Ravi}, A.~P., {Rho}, J., {Park}, S., {et~al.} 2023, \apj, 950, 14

\bibitem[{{Roming} {et~al.}(2005){Roming}, {Kennedy}, {Mason}, {Nousek}, {Ahr},
  {Bingham}, {Broos}, {Carter}, {Hancock}, {Huckle}, {Hunsberger}, {Kawakami},
  {Killough}, {Koch}, {McLelland}, {Smith}, {Smith}, {Soto}, {Boyd},
  {Breeveld}, {Holland}, {Ivanushkina}, {Pryzby}, {Still}, \&
  {Stock}}]{roming2005}
{Roming}, P. W.~A., {Kennedy}, T.~E., {Mason}, K.~O., {et~al.} 2005, \ssr, 120,
  95

\bibitem[{{Roy} {et~al.}(2016){Roy}, {Sollerman}, {Silverman}, {Pastorello},
  {Fransson}, {Drake}, {Taddia}, {Fremling}, {Kankare}, {Kumar}, {Cappellaro},
  {Bose}, {Benetti}, {Filippenko}, {Valenti}, {Nyholm}, {Ergon}, {Sutaria},
  {Kumar}, {Pandey}, {Nicholl}, {Garcia-{\'A}lvarez}, {Tomasella},
  {Karamehmetoglu}, \& {Migotto}}]{Roy2016}
{Roy}, R., {Sollerman}, J., {Silverman}, J.~M., {et~al.} 2016, \aap, 596, A67

\bibitem[{{Rybicki} \& {Lightman}(1979)}]{Rybicki1979}
{Rybicki}, G.~B. \& {Lightman}, A.~P. 1979, {Radiative processes in
  astrophysics}

\bibitem[{{Schlafly} \& {Finkbeiner}(2011)}]{schlafly11}
{Schlafly}, E.~F. \& {Finkbeiner}, D.~P. 2011, \apj, 737, 103

\bibitem[{{Schr{\o}der} {et~al.}(2020){Schr{\o}der}, {MacLeod}, {Loeb},
  {Vigna-G{\'o}mez}, \& {Mandel}}]{Schroder2020}
{Schr{\o}der}, S.~L., {MacLeod}, M., {Loeb}, A., {Vigna-G{\'o}mez}, A., \&
  {Mandel}, I. 2020, \apj, 892, 13

\bibitem[{{Sextl} {et~al.}(2023){Sextl}, {Kudritzki}, {Zahid}, \&
  {Ho}}]{Sextl2023}
{Sextl}, E., {Kudritzki}, R.-P., {Zahid}, H.~J., \& {Ho}, I.~T. 2023, \apj,
  949, 60

\bibitem[{{Shahbandeh} {et~al.}(2022){Shahbandeh}, {Hsiao}, {Ashall}, {Teffs},
  {Hoeflich}, {Morrell}, {Phillips}, {Anderson}, {Baron}, {Burns}, {Contreras},
  {Davis}, {Diamond}, {Folatelli}, {Galbany}, {Gall}, {Hachinger}, {Holmbo},
  {Karamehmetoglu}, {Kasliwal}, {Kirshner}, {Krisciunas}, {Kumar}, {Lu},
  {Marion}, {Mazzali}, {Piro}, {Sand}, {Stritzinger}, {Suntzeff}, {Taddia}, \&
  {Uddin}}]{Shahbandeh2021}
{Shahbandeh}, M., {Hsiao}, E.~Y., {Ashall}, C., {et~al.} 2022, \apj, 925, 175

\bibitem[{{Simcoe} {et~al.}(2013){Simcoe}, {Burgasser}, {Schechter}, {Fishner},
  {Bernstein}, {Bigelow}, {Pipher}, {Forrest}, {McMurtry}, {Smith}, \&
  {Bochanski}}]{simcoe13}
{Simcoe}, R.~A., {Burgasser}, A.~J., {Schechter}, P.~L., {et~al.} 2013, \pasp,
  125, 270

\bibitem[{{Simon} \& {Axford}(1966)}]{Simon1966}
{Simon}, M. \& {Axford}, W.~I. 1966, \planss, 14, 901

\bibitem[{{Skrutskie} {et~al.}(2006){Skrutskie}, {Cutri}, {Stiening},
  {Weinberg}, {Schneider}, {Carpenter}, {Beichman}, {Capps}, {Chester},
  {Elias}, {Huchra}, {Liebert}, {Lonsdale}, {Monet}, {Price}, {Seitzer},
  {Jarrett}, {Kirkpatrick}, {Gizis}, {Howard}, {Evans}, {Fowler}, {Fullmer},
  {Hurt}, {Light}, {Kopan}, {Marsh}, {McCallon}, {Tam}, {Van Dyk}, \&
  {Wheelock}}]{skrutskie2006}
{Skrutskie}, M.~F., {Cutri}, R.~M., {Stiening}, R., {et~al.} 2006, \aj, 131,
  1163

\bibitem[{{Smartt} {et~al.}(2015){Smartt}, {Valenti}, {Fraser}, {Inserra},
  {Young}, {Sullivan}, {Pastorello}, {Benetti}, {Gal-Yam}, {Knapic},
  {Molinaro}, {Smareglia}, {Smith}, {Taubenberger}, {Yaron}, {Anderson},
  {Ashall}, {Balland}, {Baltay}, {Barbarino}, {Bauer}, {Baumont}, {Bersier},
  {Blagorodnova}, {Bongard}, {Botticella}, {Bufano}, {Bulla}, {Cappellaro},
  {Campbell}, {Cellier-Holzem}, {Chen}, {Childress}, {Clocchiatti},
  {Contreras}, {Dall'Ora}, {Danziger}, {de Jaeger}, {De Cia}, {Della Valle},
  {Dennefeld}, {Elias-Rosa}, {Elman}, {Feindt}, {Fleury}, {Gall},
  {Gonzalez-Gaitan}, {Galbany}, {Morales Garoffolo}, {Greggio}, {Guillou},
  {Hachinger}, {Hadjiyska}, {Hage}, {Hillebrandt}, {Hodgkin}, {Hsiao}, {James},
  {Jerkstrand}, {Kangas}, {Kankare}, {Kotak}, {Kromer}, {Kuncarayakti},
  {Leloudas}, {Lundqvist}, {Lyman}, {Hook}, {Maguire}, {Manulis}, {Margheim},
  {Mattila}, {Maund}, {Mazzali}, {McCrum}, {McKinnon}, {Moreno-Raya},
  {Nicholl}, {Nugent}, {Pain}, {Pignata}, {Phillips}, {Polshaw}, {Pumo},
  {Rabinowitz}, {Reilly}, {Romero-Ca{\~n}izales}, {Scalzo}, {Schmidt},
  {Schulze}, {Sim}, {Sollerman}, {Taddia}, {Tartaglia}, {Terreran},
  {Tomasella}, {Turatto}, {Walker}, {Walton}, {Wyrzykowski}, {Yuan}, \&
  {Zampieri}}]{smartt2015}
{Smartt}, S.~J., {Valenti}, S., {Fraser}, M., {et~al.} 2015, \aap, 579, A40

\bibitem[{{Smith} {et~al.}(2002){Smith}, {Tucker}, {Kent}, {Richmond},
  {Fukugita}, {Ichikawa}, {Ichikawa}, {Jorgensen}, {Uomoto}, {Gunn}, {Hamabe},
  {Watanabe}, {Tolea}, {Henden}, {Annis}, {Pier}, {McKay}, {Brinkmann}, {Chen},
  {Holtzman}, {Shimasaku}, \& {York}}]{smith2002}
{Smith}, J.~A., {Tucker}, D.~L., {Kent}, S., {et~al.} 2002, \aj, 123, 2121

\bibitem[{{Smith}(1967)}]{smith1967}
{Smith}, L.~F. 1967, \aj, 72, 829

\bibitem[{{Smith} \& {Arnett}(2014)}]{Smith2014}
{Smith}, N. \& {Arnett}, W.~D. 2014, \apj, 785, 82

\bibitem[{{Stritzinger} {et~al.}(2016){Stritzinger}, {Hsiao}, {Morrell},
  {Phillips}, {Contreras}, {Castell{\'o}n}, \& {Clocchiatti}}]{stritzinger2016}
{Stritzinger}, M., {Hsiao}, E.~Y., {Morrell}, N., {et~al.} 2016, The
  Astronomer's Telegram, 8657, 1

\bibitem[{{Stritzinger} {et~al.}(2009){Stritzinger}, {Mazzali}, {Phillips},
  {Immler}, {Soderberg}, {Sollerman}, {Boldt}, {Braithwaite}, {Brown}, {Burns},
  {Contreras}, {Covarrubias}, {Folatelli}, {Freedman}, {Gonz{\'a}lez}, {Hamuy},
  {Krzeminski}, {Madore}, {Milne}, {Morrell}, {Persson}, {Roth}, {Smith}, \&
  {Suntzeff}}]{stritzinger09}
{Stritzinger}, M., {Mazzali}, P., {Phillips}, M.~M., {et~al.} 2009, \apj, 696,
  713

\bibitem[{{Stritzinger} {et~al.}(2018){Stritzinger}, {Taddia}, {Burns},
  {Phillips}, {Bersten}, {Contreras}, {Folatelli}, {Holmbo}, {Hsiao},
  {Hoeflich}, {Leloudas}, {Morrell}, {Sollerman}, \&
  {Suntzeff}}]{stritzinger2018b}
{Stritzinger}, M.~D., {Taddia}, F., {Burns}, C.~R., {et~al.} 2018, \aap, 609,
  A135

\bibitem[{{Stritzinger} {et~al.}(2022){Stritzinger}, {Taddia}, {Lawrence},
  {Patat}, {Fraser}, {Galbany}, {Holmbo}, {Hyder}, \&
  {Karamehmetoglu}}]{Stritzinger2022}
{Stritzinger}, M.~D., {Taddia}, F., {Lawrence}, S.~S., {et~al.} 2022, \apjl,
  939, L8

\bibitem[{{Taddia} {et~al.}(2016){Taddia}, {Fremling}, {Sollerman}, {Corsi},
  {Gal-Yam}, {Karamehmetoglu}, {Lunnan}, {Bue}, {Ergon}, {Kasliwal},
  {Vreeswijk}, \& {Wozniak}}]{Taddia2016}
{Taddia}, F., {Fremling}, C., {Sollerman}, J., {et~al.} 2016, \aap, 592, A89

\bibitem[{{Taddia} {et~al.}(2015){Taddia}, {Sollerman}, {Leloudas},
  {Stritzinger}, {Valenti}, {Galbany}, {Kessler}, {Schneider}, \&
  {Wheeler}}]{taddia15sdss}
{Taddia}, F., {Sollerman}, J., {Leloudas}, G., {et~al.} 2015, \aap, 574, A60

\bibitem[{{Taddia} {et~al.}(2018){Taddia}, {Stritzinger}, {Bersten}, {Baron},
  {Burns}, {Contreras}, {Holmbo}, {Hsiao}, {Morrell}, {Phillips}, {Sollerman},
  \& {Suntzeff}}]{taddia18csp}
{Taddia}, F., {Stritzinger}, M.~D., {Bersten}, M., {et~al.} 2018, \aap, 609,
  A136

\bibitem[{{Tartaglia} {et~al.}(2021){Tartaglia}, {Sollerman}, {Barbarino},
  {Taddia}, {Mason}, {Berton}, {Taggart}, {Bellm}, {De}, {Frederick},
  {Fremling}, {Gal-Yam}, {Golkhou}, {Graham}, {Ho}, {Hung}, {Kaye}, {Kim},
  {Laher}, {Masci}, {Perley}, {Porter}, {Reiley}, {Riddle}, {Rusholme},
  {Soumagnac}, \& {Walters}}]{Tartaglia2021}
{Tartaglia}, L., {Sollerman}, J., {Barbarino}, C., {et~al.} 2021, \aap, 650,
  A174

\bibitem[{{Thomas} {et~al.}(2016){Thomas}, {Tucker}, {Childress}, {Moller},
  {Ruiter}, {Schmidt}, {Seitenzahl}, {Yuan}, \& {Zhang}}]{thomas2016}
{Thomas}, A., {Tucker}, B.~E., {Childress}, M., {et~al.} 2016, The Astronomer's
  Telegram, 8664, 1

\bibitem[{{Tonry} {et~al.}(2018){Tonry}, {Denneau}, {Flewelling}, {Heinze},
  {Onken}, {Smartt}, {Stalder}, {Weiland}, \& {Wolf}}]{tonry018}
{Tonry}, J.~L., {Denneau}, L., {Flewelling}, H., {et~al.} 2018, \apj, 867, 105

\bibitem[{{Tsuna} \& {Takei}(2023)}]{Tsuna2023}
{Tsuna}, D. \& {Takei}, Y. 2023, \pasj, 75, L19

\bibitem[{{Tuthill} {et~al.}(2006){Tuthill}, {Monnier}, {Tanner}, {Figer},
  {Ghez}, \& {Danchi}}]{Tuthill2006}
{Tuthill}, P., {Monnier}, J., {Tanner}, A., {et~al.} 2006, Science, 313, 935

\bibitem[{{Valenti} {et~al.}(2008){Valenti}, {Elias-Rosa}, {Taubenberger},
  {Stanishev}, {Agnoletto}, {Sauer}, {Cappellaro}, {Pastorello}, {Benetti},
  {Riffeser}, {Hopp}, {Navasardyan}, {Tsvetkov}, {Lorenzi}, {Patat}, {Turatto},
  {Barbon}, {Ciroi}, {Di Mille}, {Frandsen}, {Fynbo}, {Laursen}, \&
  {Mazzali}}]{valenti2008}
{Valenti}, S., {Elias-Rosa}, N., {Taubenberger}, S., {et~al.} 2008, \apjl, 673,
  L155

\bibitem[{{Valenti} {et~al.}(2012){Valenti}, {Taubenberger}, {Pastorello},
  {Aramyan}, {Botticella}, {Fraser}, {Benetti}, {Smartt}, {Cappellaro},
  {Elias-Rosa}, {Ergon}, {Magill}, {Magnier}, {Kotak}, {Price}, {Sollerman},
  {Tomasella}, {Turatto}, \& {Wright}}]{Valenti2012}
{Valenti}, S., {Taubenberger}, S., {Pastorello}, A., {et~al.} 2012, \apjl, 749,
  L28

\bibitem[{{Vink}(2017)}]{Vink2017}
{Vink}, J.~S. 2017, Philosophical Transactions of the Royal Society of London
  Series A, 375, 20160269

\bibitem[{{Williams} {et~al.}(2021){Williams}, {Morrell}, {Boutsia}, \&
  {Massey}}]{Williams2021}
{Williams}, P.~M., {Morrell}, N.~I., {Boutsia}, K., \& {Massey}, P. 2021,
  \mnras, 505, 5029

\bibitem[{{Yi} {et~al.}(2016){Yi}, {Zhang}, {Wu}, {Shappee}, {Prieto}, \&
  {Dong}}]{yi2016}
{Yi}, W., {Zhang}, J.-J., {Wu}, X.-B., {et~al.} 2016, The Astronomer's
  Telegram, 8655, 1

\bibitem[{{Yoon}(2017)}]{Yoon2017}
{Yoon}, S.-C. 2017, \mnras, 470, 3970

\bibitem[{{Zavala} {et~al.}(2022){Zavala}, {Toal{\'a}}, {Santamar{\'\i}a},
  {Ramos-Larios}, {Sabin}, {Quino-Mendoza}, {Rubio}, \&
  {Guerrero}}]{Zavala2022}
{Zavala}, S., {Toal{\'a}}, J.~A., {Santamar{\'\i}a}, E., {et~al.} 2022, \mnras,
  513, 3317

\bibitem[{{Zenati} {et~al.}(2022){Zenati}, {Wang}, {Bobrick}, {DeMarchi},
  {Glanz}, {Rozner}, {Rest}, {Metzger}, {Margutti}, {Gomez}, {Smith}, {Toonen},
  {Bright}, {Norman}, {Foley}, {Gagliano}, {Krolik}, {Smartt}, {Villar},
  {Narayan}, {Fox}, {Auchettl}, {Brethauer}, {Clocchiatti}, {Coelln},
  {Coppejans}, {Dimitriadis}, {Doroszmai}, {Drout}, {Jacobson-Galan}, {Gao},
  {Ridden-Harper}, {Kilpatrick}, {Laskar}, {Matthews}, {Rest}, {Smith},
  {McKenzie Stauffer}, {Stroh}, {Strolger}, {Terreran}, {Pierel}, \&
  {Piro}}]{Zenati2022}
{Zenati}, Y., {Wang}, Q., {Bobrick}, A., {et~al.} 2022, arXiv e-prints,
  arXiv:2207.07146

\end{thebibliography}

\clearpage
\begin{table}
\centering
\caption{Journal of Spectroscopic Observations.\label{tab:specjor}}
\begin{tabular}{lllll}
\hline
\hline 
Date & JD-2,457,000+ & Phase\tablefootmark{a} & Telescope & Instrument \\
\hline
\multicolumn{5}{c}{\bf Optical}\\
2016 Feb 08   &  426.60 & $-$6.9   & Lijiang-2.4m & YFOSC          \\
2016 Feb 09   &  427.68 & $-$5.8   & du Pont & WFCCD                  \\
2016 Feb 09   &  427.7  & $-$5.8   & ANU-2.3m & WiFeS                 \\
2016 Feb 10   &  428.78 & $-$4.7   & du Pont & WFCCD                 \\
2016 Feb 12   &  430.86 & $-$2.6   & du Pont & WFCCD                  \\
2016 Feb 13   &  431.85 & $-$1.6   & NTT     & EFOSC                \\
2016 Feb 16   &  435.81 & $+$2.3   & NTT     & EFOSC                \\
2016 Feb 21   &  435.82 & $+$2.4   & BAADE   & MagE                  \\
2016 Feb 23   &  441.85 & $+$8.4   & NTT     & EFOSC                \\
2016 Mar 08   &  456.90 & $+$23.4  & NTT     & EFOSC                \\
2016 Mar 11   &  458.84 & $+$25.4  & CLAY    & LDSS3                 \\
2016 Mar 17   &  464.73 & $+$31.3  & du Pont & B\&C                  \\
2016 Mar 29   &  476.64 & $+$43.2  & BAADE   & IMACS                 \\
2016 Apr 04   &  482.82 & $+$49.4  & NTT     & EFOSC                \\ 
2016 Jun 10   &  549.68 & $+$116.2 & du Pont & WFCCD                 \\
2016 Jul 20   &  589.49 & $+$156.0 & CLAY    & LDSS3                 \\
\multicolumn{5}{c}{\bf NIR}\\       
2016 Feb 15   &  433.80 & $+$0.3   & NTT    & Sofi                   \\
2016 Feb 18   &  436.83 & $+$3.4   & NTT    & Sofi                   \\
2016 Feb 21   &  439.74 & $+$6.3   & BAADE  & FIRE                   \\
2016 Feb 24   &  442.82 & $+$9.4   & NTT    & Sofi      \\
2016 Mar 08   &  455.82 & $+$22.4  & NTT    & Sofi      \\
2016 Mar 18   &  465.89 & $+$32.4  & BAADE  & FIRE      \\
2016 Mar 27   &  474.81 & $+$41.3  & BAADE  & FIRE      \\ 
2016 Apr 12   &  491.82 & $+$58.4  & NTT    & Sofi      \\
2016 May 22   &  530.60 & $+$97.1  & BAADE  & FIRE\tablefootmark{b}\\ 
2016 Jul 15   &  584.47 & $+$151.0 & BAADE  & FIRE      \\
2016 Jul 22   &  592.45 & $+$159.0 & BAADE  & FIRE\tablefootmark{b}\\ 
2016 Aug 01   &  602.49 & $+$169.0 & NTT    & Sofi      \\ 
2016 Aug 09   &  609.50 & $+$176.0 & NTT    & Sofi      \\
\hline 
\end{tabular}
\tablefoot{\\
\tablefoottext{a}{Days relative to the epoch of $B$-band maximum, i.e., JD-2,457,433.47.}
\tablefoottext{b}{Spectrum obtained in echelle mode.}
}
\end{table}

\clearpage 

\begin{table}
\centering
\scriptsize
\caption{Spectral line IDs in the optical.\label{tab:linelogopt}}
\begin{tabular}{c|cc|c|cc|cc|c|cc|c|c|c|cc}
\hline
\hline
Phase\tablefootmark{a} & \ion{C}{ii}  & &
[\ion{O}{i}] & \ion{O}{i} & &\ion{Ca}{ii}& & [\ion{Ca}{ii}]&\ion{C}{i} & &
\ion{Sc}{ii} &\ion{Ba}{ii} &
\ion{Fe}{ii} &
\ion{Na}{i} \\
        & 6580        &7234 & 6300        & 7774        & 9263   &   8542 & H$\&$K &  7291,324     &  9086,95      & 9406   &  5531,5663,6246  &   6142,6497    &  multiplet 42\tablefootmark{*}    &  5890,96\tablefootmark{**}         \\
        \hline 
$-$6.9   & Y            & Y    &  N        &  Y          &  $-$   &   N            & $-$ &   N           &  N            &  $-$   &    N             &    N           &   N                &   ?               \\
$-$5.8     & Y            & Y    &  N        &  Y          &  $-$   &   Y            & N   &   N           &  N            &  $-$   &    N             &    N           &   N                &   ?               \\
$-$5.8    & Y            & Y    &  N        &  Y          &  Y     &   Y            & N   &   N           &  N            &  $-$   &    N             &    N           &   N                &   ?               \\
$-$4.7     & Y            & Y    &  N        &  Y          &  $-$   &   Y            & N   &   N           &  N            &  $-$   &    N             &    N           &   N                &   ?               \\
$-$2.6     & Y            & Y    &  N        &  Y          &  $-$   &   Y            & N   &   N           &  N            &  $-$   &    N             &    N           &   N                &   ?               \\
$-$1.6    & Y            & Y    &  N        &  Y          &  Y     &   Y            & ?   &   N           &  N            &  N     &    N             &    N           &   N                &   ?               \\
$+$2.3     & Y            & Y    &  N        &  Y          &  Y     &   Y            & ?   &   N           &  N            &  N     &    N             &    N           &   Y                &   ?              \\
$+$2.4   & Y            & Y    &  N        &  Y          &  Y     &   Y            & Y   &   N           &  N            &  N     &    N             &    N           &   Y                &   ?               \\
$+$8.4     & Y            & ?    &  N        &  Y          &  Y     &   Y            & Y   &   N           &  N            &  Y     &    N             &    N           &   Y                &   ?               \\
$+$23.4     & N            &  N   &  N        &  Y          &  Y     &   Y            & Y   &   N           &  Y            &  Y     &    Y             &    Y           &   Y                &   Y               \\
$+$25.4     & N            &  N   &  N        &  Y          &  Y     &   Y            & ?   &   N           &  Y            &  $-$   &    Y             &    Y           &   Y                &   Y              \\
$+$31.3     & N            &  N   &  N        &  Y          &  Y     &   Y            & $-$ &   N           &  Y            &  Y     &    Y             &    Y           &   Y                &   Y                \\
$+$43.2     & N            &  N   &  N        &  Y          &  Y     &   Y            & $-$ &   N           &  Y            &  Y     &    Y             &    Y           &   Y                &   Y                \\
$+$49.4    & N            &  N   &  N        &  Y          &  Y     &   Y            & ?   &   N           &  Y            &  Y     &    Y             &    Y           &   Y                &   Y               \\
$+$116.2  & N            &  N   &  Y        &  Y          &  $-$   &   Y            & $-$ &   Y           &  Y            &  $-$   &    N             &    N           &   N                &   Y                \\
$+$156.0 & N            &  N   &  Y        &  Y          &  $-$   &   Y            & $-$ &   Y           &  Y            &  $-$   &    N             &    N           &   N                &   Y                \\
\hline 
\end{tabular}
\end{table}
\tablefoot{Y $=$ detection, N $=$ non detection, $-$ $=$ line wavelength not covered, ? $=$ uncertain detection.\\
\tablefoottext{a}{Days relative to the epoch of $B$-band maximum, i.e., JD-2,457,433.47.}
\tablefoottext{*}{Same evolution for \ion{Fe}{ii} $+$ \ion{Co}{ii}.}
\tablefoottext{**}{Narrow \ion{Na}{i}~D present at all epochs.}
}

\clearpage 

\begin{sidewaystable}
\tiny
\centering
\caption{Spectral line IDs in the NIR.\label{tab:linelognir}}
\begin{tabular}{l|llllllllll|llllll|ll|l}
\hline\hline             
Phase\tablefootmark{a} & \ion{C}{i} & &&&&&&&&&\ion{H}{i} &&&&&&\ion{Mg}{i}&&\ion{Ca}{ii} \\
 &&&&&&&&&&&Pa-$\zeta$ &Pa-$\delta$ &Pa-$\gamma$ &Pa-$\beta$ &Br-$\zeta$ &Br-$\gamma$ && &\\
\hline
   &  9086,95 & 9406  &9605,23 & 10695   &11330\tablefootmark{*} & 11848 & 12614& 13743 & 14540& 16890&  9231.5\tablefootmark{**} &  10049   & 10938& 12822 &  17367   &21661&  15025         &  21465  & 8542  \\ 
   
\hline
  $+$0.33    &   $-$      &  N    &   Y      & Y       &Y                        &  Y    &  Y   &   Y   &   Y  & Y    &  $-$                                &  N                &   Y?       &   N           &   N               &    N            &  N             &  Y      & $-$ \\
  $+$3.36    &   $-$      &  N    &   Y      & Y       &Y                        &  Y    &  Y   &   ?   &   Y  & Y    &  $-$                                &  N                &   Y?           &   N           &   N               &    N            &  N             &  Y      & $-$ \\
$+$6.27    &   N        &  N    &   Y      & Y       &Y                        &  Y    &  Y   &   Y   &   Y  & Y    &  Y                                  &  N                &   Y?           &   N           &   N               &    N            &  N             &  Y      & Y \\
$+$9.35    &   $-$      &  Y?   &   Y      & Y       &Y                        &  Y    &  Y   &   ?   &   Y  & Y    &  $-$                                &  N                &   Y?           &   N           &   N               &    N            &  N             &  Y      & $-$ \\
$+$22.35   &   $-$      &  Y    &   Y      & Y       &Y                        &  Y    &  Y   &   ?   &   Y  & Y    &  $-$                                &  N                &   Y?          &   N           &   N               &    N            &  Y             &  Y      & $-$ \\
$+$32.42   &   Y        &  Y    &   Y      & Y       &Y                        &  Y    &  Y   &   Y   &   Y  & Y    &  Y                                  &  N                &   Y?           &   N           &   N               &    N            &  Y             &  Y      & Y\\
$+$41.34   &   Y        &  Y    &   Y      & Y       &Y                        &  Y    &  Y   &   Y   &   Y  & Y    &  Y                                  &  N                &   Y?           &   N           &    N              &    N            &  Y             &  Y      & Y \\ 
$+$58.35   &   $-$      &  Y    &   Y      & Y       &N                        &  Y    &  Y   &   ?   &   Y  & Y    &  $-$                                &  Y                &   Y           &   Y           &    N              &    N            &  Y             &  N      & $-$  \\
$+$97.13   &   Y        &  N    &   N      & Y       &N                        &  Y    &  N   &   $-$ &   N  & N    &  ?(Y)                               &  Y                &   Y           &   Y           &    Y              &    Y            &  Y             &  N      & Y \\ 
$+$151.00 &   Y        &  N    &   N      & Y       &N                        &  Y    &  N   &   N   &   N  & N    &  N                                  &  N                &   Y           &   Y           &    N              &    ?            &  Y             &  N      & Y \\
$+$158.98&   Y        &  N    &   N      & Y       &N                        &  Y    &  N   &   $-$ &   N  & N    &  N                                  &  N                &   Y           &   Y           &    N              &    Y            &  Y             &  N      & Y \\ 
$+$169.02&   $-$      &  N    &   N      & Y       &N                        &  Y    &  N   &   N   &   N  & N    &  N                                  &  N                &   Y           &   Y           &    N              &    N            &  Y             &  N      & $-$ \\ 
$+$176.03&   $-$      &  N    &   N      & Y       &N                        &  Y    &  N   &   N   &   N  & N    &  N                                  &  N                &   Y           &   Y           &    N              &    N            &  Y             &  N      & $-$ \\
\hline
\end{tabular}
\tablefoot{Y $=$ detection, N $=$ non detection, $-$ $=$ line wavelength not covered, ? $=$ uncertain detection.\\
\tablefoottext{a}{Days relative to the epoch of $B$-band maximum, i.e., JD$-$2,457,433.47.}
\tablefoottext{$*$}{Line blended with \ion{O}{i}~$\lambda$11287.}
\tablefoottext{$**$}{Line blended with \ion{O}{i}~$\lambda$9263.}
}
\end{sidewaystable}

\clearpage 
\begin{table}
\centering
\tiny
\caption{\ion{H}{i} line fluxes measured from the  $ +97$~d NIR spectrum of SN~2016adj (see Fig.~\ref{fig:Hlines}).\label{tab:H1fluxes}}
\begin{tabular}{lll}
\hline
\hline 
Line ID & Absorption & Emission \\
\hline
Pa-$\delta$ & $(2.47\pm0.20)\times10^{-13}$ & $(1.29\pm0.15)\times10^{-13}$ \\
Pa-$\gamma$ & $(7.33\pm1.56)\times10^{-14}$ & $(1.53\pm0.27)\times10^{-13}$ \\
Pa-$\beta$  & $(7.12\pm1.13)\times10^{-14}$ & $(7.88\pm1.16)\times10^{-14}$ \\
Br-$\zeta$  &   $\cdots$                    & $(1.07\pm0.41)\times10^{-14}$ \\
Br-$\gamma$ & $(5.98\pm3.73)\times10^{-15}$ & $(2.18\pm0.46)\times10^{-14}$\\
\hline 
\end{tabular}
\end{table}

\begin{table}
\centering
\tiny
\caption{Light curve parameters.\label{tab:peak}}
\begin{tabular}{llllll} 
\hline
\hline 
Filter & $t_{max}\tablefootmark{a}$ & $m_{max}$ & $M_{max} (A_{V}^{MW})$ & $M_{max} (A_{V}^{MW}+ A_{V}^{host})$ & $\Delta$m$_{\rm 15}(X)$\\
($X$) & (days) & (mag) & (mag) & (mag) & (mag) \\
\hline 
$u$ & $-$2.74      & 19.00    &$-$9.14     &$-$17.15 &1.32\\
$B$ & $+$0.00      & 17.48    &$-$10.61    &$-$18.05 &0.85\\
$g$ & $-$1.53      & 16.66    &$-$11.40    &$-$18.48 &0.60\\
$V$ & $+$0.00      & 15.76    &$-$12.23    &$-$18.58 &0.54\\
$r$ & $-$2.36      & 14.59    &$-$13.34    &$-$19.08 &0.41\\
$i$ & $-$1.71      & 13.26    &$-$14.60    &$-$18.95 &0.38\\
$Y$ & $<$~$+$11.24 & $<$11.33 &$<-$16.45   &$<-$18.85 & $\ldots$\\
$J$ & $<$~$+$0.31  & $<$10.49 &$<-$17.26   &$<-$18.91 & $\ldots$\\
$H$ & $<$~$+$3.36  & $<$9.88  &$<-$17.84   &$<-$18.88 & $\ldots$\\
$K$ & $<$~$+$0.42  & $<$9.29  &$<-$18.41   &$<-$19.10 & $\ldots$\\
\hline 
\end{tabular}
\tablefoot{\\
\tablefoottext{a}{Errors on peak epoch, apparent peak magnitude and $\Delta$m$_{\rm 15}(X)$ are  $\approx2$ days, $\approx0.05$ mag and $\approx0.1$ mag, respectively. The extinction correction is based on the best-fit FTZ99 reddening law  to the color excesses of SN~2016adj  with $R_V^{host}=5.7\pm0.7$  and $A_V^{host}=6.3\pm0.2$ mag   (see Fig.~\ref{fig:colorexcess}). Finally, the uncertainty on the distance is $\approx$0.31 Mpc.}
}
\end{table}

\begin{table}
\centering
\caption{Hydrogen emission-line ratios.  \label{tab:lineratios}}
\begin{tabular}{lc}
\hline
\hline 
Line &  Ratio with Pa-$\beta$ \\
\hline 
Pa-$\beta$   &  1.00 \\
Pa-$\gamma$  &  1.94\\
Pa-$\delta$  &  1.64\\
Br-$\gamma$  &  0.28\\
Br-$\zeta$   &  0.14\\
\hline
\end{tabular}
\tablefoot{Line ratios computed using emission line flux values listed in Table~\ref{tab:H1fluxes}.}
\end{table}

\begin{appendix}

\section{Observations and data reductions}
\label{sec:appendixobs}

\subsection{Ground-based observations}

We collected an  extensive set of ground-based optical and NIR  photometry and spectroscopy of  SN~2016adj. The bulk of the data were obtained by the Carnegie Supernova Project-II \citep[][hereafter CSP-II]{Phillips2019} and the Public ESO Spectroscopic Survey of Transient Objects  \citep[][hereafter PESSTO]{smartt2015}. 
Finally, we identified a  detection of SN~2016adj recovered in a single VST $i$-band image obtained nearly two weeks prior to the original discovery by BOSS.

\subsubsection{CSP-II observations}
\label{sec:cspIIobs}

 The CSP-II obtained detailed followup spectroscopy and photometry  using key facilities located on the
 Las Campanas Observatory (LCO). 
  As summarized in Table~\ref{tab:specjor}, CSP-II obtained 9 epochs of  optical spectra with  the du Pont telescope equipped with the  B\&C  and WFCCD  spectrographs, the Magellan BAADE telescope equipped with IMACS (Inamori-Magellan Areal Camera \& Spectrograph; \citealt{dressler2011}), and the Magellan Baade telescope equipped with MagE (Magellan Echellette) and LDSS3 (Low Dispersion Survey Spectrograph). These data were reduced in the IRAF environment following standard procedures as outlined by \citet{hamuy2006}. 
 
 The CSP-II also obtained 6 epochs of NIR spectroscopy   with the FIRE (Folded port InfraRed Echellette; \citealt{simcoe13}) spectrograph   attached to the Magellan BAADE telescope. Four of these spectra were obtained in low-resolution mode, while two were obtained in echelle mode providing higher spectral resolution. The lower resolution data were reduced following procedures  outlined by  \citet{hsiao2019}, while the FIRE/echelle spectra were reduce  using the \texttt{FIREHOSE IDL} pipeline provided by the instrument developer R. Simcoe \citep{simcoe13}.

 As the  detection  of hydrogen features in the NIR spectra of SN~2016adj are of significant interest,   particular attention was paid to  investigate whether or not the purported P-Cygni hydrogen features detected in the $+$97~d medium-resolution  spectrum (see Fig.~\ref{fig:Hlines}) are intrinsic and not simply an artifact of the telluric correction. 

The $+$97~d NIR spectrum was telluric corrected using observations of the telluric standard star HIP 64381 obtained just after the science data. Indeed, at times the kernel constructed from the Pa-$\delta$ feature of this star can lead to corrections that leave residual artefacts, however, not with the relative intensities of the features present in the science spectrum.  
Turning to the $+$159~d echelle spectrum unfortunately no telluric spectrum was obtained, and therefore that used in the previous echelle spectral data reductions was used, which provided adequate telluric removal. Visual inspection of the 2-D science images of both echelle nights reveal obvious features at the location of the detected hydrogen features. Furthermore, any artefacts related to telluric removal would not be accompanied by P-Cygni profiles, and if present  the artefacts would be located at zero red-shift, which in the rest-frame of Centaurus~A would be around $-500$~km~s$^{-1}$. This is inconsistent with the line velocities of the NIR lines highlighted  in Fig.~\ref{fig:Hlines}. 

\begin{figure}
\centering
\resizebox{\hsize}{!}
{\includegraphics{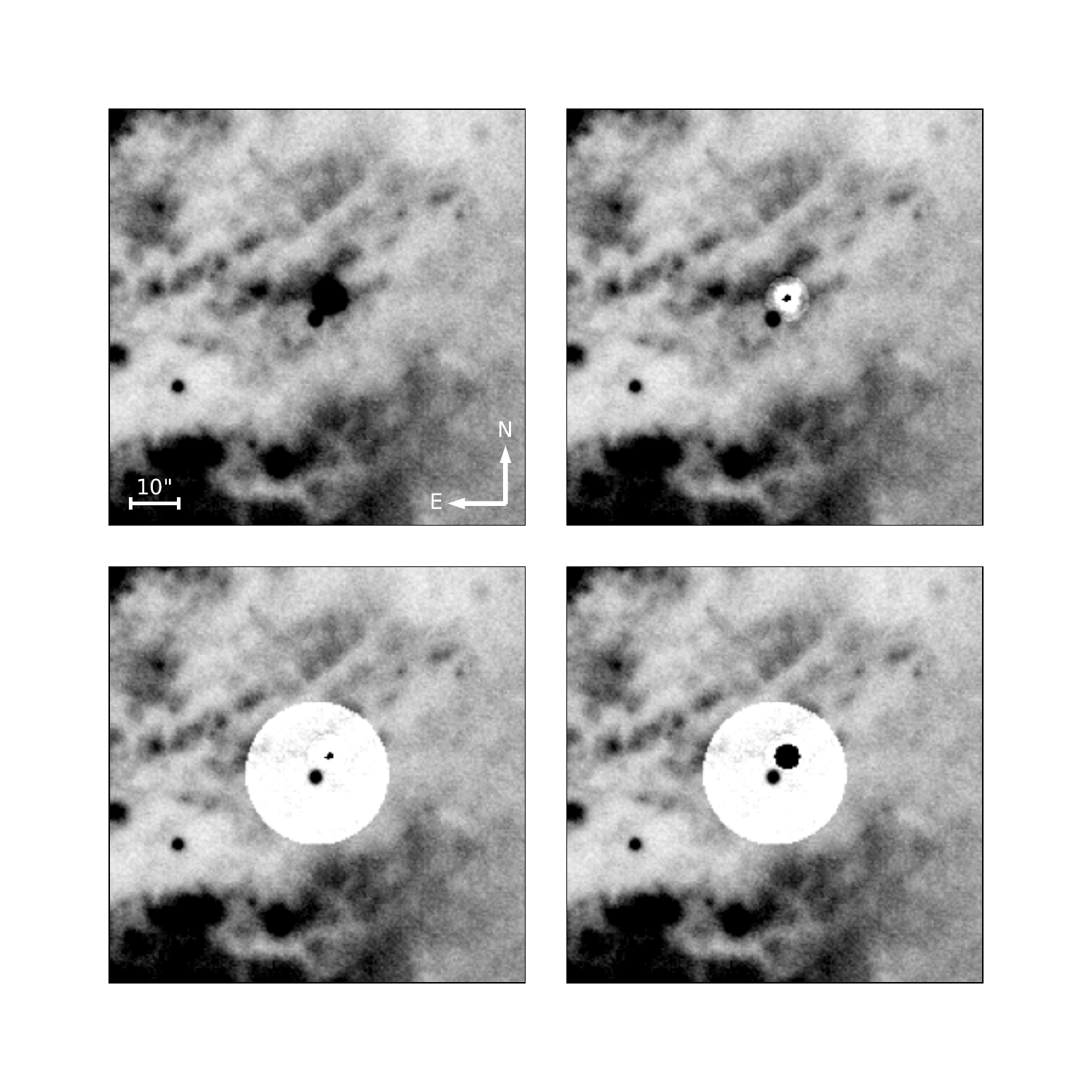}}
\caption{Steps of host-galaxy template subtraction. \textit{Top left:} Optical-band science image created from stacking multiple images and zoomed in on the location of SN~2016adj. North is up and East is Left.  \textit{Top right:} Residual obtained after subtraction of a modeled PSF from the PSF of the bright foreground star located in close proximity to SN~2016adj. \textit{Bottom right:} Mask applied to the residual of the bright star. \textit{Bottom left:} Region surrounding SN~2016adj copied into the initial science image after performing host-galaxy template subtraction of the masked science image.}
\label{fig:tempsub}
\end{figure}

Turning to the broadband imaging of SN~2016adj, depending on the particular band, up to 30  epochs of optical ($uBgVri$) photometry  were obtained with the 1-m Swope telescope equipped with a  e2v CCD. The CSP-II also obtained 8 epochs of broad-band NIR  ($YJH$) imaging with the 2.5-m du Pont telescope equipped with the NIR imager RetroCam  containing a Rochwell HAWAII-1 HgCdTe detector. These data were reduced as described by \citet[][and references therein]{krisciunas2017}.

The bright star in close vicinity of the position of SN~2016adj posed significant issues to standard template subtraction and PSF photometry techniques.  After extensive trial and error, optimal results were obtained  following the approach illustrated in Fig.~\ref{fig:tempsub}. 
First, for each stacked science image (top left) a PSF model  constructed from field stars was subtracted from the PSF of the bright star,  resulting in a residual of it (top right). This residual is then masked (bottom right), and then host-galaxy template subtraction is performed  yielding a suitable data product (bottom left) to measure the brightness of the SN free of both contamination by the bright star and the underlying host-galaxy light.
 
 PSF photometry of SN~2016adj was  computed    differential relative to a local sequence of secondary standard stars. These secondary standards were themselves calibrated relative to standard star fields  observed over multiple photometric nights and with  photometry presented by \citet{landolt1992} ($BV$), \citet{smith2002} ($ugri$), \citet{krisciunas2017} ($Y$), and \citet{persson1998} ($JH$). In order to compute photometry of the SN in the natural photometric system, which represents the purest form of the data, the published standard star  magnitudes were converted to the CSP-II \textit{natural} system following the relation (using the $B$ band as an example): $B_{nat} = B_{STD} + CT_{B}(B_{STD}-V_{STD})$, where  color term values are taken from Table~4 of \citet{krisciunas2017}. 
 
 For reference, photometry of the optical ($uBgVri$) and NIR ($YJH$) secondary standards in the standard system is provided  in Tables~\ref{tab:optlocseq} and \ref{tab:nirlocseq}, respectively.
 All photometry presented in this work was computed using the \texttt{PYTHON}-based photometry  pipeline developed by the Aarhus-Barcelona FLOWS\footnote{\url{https://flows.phys.au.dk/}}  project  and  available on \texttt{GITHUB}\footnote{\url{https://github.com/SNflows/}}.  

Final natural system optical and NIR  photometry of SN~2016adj obtained by CSP-II  is listed in Table~\ref{tab:phot} with accompanying 1-$\sigma$ uncertainties  computed by adding in quadrature the PSF-photometry error and the nightly zero-point error.


 \subsubsection{PESSTO observations}
 
 PESSTO obtained optical-NIR photometry and spectroscopy  with the ESO-La Silla Observatory's 3.4-m New Technology Telescope (NTT), equipped with EFOSC2 (ESO Faint Object Spectrograph and Camera; \citealt{buzzoni1984})  and  Sofi (Son Of ISAAC; \citealt{moorwood1998}).  The data set includes 5 optical spectra, 7 NIR spectra, several epochs of $uBgVri$-band photometry, and up to 7 epochs of  NIR photometry.
 Reductions of the data were performed   following the procedures outlined in the PESSTO survey paper  \citep{smartt2015}. 

Prior to computing photometry of the EFOSC2 science images, each image went through the same arduous template-subtraction process described in Sect.~\ref{sec:cspIIobs} and summarized in Fig.~\ref{fig:tempsub}. Due to the limited field-of-view afforded by EFOSC2 and Sofi the   local sequences defined in Sect.~\ref{sec:cspIIobs} and used with the CSP-II images were not suitable to compute nightly zero-points. Nightly optical and NIR zeropoints were therefore computed relative to ATLAS-REFCAT2 \citep{tonry018} and 2MASS \citep{skrutskie2006} stars contained within each of the PESSTO science images. In the optical the number of calibration stars typically consisted of (depending on the filter and exposure time) between $\sim$30-50, while in the NIR the number of calibrating stars was typically $\gtrsim30$. Prior to computing the optical local night zeropoints the photometry of the calibrating stars was transformed to the NTT (+EFOSC2) natural system using color terms provided by \citet{smartt2015}. The resulting photometry is listed in Table~\ref{tab:phot}.


 \subsubsection{ESO VST early $i$-band image of SN~2016adj}

SN~2016adj was recovered in a single $i$-band image taken on 2016 January 26.3 UT with the ESO-Paranal VLT Survey Telescope (VST), equipped with the wide-field imaging camera OmegaCAM \citep{arnaboldi1998,kuijken2002,kuijken2011}.
The date of this observation turns out to be  $-$19.9~d prior to the epoch of $B$-band maximum, and only a  mere $+$2.5~d  past our inferred explosion epoch (see Sect.~\ref{sec:explosionparameters}).  This epoch is also  13.28 days  prior to the reported date of discovery \citep{marples2016}. 
Template subtraction  was performed on the science image as described above and  PSF photometry  was  computed deferentially  relative to 100 ATLAS-REFCAT2 stars  contained within the field of the science image. 

\subsubsection{ESO VLT X-shooter observations}

The X-Shooter medium-resolution spectrograph attempted to obtain  late-phase spectroscopy (ESO program 098.D-0540(A)) of SN~2016adj at late phases. 
Unfortunately, the  four 1-hour observational blocks  executed between 2017 February 03.2 UT and 2017 April 01.1 UT contained  no useful science data.



\subsection{Pre-maximum Swift/UVOT followup}

Over the first twenty-four hours after the reported discovery  several snapshot observations of SN~2016adj were obtained  with the UVOT camera \citep{roming2005} 
onboard the \textit{Neil Gehrels Swift Observatory} \citep{gehrels04}. Due to the significant host reddening, SN~2016adj was only detected in the  $uvw1$, $U$, $B$, and $V$ bands.
However, due to the high-galaxy background it was not possible to obtain reliable photometry  for the optical detections. Furthermore,  $uvw2$ was not considered because reddened sources like SN~2016adj are dominated by the red tail of this UV filter.
In summary, out of the initial \textit{Swift} observations, only  a single $uvw1$ point a few days before $B$-band maximum is recovered, as well as two upper limits  for both the $uvw1$ and  $uvm2$ bands around $-$6~d.
Aperture photometry  was  computed at the position of SN~2016adj following the method used for the Swift Optical Ultraviolet Supernova Archive (SOUSA;  \citealt{brown2014}). The underlying galaxy count rates were subtracted from the science flux measurements and  photometry was computed using a 5\arcsec\ aperture along with the \textit{Swift}/UVOT Vega-based photometric zeropoints  \citep{breeveld2011}  and the most recent sensitivity corrections.\footnote{\url{https://heasarc.gsfc.nasa.gov/docs/heasarc/caldb/swift/docs/uvot/uvotcaldb\_throughput\_06.pdf}} 
 The single UVOT photometry measurement as well as several limits are listed in Table~\ref{tab:swiftphot}. 
 The $uvw1$ measurement must be used with the understanding that the effective wavelength is shifted redward as SN~2016adj is such a red and reddened source  \citep{Brown2015}.

\clearpage 
\begin{table}
\centering
\tiny
\caption{Optical photometry of the local sequence  in the Swope `natural' system.\label{tab:optlocseq}}
\begin{tabular}{lllllllll}
\hline
\hline 
OBJ &
RA & 
DEC & 
$u$ &
$g$ &
$r$ &
$i$ & 
$B$ & 
$V$ \\
      &
(deg) & 
(deg) & 
(mag) & 
(mag) & 
(mag) & 
(mag) & 
(mag) & 
(mag) \\
\hline 
  1 & 201.522629 & $-$42.996449 & 12.491(006)& \ldots     & \ldots     & 10.889(005)& 11.671(017)& 11.131(007)\\
  2 & 201.369522 & $-$42.935387 & 14.196(002)& 12.555(003)& 11.821(003)& 11.501(002)& 12.975(002)& 12.184(002)\\
  3 & 201.462646 & $-$42.900326 & 13.600(002)& 12.411(003)& 12.038(002)& 11.949(002)& 12.694(003)& 12.244(002)\\
  4 & 201.262741 & $-$43.087578 & 15.933(003)& 13.298(002)& 12.283(002)& 11.881(002)& 13.833(002)& 12.776(002)\\
  5 & 201.218903 & $-$43.022713 & 15.915(003)& 13.378(002)& 12.447(002)& 12.093(002)& 13.883(002)& 12.900(002)\\
  6 & 201.331085 & $-$43.120270 & 14.538(002)& 13.205(002)& 13.013(002)& 13.037(002)& 13.415(002)& 13.081(002)\\
  7 & 201.451691 & $-$43.121784 & 14.639(002)& 13.369(002)& 12.890(002)& 12.739(002)& 13.680(002)& 13.112(002)\\
  8 & 201.387848 & $-$42.896599 & 14.583(002)& 13.357(002)& 12.955(002)& 12.846(002)& 13.640(002)& 13.151(002)\\
  9 & 201.292801 & $-$42.941120 & 14.698(002)& 13.423(002)& 12.947(002)& 12.799(002)& 13.729(002)& 13.177(002)\\
 10 & 201.314255 & $-$42.894527 & 14.773(003)& 13.549(003)& 13.120(003)& 12.980(002)& 13.842(002)& 13.338(002)\\
 11 & 201.375595 & $-$42.975994 & 16.466(004)& 13.966(002)& 12.969(002)& 12.546(002)& 14.498(002)& 13.447(002)\\
 12 & 201.513779 & $-$43.127510 & 14.941(002)& 13.706(003)& 13.269(003)& 13.157(002)& 13.995(002)& 13.463(002)\\
 13 & 201.190826 & $-$42.965912 & 15.272(003)& 13.774(002)& 13.204(003)& 13.027(002)& 14.112(002)& 13.464(002)\\
 14 & 201.294189 & $-$43.079380 & 15.199(002)& 13.908(002)& 13.419(002)& 13.266(002)& 14.207(002)& 13.648(002)\\
 15 & 201.376205 & $-$43.076775 & 15.452(003)& 14.049(002)& 13.497(002)& 13.315(002)& 14.376(002)& 13.752(002)\\
 16 & 201.389236 & $-$42.959129 & 16.094(003)& 14.187(002)& 13.457(002)& 13.211(002)& 14.599(002)& 13.802(002)\\
 17 & 201.209183 & $-$42.978321 & 15.301(003)& 14.059(002)& 13.565(002)& 13.401(002)& 14.368(002)& 13.790(002)\\
 18 & 201.193161 & $-$42.954117 & 15.450(003)& 14.089(002)& 13.563(003)& 13.405(003)& 14.401(002)& 13.798(002)\\
 19 & 201.406769 & $-$42.911167 & 15.945(003)& 14.219(002)& 13.601(002)& 13.417(002)& 14.593(002)& 13.896(002)\\
 20 & 201.482773 & $-$42.903744 & 15.456(003)& 14.239(002)& 13.778(002)& 13.644(002)& 14.534(002)& 13.988(002)\\
 21 & 201.281021 & $-$43.069424 & 17.379(006)& 14.719(002)& 13.347(002)& 12.269(002)& 15.306(002)& 14.046(002)\\
 22 & 201.417130 & $-$43.088680 & 15.666(003)& 14.356(002)& 13.836(002)& 13.666(002)& 14.676(002)& 14.074(002)\\
 23 & 201.307526 & $-$43.118488 & 15.642(003)& 14.361(002)& 13.850(002)& 13.666(002)& 14.670(002)& 14.088(002)\\
 24 & 201.261368 & $-$42.903248 & 17.017(005)& 14.672(002)& 13.687(002)& 13.248(002)& 15.200(002)& 14.185(002)\\
 25 & 201.480774 & $-$43.129978 & 17.556(008)& 14.843(002)& 13.511(002)& 12.829(002)& 15.410(002)& 14.198(002)\\
 26 & 201.341858 & $-$42.957169 & 15.841(003)& 14.514(002)& 13.991(002)& 13.814(003)& 14.830(002)& 14.233(002)\\
 27 & 201.474945 & $-$42.901730 & 16.374(004)& 14.573(002)& 13.905(002)& 13.717(002)& 14.959(002)& 14.222(002)\\
 28 & 201.294952 & $-$42.982819 & 16.316(004)& 14.737(002)& 14.031(002)& 13.718(002)& 15.122(002)& 14.366(002)\\
 29 & 201.206558 & $-$42.920586 & 17.379(006)& 14.918(002)& 13.918(003)& 13.509(002)& 15.422(002)& 14.403(002)\\
 30 & 201.289230 & $-$42.897415 & 15.904(003)& 14.713(002)& 14.258(003)& 14.113(003)& 15.002(002)& 14.462(002)\\
 31 & 201.218689 & $-$43.103252 & 16.832(005)& 14.933(002)& 14.128(002)& 13.780(002)& 15.375(002)& 14.508(002)\\
 32 & 201.249344 & $-$43.063400 & 16.922(005)& 14.932(002)& 14.190(002)& 13.949(003)& 15.335(002)& 14.545(002)\\
 33 & 201.291397 & $-$43.074680 & 16.467(004)& 14.898(002)& 14.276(003)& 14.076(003)& 15.258(002)& 14.565(002)\\
 34 & 201.472443 & $-$42.941597 & 16.455(004)& 14.913(002)& 14.269(003)& 14.041(003)& 15.266(002)& 14.577(002)\\
 35 & 201.334961 & $-$42.943768 & 16.607(004)& 14.967(002)& 14.303(003)& 14.067(003)& 15.336(002)& 14.620(002)\\
 36 & 201.387497 & $-$43.083870 & 16.459(004)& 15.023(002)& 14.401(003)& 14.165(003)& 15.374(002)& 14.689(002)\\
 37 & 201.398911 & $-$43.061455 & 17.876(009)& 15.232(002)& 14.064(002)& 13.533(002)& 15.750(002)& 14.671(002)\\
 38 & 201.267929 & $-$42.961002 & 16.731(004)& 15.010(002)& 14.310(003)& 14.027(003)& 15.394(002)& 14.648(002)\\
 39 & 201.466736 & $-$43.082668 & 17.579(007)& 15.165(002)& 14.257(003)& 13.941(003)& 15.648(002)& 14.690(002)\\
 40 & 201.452026 & $-$42.920498 & 16.340(004)& 14.981(002)& 14.438(003)& 14.254(003)& 15.310(002)& 14.688(002)\\
 41 & 201.463058 & $-$42.964729 & 16.615(004)& 15.037(002)& 14.434(003)& 14.241(003)& 15.392(002)& 14.715(002)\\
 \hline 
 \end{tabular}
\end{table}

\begin{table}
\centering
\tiny
\caption{CSP-II NIR local sequence photometry of SN~2016adj in the `standard' system. \label{tab:nirlocseq}}
\begin{tabular}{lllllllll}
\hline
\hline 
OBJ & 
RA*15 &
DEC   &
$Y$ &
$err$ &
$J$    &
$err$  & 
$H$    &
$err$  \\
 & (deg) & (deg) & (mag) & (mag) & (mag) & (mag) & (mag) & (mag) \\
\hline 
101 & 201.353500 & $-$43.025 & 12.150 & 0.613 & 11.708 & 0.664 & 11.088 & 0.670 \\
102 & 201.351501 & $-$43.038 & 12.101 & 0.008 & 11.679 & 1.486 & 11.032 & 0.007 \\
105 & 201.349304 & $-$43.044 & 14.845 & 0.045 & 14.395 & 0.031 & 13.768 & 0.034 \\
106 & 201.366898 & $-$43.043 & 16.063 & 0.115 & 15.539 & 0.085 & 14.987 & 0.096 \\
107 & 201.372406 & $-$43.001 & 14.581 & 0.689 & 14.151 & 0.756 & 13.860 & 0.730 \\
108 & 201.390488 & $-$43.026 & 14.555 & 0.625 & 14.286 & 0.713 & 14.029 & 0.724 \\
\hline 
 \end{tabular}
\end{table}

 \newpage 
 \onecolumn 
 
\begin{longtable}{llcccc}
\caption{Ground-based photometry of SN~2016adj.\label{tab:phot}}\\
\hline\hline 
MJD & phase\tablefootmark{a} &  mag &  errmag & filter & source \\
\hline 
\endfirsthead
\caption{continued.}\\
\hline\hline 
MJD & phase &  mag &  errmag & filter & source \\
\hline
\endhead
\hline
\endfoot
57427.27 &$-$5.70 & 19.05 &    0.04 &     $u$&    CSP \\
57429.39 &$-$3.58 & 19.14 &    0.05 &     $u$&    CSP \\
57430.37 &$-$2.60 & 19.08 &    0.06 &     $u$&    CSP \\
57434.29 &   1.32 & 19.07 &    0.04 &     $u$&    CSP \\
57435.23 &   2.26 & 19.19 &    0.04 &     $u$&    CSP \\
57437.30 &   4.34 & 19.36 &    0.06 &     $u$&    CSP \\
57439.24 &   6.27 & 19.53 &    0.11 &     $u$&    CSP \\
57441.23 &   8.26 & 20.03 &    0.25 &     $u$&    CSP \\
57442.26 &   9.30 & 20.17 &    0.21 &     $u$&    CSP \\
57443.25 &  10.28 & 20.18 &    0.22 &     $u$&    CSP \\
57444.31 &  11.35 & 20.35 &    0.23 &     $u$&    CSP \\
57446.33 &  13.36 & 20.39 &    0.18 &     $u$&    CSP \\
57450.31 &  17.34 & 20.36 &    0.16 &     $u$&    CSP \\
57451.38 &  18.41 & 20.98 &    0.23 &     $u$&    CSP \\
57464.28 &  31.31 & 20.99 &    0.28 &     $u$&    CSP \\
57465.26 &  32.30 & 20.78 &    0.20 &     $u$&    CSP \\
57474.13 &  41.16 & 21.28 &    0.49 &     $u$&    CSP \\
57427.24 &  $-$5.73 & 17.52 &    0.02 &     $B$&    CSP \\
57429.39 &  $-$3.57 & 17.51 &    0.03 &     $B$&    CSP \\
57430.38 &  $-$2.58 & 17.49 &    0.03 &     $B$&    CSP \\
57434.27 &   1.30 & 17.51 &    0.03 &     $B$&    CSP \\
57435.20 &   2.24 & 17.52 &    0.02 &     $B$&    CSP \\
57437.26 &   4.30 & 17.56 &    0.03 &     $B$&    CSP \\
57439.22 &   6.25 & 17.68 &    0.03 &     $B$&    CSP \\
57441.21 &   8.24 & 17.79 &    0.05 &     $B$&    CSP \\
57442.29 &   9.32 & 17.88 &    0.04 &     $B$&    CSP \\
57443.24 &  10.27 & 17.98 &    0.04 &     $B$&    CSP \\
57444.33 &  11.37 & 18.05 &    0.04 &     $B$&    CSP \\
57446.33 &  13.37 & 18.26 &    0.04 &     $B$&    CSP \\
57450.33 &  17.36 & 18.51 &    0.05 &     $B$&    CSP \\
57451.39 &  18.42 & 18.54 &    0.06 &     $B$&    CSP \\
57456.33 &  23.36 & 18.87 &    0.03 &     $B$&    CSP \\
57456.39 &  23.42 & 18.90 &    0.07 &     $B$&    CSP \\
57457.30 &  24.34 & 18.84 &    0.03 &     $B$&    CSP \\
57458.29 &  25.32 & 18.97 &    0.03 &     $B$&    CSP \\
57464.25 &  31.29 & 19.52 &    0.09 &     $B$&    CSP \\
57465.25 &  32.28 & 19.34 &    0.09 &     $B$&    CSP \\
57473.18 &  40.21 & 20.19 &    0.25 &     $B$&    CSP \\
57478.19 &  45.23 & 20.60 &    0.25 &     $B$&    CSP \\
57498.13 &  65.17 & 20.56 &    0.33 &     $B$&    CSP \\
57427.24 &  $-$5.73 & 16.67 &    0.02 &     $g$&    CSP \\
57428.39 &  $-$4.58 & 16.68 &    0.02 &     $g$&    CSP \\
57429.38 &  $-$3.58 & 16.70 &    0.02 &     $g$&    CSP \\
57430.37 &  $-$2.60 & 16.68 &    0.02 &     $g$&    CSP \\
57434.27 &   1.30 & 16.69 &    0.02 &     $g$&    CSP \\
57435.21 &   2.24 & 16.71 &    0.02 &     $g$&    CSP \\
57437.27 &   4.31 & 16.76 &    0.02 &     $g$&    CSP \\
57439.22 &   6.26 & 16.86 &    0.02 &     $g$&    CSP \\
57441.24 &   8.27 & 16.98 &    0.02 &     $g$&    CSP \\
57442.26 &   9.29 & 17.00 &    0.02 &     $g$&    CSP \\
57443.27 &  10.30 & 17.13 &    0.02 &     $g$&    CSP \\
57444.30 &  11.34 & 17.19 &    0.03 &     $g$&    CSP \\
57446.31 &  13.34 & 17.27 &    0.03 &     $g$&    CSP \\
57450.30 &  17.33 & 17.48 &    0.03 &     $g$&    CSP \\
57451.37 &  18.40 & 17.47 &    0.03 &     $g$&    CSP \\
57456.33 &  23.36 & 17.72 &    0.03 &     $g$&    CSP \\
57456.40 &  23.43 & 17.65 &    0.04 &     $g$&    CSP \\
57457.30 &  24.34 & 17.82 &    0.04 &     $g$&    CSP \\
57458.29 &  25.32 & 17.86 &    0.03 &     $g$&    CSP \\
57464.30 &  31.34 & 18.25 &    0.05 &     $g$&    CSP \\
57465.28 &  32.31 & 18.26 &    0.05 &     $g$&    CSP \\
57473.20 &  40.24 & 18.98 &    0.11 &     $g$&    CSP \\
57474.14 &  41.18 & 18.87 &    0.10 &     $g$&    CSP \\
57478.16 &  45.19 & 19.15 &    0.08 &     $g$&    CSP \\
57479.17 &  46.20 & 19.06 &    0.08 &     $g$&    CSP \\
57481.20 &  48.24 & 19.35 &    0.08 &     $g$&    CSP \\
57498.17 &  65.20 & 19.80 &    0.17 &     $g$&    CSP \\
57499.13 &  66.16 & 19.93 &    0.29 &     $g$&    CSP \\
57493.37 &  60.40 & 19.64 &    0.11 &     $g$& PESSTO \\
57427.22 &  $-$5.74 & 15.80 &    0.02 &     $V$&    CSP \\
57428.39 &  $-$4.58 & 15.79 &    0.02 &     $V$&    CSP \\
57429.39 &  $-$3.58 & 15.76 &    0.02 &     $V$&    CSP \\
57430.38 &  $-$2.59 & 15.76 &    0.02 &     $V$&    CSP \\
57434.26 &   1.29 & 15.78 &    0.02 &     $V$&    CSP \\
57435.19 &   2.23 & 15.79 &    0.02 &     $V$&    CSP \\
57437.26 &   4.29 & 15.88 &    0.02 &     $V$&    CSP \\
57446.33 &  13.36 & 16.32 &    0.02 &     $V$&    CSP \\
57450.32 &  17.36 & 16.42 &    0.02 &     $V$&    CSP \\
57451.39 &  18.42 & 16.50 &    0.03 &     $V$&    CSP \\
57456.32 &  23.36 & 16.61 &    0.03 &     $V$&    CSP \\
57456.37 &  23.41 & 16.72 &    0.03 &     $V$&    CSP \\
57457.30 &  24.33 & 16.60 &    0.03 &     $V$&    CSP \\
57458.29 &  25.32 & 16.65 &    0.02 &     $V$&    CSP \\
57464.28 &  31.31 & 17.09 &    0.04 &     $V$&    CSP \\
57465.25 &  32.29 & 17.17 &    0.04 &     $V$&    CSP \\
57473.18 &  40.21 & 17.91 &    0.07 &     $V$&    CSP \\
57474.12 &  41.15 & 17.89 &    0.07 &     $V$&    CSP \\
57478.18 &  45.22 & 18.03 &    0.06 &     $V$&    CSP \\
57479.17 &  46.21 & 17.97 &    0.05 &     $V$&    CSP \\
57481.19 &  48.23 & 18.04 &    0.04 &     $V$&    CSP \\
57482.24 &  49.28 & 17.82 &    0.05 &     $V$&    CSP \\
57493.10 &  60.14 & 18.59 &    0.11 &     $V$&    CSP \\
57494.24 &  61.27 & 18.81 &    0.16 &     $V$&    CSP \\
57498.14 &  65.17 & 18.54 &    0.11 &     $V$&    CSP \\
57499.10 &  66.14 & 19.07 &    0.19 &     $V$&    CSP \\
57482.37 &  49.41 & 17.81 &    0.03 &     $V$& PESSTO \\
57488.28 &  55.31 & 18.74 &    0.03 &     $V$& PESSTO \\
57493.37 &  60.40 & 18.73 &    0.03 &     $V$& PESSTO \\
57427.25 &  $-$5.72 & 14.63 &    0.01 &     $r$&    CSP \\
57428.39 &  $-$4.58 & 14.64 &    0.02 &     $r$&    CSP \\
57429.38 &  $-$3.59 & 14.61 &    0.01 &     $r$&    CSP \\
57430.36 &  $-$2.60 & 14.59 &    0.02 &     $r$&    CSP \\
57434.27 &   1.31 & 14.62 &    0.01 &     $r$&    CSP \\
57435.21 &   2.25 & 14.60 &    0.02 &     $r$&    CSP \\
57437.28 &   4.31 & 14.71 &    0.01 &     $r$&    CSP \\
57439.22 &   6.26 & 14.77 &    0.01 &     $r$&    CSP \\
57441.25 &   8.28 & 14.87 &    0.02 &     $r$&    CSP \\
57442.25 &   9.29 & 14.89 &    0.01 &     $r$&    CSP \\
57443.27 &  10.30 & 14.93 &    0.02 &     $r$&    CSP \\
57444.30 &  11.34 & 14.99 &    0.02 &     $r$&    CSP \\
57446.30 &  13.34 & 15.07 &    0.02 &     $r$&    CSP \\
57450.29 &  17.33 & 15.16 &    0.02 &     $r$&    CSP \\
57451.36 &  18.40 & 15.21 &    0.02 &     $r$&    CSP \\
57456.33 &  23.37 & 15.42 &    0.03 &     $r$&    CSP \\
57456.39 &  23.42 & 15.39 &    0.02 &     $r$&    CSP \\
57457.31 &  24.34 & 15.41 &    0.04 &     $r$&    CSP \\
57458.29 &  25.33 & 15.47 &    0.02 &     $r$&    CSP \\
57464.31 &  31.34 & 15.75 &    0.02 &     $r$&    CSP \\
57465.29 &  32.32 & 15.81 &    0.02 &     $r$&    CSP \\
57473.21 &  40.24 & 16.18 &    0.02 &     $r$&    CSP \\
57474.15 &  41.18 & 16.22 &    0.02 &     $r$&    CSP \\
57478.15 &  45.18 & 16.40 &    0.03 &     $r$&    CSP \\
57481.20 &  48.23 & 16.56 &    0.02 &     $r$&    CSP \\
57493.11 &  60.14 & 16.92 &    0.04 &     $r$&    CSP \\
57494.24 &  61.27 & 16.91 &    0.05 &     $r$&    CSP \\
57498.17 &  65.21 & 17.12 &    0.04 &     $r$&    CSP \\
57499.13 &  66.17 & 17.10 &    0.05 &     $r$&    CSP \\
57508.29 &  75.33 & 17.31 &    0.06 &     $r$&    CSP \\
57488.28 &  55.31 & 16.70 &    0.05 &     $r$& PESSTO \\
57493.37 &  60.40 & 16.77 &    0.07 &     $r$& PESSTO \\
57608.04 & 175.07 & 18.94 &    0.64 &     $r$& PESSTO \\
57642.00 & 209.03 & 19.40 &    0.09 &     $r$& PESSTO \\
57427.25 &  $-$5.71 & 13.33 &    0.02 &     $i$&    CSP \\
57428.39 &  $-$4.58 & 13.32 &    0.02 &     $i$&    CSP \\
57429.38 &  $-$3.59 & 13.27 &    0.01 &     $i$&    CSP \\
57430.36 &  $-$2.61 & 13.27 &    0.02 &     $i$&    CSP \\
57434.28 &   1.31 & 13.24 &    0.02 &     $i$&    CSP \\
57435.22 &   2.25 & 13.27 &    0.01 &     $i$&    CSP \\
57437.28 &   4.31 & 13.33 &    0.02 &     $i$&    CSP \\
57439.23 &   6.26 & 13.41 &    0.02 &     $i$&    CSP \\
57441.26 &   8.29 & 13.46 &    0.02 &     $i$&    CSP \\
57442.25 &   9.28 & 13.50 &    0.01 &     $i$&    CSP \\
57443.27 &  10.30 & 13.54 &    0.01 &     $i$&    CSP \\
57444.30 &  11.33 & 13.60 &    0.01 &     $i$&    CSP \\
57446.30 &  13.33 & 13.66 &    0.02 &     $i$&    CSP \\
57450.29 &  17.33 & 13.83 &    0.02 &     $i$&    CSP \\
57451.36 &  18.39 & 13.87 &    0.02 &     $i$&    CSP \\
57456.33 &  23.37 & 14.06 &    0.03 &     $i$&    CSP \\
57456.33 &  23.37 & 14.10 &    0.10 &     $i$&    CSP \\
57457.31 &  24.34 & 14.10 &    0.03 &     $i$&    CSP \\
57458.30 &  25.33 & 14.18 &    0.02 &     $i$&    CSP \\
57464.31 &  31.35 & 14.45 &    0.01 &     $i$&    CSP \\
57465.29 &  32.33 & 14.50 &    0.02 &     $i$&    CSP \\
57473.21 &  40.24 & 14.88 &    0.02 &     $i$&    CSP \\
57474.15 &  41.18 & 14.93 &    0.02 &     $i$&    CSP \\
57478.15 &  45.19 & 15.09 &    0.03 &     $i$&    CSP \\
57479.14 &  46.18 & 15.09 &    0.02 &     $i$&    CSP \\
57481.20 &  48.24 & 15.18 &    0.03 &     $i$&    CSP \\
57482.26 &  49.29 & 15.19 &    0.05 &     $i$&    CSP \\
57493.11 &  60.15 & 15.48 &    0.03 &     $i$&    CSP \\
57494.24 &  61.28 & 15.54 &    0.03 &     $i$&    CSP \\
57498.18 &  65.21 & 15.64 &    0.02 &     $i$&    CSP \\
57499.14 &  66.17 & 15.71 &    0.03 &     $i$&    CSP \\
57508.30 &  75.34 & 15.97 &    0.04 &     $i$&    CSP \\
57441.33 &   8.37 & 13.51 &    0.05 &     $i$& PESSTO \\
57482.37 &  49.41 & 15.14 &    0.04 &     $i$& PESSTO \\
57488.28 &  55.31 & 15.36 &    0.05 &     $i$& PESSTO \\
57493.37 &  60.40 & 15.46 &    0.05 &     $i$& PESSTO \\
57600.08 & 167.11 & 17.62 &    0.09 &     $i$& PESSTO \\
57413.29 & $-$19.68 & 15.67&   0.01 &     $i$&    VST \\
57444.20 &  11.23 & 11.33 &    0.09 &     $Y$&    CSP \\
57445.20 &  12.23 & 11.39 &    0.07 &     $Y$&    CSP \\
57445.22 &  12.26 & 11.37 &    0.06 &     $Y$&    CSP \\
57447.32 &  14.36 & 11.48 &    0.07 &     $Y$&    CSP \\
57449.33 &  16.37 & 11.50 &    0.11 &     $Y$&    CSP \\
57474.24 &  41.27 & 12.49 &    0.06 &     $Y$&    CSP \\
57474.26 &  41.29 & 12.52 &    0.07 &     $Y$&    CSP \\
57475.23 &  42.27 & 12.53 &    0.06 &     $Y$&    CSP \\
57475.25 &  42.29 & 12.55 &    0.04 &     $Y$&    CSP \\
57476.22 &  43.25 & 12.58 &    0.04 &     $Y$&    CSP \\
57476.24 &  43.27 & 12.61 &    0.06 &     $Y$&    CSP \\
57477.24 &  44.27 & 12.57 &    0.06 &     $Y$&    CSP \\
57477.25 &  44.29 & 12.64 &    0.06 &     $Y$&    CSP \\
57444.21 &  11.25 & 10.80 &    0.08 &     $J$&    CSP \\
57445.20 &  12.24 & 10.72 &    0.06 &     $J$&    CSP \\
57445.24 &  12.27 & 10.69 &    0.07 &     $J$&    CSP \\
57447.33 &  14.37 & 10.79 &    0.06 &     $J$&    CSP \\
57449.34 &  16.38 & 10.79 &    0.06 &     $J$&    CSP \\
57474.25 &  41.28 & 11.71 &    0.08 &     $J$&    CSP \\
57474.27 &  41.30 & 11.70 &    0.06 &     $J$&    CSP \\
57475.24 &  42.28 & 11.73 &    0.07 &     $J$&    CSP \\
57475.26 &  42.30 & 11.70 &    0.28 &     $J$&    CSP \\
57476.23 &  43.26 & 11.76 &    0.04 &     $J$&    CSP \\
57476.25 &  43.28 & 11.85 &    0.11 &     $J$&    CSP \\
57477.25 &  44.28 & 11.64 &    0.03 &     $J$&    CSP \\
57477.26 &  44.30 & 11.82 &    0.05 &     $J$&    CSP \\
57433.28 &   0.31 & 10.49 &    0.02 &     $J$& PESSTO \\
57436.33 &   3.36 & 10.58 &    0.02 &     $J$& PESSTO \\
57442.35 &   9.39 & 10.67 &    0.02 &     $J$& PESSTO \\
57455.41 &  22.44 & 10.97 &    0.01 &     $J$& PESSTO \\
57491.32 &  58.35 & 12.03 &    0.02 &     $J$& PESSTO \\
57608.99 & 176.03 & 14.88 &    0.03 &     $J$& PESSTO \\
57640.99 & 208.03 & 15.61 &    0.03 &     $J$& PESSTO \\
57445.23 &  12.26 &  9.94 &    0.05 &     $H$&    CSP \\
57447.33 &  14.36 &  9.92 &    0.07 &     $H$&    CSP \\
57449.34 &  16.37 &  9.89 &    0.06 &     $H$&    CSP \\
57474.25 &  41.28 & 10.66 &    0.10 &     $H$&    CSP \\
57474.27 &  41.30 & 10.80 &    0.05 &     $H$&    CSP \\
57475.24 &  42.27 & 10.73 &    0.10 &     $H$&    CSP \\
57476.23 &  43.26 & 10.90 &    0.03 &     $H$&    CSP \\
57476.25 &  43.28 & 10.90 &    0.06 &     $H$&    CSP \\
57477.24 &  44.28 & 10.88 &    0.08 &     $H$&    CSP \\
57477.26 &  44.29 & 10.87 &    0.08 &     $H$&    CSP \\
57436.33 &   3.36 &  9.88 &    0.09 &     $H$& PESSTO \\
57455.41 &  22.44 & 10.17 &    0.09 &     $H$& PESSTO \\
57491.32 &  58.35 & 11.08 &    0.13 &     $H$& PESSTO \\
57608.99 & 176.02 & 13.60 &    0.06 &     $H$& PESSTO \\
57640.99 & 208.02 & 14.14 &    0.05 &     $H$& PESSTO \\
57433.39 &   0.42 &  9.31 &    0.01 &     $K$& PESSTO \\
57436.33 &   3.36 &  9.30 &    0.02 &     $K$& PESSTO \\
57442.36 &   9.40 &  9.41 &    0.02 &     $K$& PESSTO \\
57455.41 &  22.45 &  9.53 &    0.03 &     $K$& PESSTO \\
57491.32 &  58.35 & 10.36 &    0.02 &     $K$& PESSTO \\
57608.98 & 176.01 & 12.04 &    0.03 &     $K$& PESSTO \\
57640.98 & 208.02 & 12.45 &    0.02 &     $K$& PESSTO \\
\end{longtable}

\twocolumn

\begin{table}
\centering
\tiny
\caption{\textit{Swift} photometry of SN~2016adj.\label{tab:swiftphot}}
\begin{tabular}{llllll}
\hline
\hline 
JD-2457000 & 
phase &
mag & 
error mag &
filter \\
(days) &
(days) &
(mag) &
(mag) \\
\hline
427.27 & $-$6.20 & $>$18.726  &  $\cdots$  & $uvm2$\\  
427.67 & $-$5.80 & $>$20.050  &  $\cdots$  & $uvm2$\\  
428.00 & $-$5.47 & $>$18.726  &  $\cdots$  & $uvm2$\\      
427.26 & $-$6.21 & $>$18.522  &  $\cdots$  & $uvw1$\\  
427.73 & $-$5.74 & 18.749     &  0.355   & $uvw1$\\  
427.99 & $-$5.48 & $>$18.527  &  $\cdots$  & $uvw1$\\  
\hline 
\end{tabular}
\end{table}


\section{On the local environment of SN~2016adj in the dust lane of Centaurus~A}
\label{sec:appendixMUSE}

The position of SN~2016adj was previously observed with the ESO-Paranal VLT equipped with the MUSE (Multi Unit Spectroscopic Explorer; \citealt{bacon2014}) instrument on 2015 January 25 (PI J. Walcher, ESO Program ID 094.B-0298). 
A single pointing of  295 seconds  covering the coordinates of SN~2016adj was identified within the ESO archive providing a field-of-view (FOV) of 1 arcmin.  We now summarize the data reduction and analysis techniques used to estimate  values related to the gas phase $E(B-V)_{host}^{gas}$ color excess, the metallicity, and star formation rate. 
 
As a first step the MUSE data cube was downloaded from the ESO archive and basic data reductions were performed following the procedures presented by \citet{kruehler2017}.
To analyze the data cube we followed the method described by \citet[][see also  \citealt{galbany2018} and references therein]{lyman2018} making extensive use of the  \texttt{PYTHON} package \texttt{IFUANAL}\footnote{\url{https://github.com/Lyalpha/ifuanal}}. \texttt{IFUANAL} performs spaxel binning by growing bins from initial peaks in a H$\alpha$ map based on some constraints on pixel values and weighted distances to their respective peaks.
Once all individual spectra in the data cube were combined in the bins provided by the segmentation routine, the associated stellar continuum and emission lines were fit   with the stellar population package \texttt{STARLIGHT} \citep{2005MNRAS.358..363C}, revealing a best-fit stellar population model (SSP). The stellar population model was then removed from each spaxel spectrum, resulting in a pure gas-emission spectrum.

Armed with the pure gas phase spectra for each spaxel region,  a Gaussian function was fit to the strongest emission lines, including H$\alpha$, H$\beta$, [\ion{O}{iii}] $\lambda$5007, [\ion{N}{ii}] $\lambda$6583, and the [\ion{S}{ii}] $\lambda\lambda$6719,6731 doublet.

The Balmer decrement was measured and assuming case~B recombination with the canonical $H\alpha$-to-$H\beta$ ratio of 2.86 \citep{osterbrock2006},   $E(B-V)_{host}^{gas}$ was inferred.
The extinction-corrected line flux measurements were then used to estimate both the star-formation rate density ($\Sigma$SFR) and the gas-phase metallicity (12 + log(O/H)) using published O3N2 calibrations \citep[e.g.,][]{marino2013,dopita2016}. 
Finally, the stellar population model is also used to normalize the extracted spectra  to enable an H$\alpha$ equivalent width  measurement, which  provides an assessment of the prevalence of young stellar populations relative to the older stellar content.

\begin{figure*}
\centering
\resizebox{\hsize}{!}
{\includegraphics{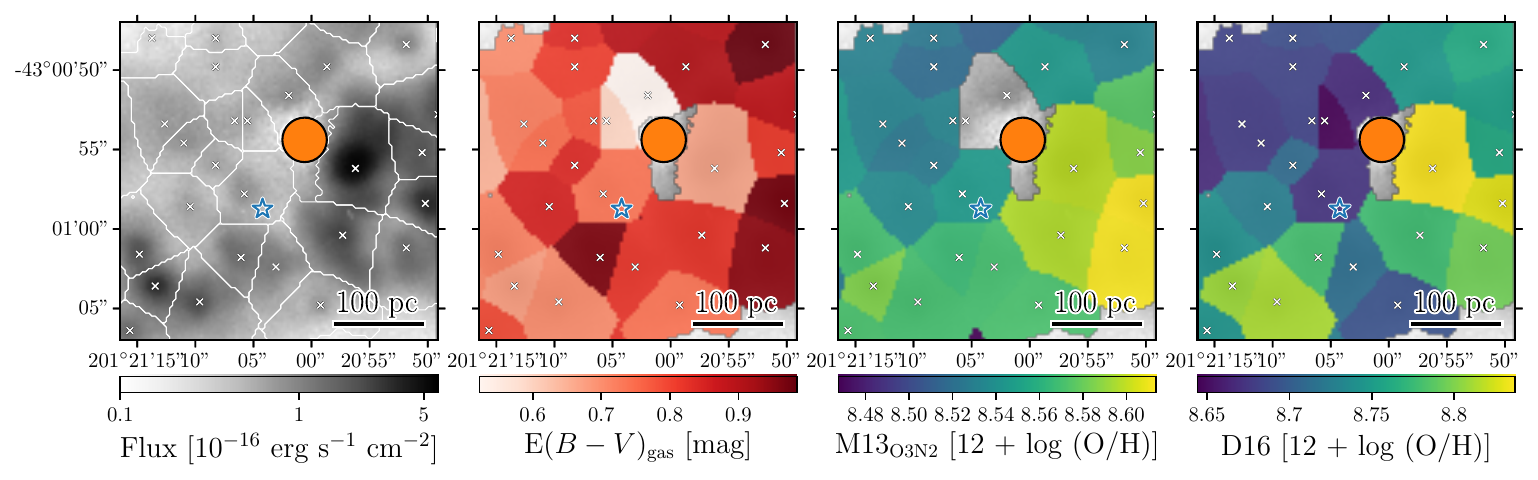}}
\caption{ Spaxel binned MUSE data cube revealing  (from left to right)  the $H\alpha$ flux (panel 1), the Balmer decrement-based $E(B-V)_{host}^{gas}$ color excess map (panel 2), and the  O3N2 gas phase metallicity maps  computed using the   \citet[][panel 3]{marino2013}  and \citet[][panel 4]{dopita2016} calibrations.
The position of SN~2016adj is indicated in each panel by a star, crosses reveal \ion{H}{ii}-region peaks, and the bright foreground star in close proximity of the position of SN~2016adj  is masked  by a orange circle. 
}
\label{fig:museplot1}
\end{figure*}

The results of this analysis are summarized in the multi-panel plot shown in Fig.~\ref{fig:museplot1}.
Contained within the various panels are  the spatial segmentation regions  overlaid with the analysis mappings including:   the observed $H\alpha$ flux  (panel 1),   the $E(B-V)^{gas}_{host}$ map (panel 2), and  the O3N2 gas-phase metallicity maps inferred from the \citet{marino2013} (panel 3) and \citet{dopita2016} (panel 4) calibrations.
We note that in each panel the location of the SN is marked  with a star symbol, while the center of the  \ion{H}{ii}-region's peak emission   within  each  segmented region is marked with a  cross. Also, the   location of the bright foreground star  in close proximity of the location of SN~2016adj   is masked  with a filled orange circle, and this region was excluded from the entire analysis. 

The extinction map suggests  for the bin containing the location of  SN~2016adj a color excess $E(B-V)_{host}^{gas} \approx 0.74\pm0.36$ mag.
Moreover, the O3N2 gas phase metallicity of this bin is measured to be (12 + log(O/H)) $= 8.55\pm0.18$ dex (panel 3) and $8.67\pm0.18$ dex (panel 4). These estimates are consistent with the solar value of (12 + log(O/H)) $= 8.7\pm0.2$ dex \citep{asplund2009}.
We also obtained a H$\alpha$ equivalent width of 54$\pm$2 \AA, and a $\Sigma$SFR = 3.47$\pm$0.13$\times$10$^{-4}$ M$_\odot$ yr$^{-1}$ kpc$^{-2}$ using the \citet{1998ApJ...498..541K} relation. 
This $\Sigma$SFR lays in the bottom 10\% of the 66 stripped-envelope SN environments in the PISCO compilation \citep{galbany2018}.

\begin{figure*}
\centering
\resizebox{\hsize}{!}
{\includegraphics{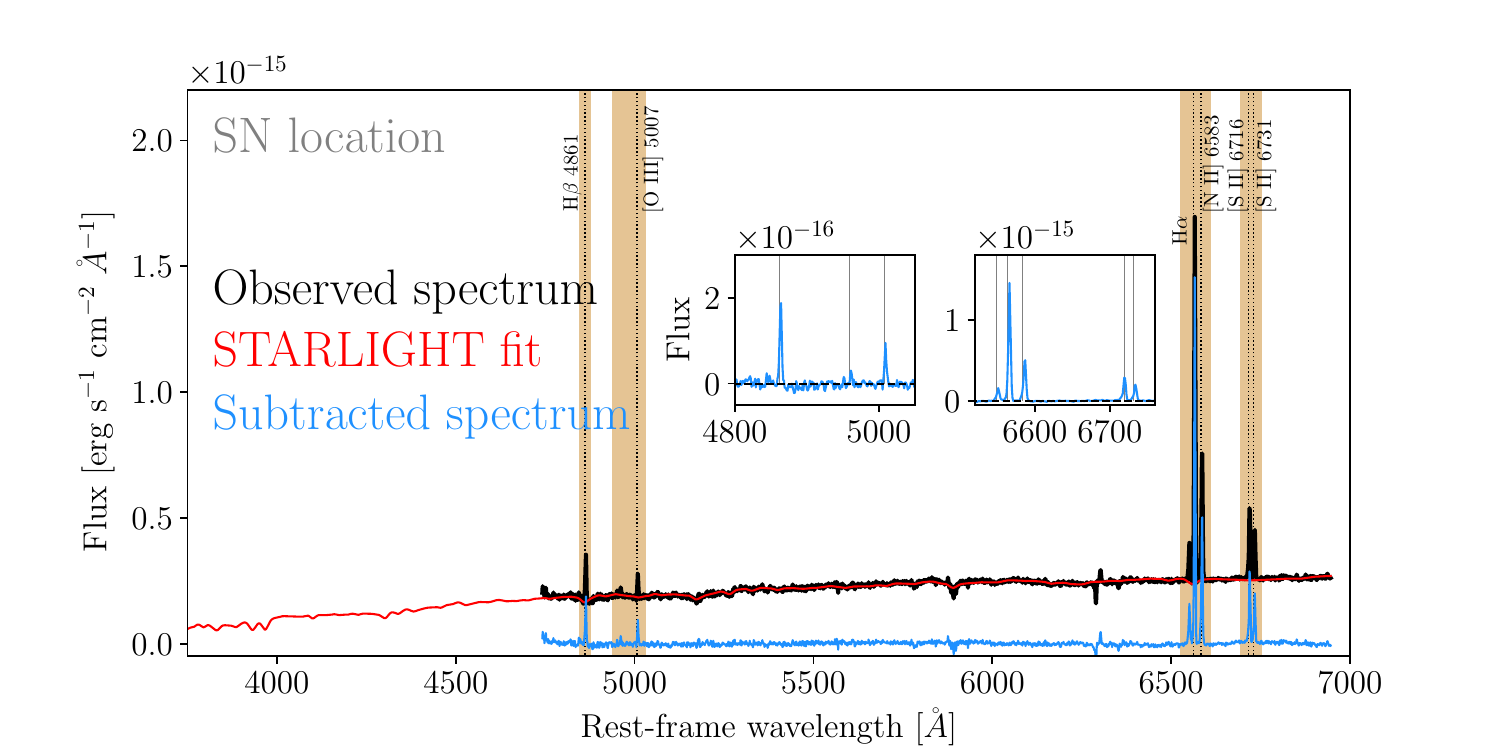}}
\caption{The 1-D spectrum extracted from the MUSE observations at the location of SN~2016adj with prominent nebular emission features from the host labeled. 
The underlying stellar population was estimated using \texttt{STARLIGHT} (red line) and subtracted from the observed spectrum. The resulting gas-phase spectrum is shown in blue. 
A single Gaussian function was then fit to the nebular lines plotted within the insets. Flux ratios of these lines combined with the calibrations of \citet{marino2013} and \citet{dopita2016} indicate gas-phase 03N2 meallicities of (12 + log$_{10}$(O/H)) $= 8.55\pm0.18$ dex and 8.71$\pm$0.18 dex, respectively.}
\label{fig:museplot2}
\end{figure*}

In addition to the  segmented regions, the same analysis was  applied to a spectrum extracted from the data cube using a seeing-diameter aperture centered on the coordinates of SN~2016adj.
The extracted 1-D spectrum is plotted  in Fig.~\ref{fig:museplot2}, together with the best \texttt{STARLIGHT} fit, and the subtraction of the two, revealing  the pure gas-phase spectrum.
Following the techniques above, the Balmer decrement implies   $E(B-V)^{gas}_{host} = 0.92\pm0.37$ mag, and assuming the \citet{marino2013} O3N2 and \citet{dopita2016} calibrations, gas phase metallicities of (12 + log(O/H)) $= 8.55\pm0.18$  dex  and   $8.71\pm0.18$ dex, respectively. 
The estimate of the color excess at the location of SN~2016adj is somewhat higher than that obtained for the segmented region it is contained within, while being consistent with the value implied from the colors of SN~2016adj. 
The metallicity estimates  are consistent (within the uncertainties) with the value inferred for the spaxel region containing the location of SN~2016adj.

Finally, the H$\alpha$ equivalent width measurement at the location is  171$\pm$5 \AA, which is higher than the value measured at the segmented region, and is positioned within the top 10\% of the PISCO sample. The $\Sigma$SFR = (4.94$\pm$0.17) $\times$10$^{-4}$ M$_\odot$ yr$^{-1}$ kpc$^{-2}$,  is also higher than the average of the PISCO sample, and completely consistent with the values inferred above. 

On the other hand, the mere observation of a CC SN in Centaurus~A is consistent with the idea  presented by \citet{Dellavalle2003} that repeated episodes of interaction and/or merging of early-type galaxies with dwarf companions, on time scales of about 1~Gyr, are responsible both for inducing a strong radio activity observed in about 10\% early-type galaxies, and for supplying a consistent number of CC SN progenitors to the stellar population of elliptical galaxies. \citet{Dellavalle2005} estimated that CC events should occur in these systems at a rate of up to 10\% SNe-Ia. However, this is still not the probability of detecting CC-SNe in radio-galaxies because, on average, they are fainter at peak luminosity than SNe-Ia and therefore suffer from an obvious observational bias emphasized by the presence of a massive dust lane as observed, for example, in Centaurus A.

\newpage  

\section{Pre-explosion progenitor analysis}
\label{sec:progenitoranalysis-appendix}
Here we detail our efforts in analyzing  pre-explosion images of Centaurus~A.

\subsection{AO imaging}

The ESO archive  was used to identify deep images of Centaurus~A obtained prior to the discovery of SN~2016adj. Fortuitously, the galaxy was observed on 2008 June 30 with the VLT (+NACO) equipped with  $J$, $H$ and $K_s$ filters.  The data were taken with the S27 camera  (FOV is 28$\times$28 arcsec with 27 mas pixels), and reduced reduced using ESOREX scripts. In brief, the reduction process consisted of dark subtraction, flat fielding, and sky subtraction using off target sky frames. Given that in most frames the site of SN~2016adj fell outside the FOV, we were left with only a single 180~s exposure in $H$ and $K_s$. In the case of $J$ band, where there were 2x180~s images, we co-added these into a single deeper image. We measured the FWHM of the $K_s$-band image to be 2.8~pixels (0\farcs08).

We also retrieved from the ESO archive  adaptive optics (AO) $K_s$-band imaging of SN~2016adj taken with the VLT+NACO on the night of 2016 February 28. The S54 camera was used, which samples a 56$\times$56\arcsec\ field of view (FOV) with 54 mas pixels. The AO correction was obtained using the SN itself as a natural guide star. As SN~2016adj was extremely bright at the time of these observations, short 60~s exposures (DIT=0.4~s, NDIT=150)
off-source sky frames were interleaved with on-source observations in order to accurately remove the sky background. The raw data were reduced using ESOREX, to give a final deep stacked image with exposure time 1920~s.

Unfortunately, NACO exhibited additional noise at the time of these observations that resulted in a pattern of horizontal stripes superimposed on each image. As this pattern is different in each quadrant of the detector, we only concern ourselves with the quadrant containing the position of SN~2016adj. We used a custom script to model the pattern of stripes and remove them  from the stacked image. The final image therefore has  a smaller FOV than the full S54 camera, but contains only minimal residual signs of the pattern of stripes. The FWHM of several point sources in the $K_s$-band image reveal an impressive image quality measurement of $0\farcs1$.

In order to locate the position of SN~2016adj on the pre-explosion images, we aligned the pre-explosion $K_s$-band image to the post-explosion $K_s$-band image. Fourteen point sources common to both frames were identified within a 5\arcsec\ radius of SN~2016adj. Their positions were measured, and a geometric transformation was derived between the two images. The root-mean-square (rms) uncertainty on the transformation was 8 and 6 mas in the x and y direction, respectively ($\sim$0.3 pixels on the pre-explosion frame). This lies close to, but offset from a nearby source located approximately 2 pixels to the N.

\begin{figure}[!htbp]
\centering
\resizebox{\hsize}{!}
{\includegraphics[width=0.5\textwidth]{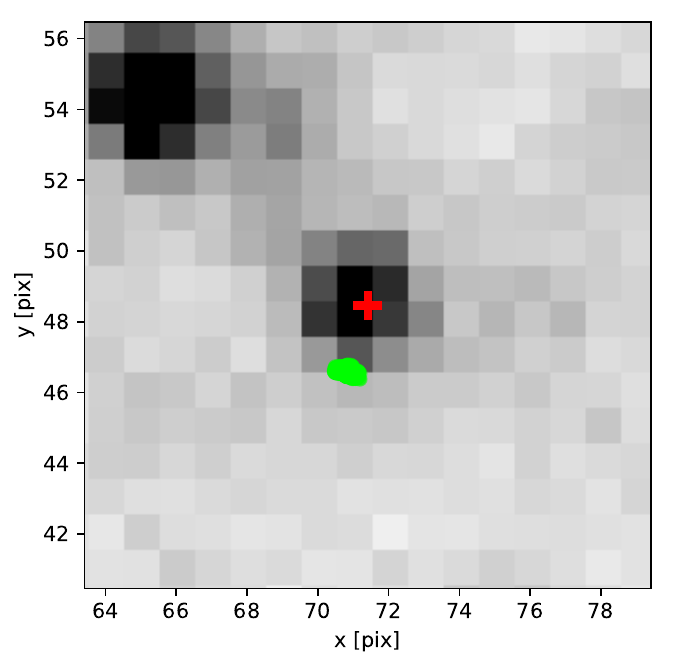}}
   \caption{Pre-explosion NACO $K_s$-band image covering the position of SN~2016adj. The centroid of the nearby source is marked with a red cross, the transformed positions of the SN from our bootstrap analysis are plotted in green.}
   \label{fig:K_astrometry}
\end{figure}

The proximity of the source to the N leads us to consider whether the uncertainty of the transformation might be underestimated. To test this further, a bootstrap calculation was performed. First, 10 of the 14 reference sources used for the image alignment were randomly selected, and then a geometric transformation was calculated for only this subset. We first tested a transformation that allowed for translation, rotation and separate scaling factors in both the x and y directions. We also tested a more flexible geometric transformation that included an additional polynomial term to account for any distortion between the two frames. In both cases, we drew a different random set of 10 sources 1000 times.  The results of this test are presented in Fig.~\ref{fig:K_astrometry}. In all cases, we find the transformed position of SN~2016adj lies offset from the nearby source. The 1$\sigma$ uncertainty in position from our bootstrap analysis is 
x = 0.12, y = 0.07 pixels, for the transformation with translation, rotation and scaling; and 
x = 0.15, y = 0.07 pixels, for the transformation with an additional polynomial term. We note that these uncertainties are smaller than those found originally when deriving the transformation.

From the preceding analysis, we conclude  that the nearby source is likely not the progenitor of SN~2016adj. While the offset is very small (2 pixels, which is 0.9 pc at the distance of Centaurus~A), this is more than 6$\sigma$ from the transformed position of the SN. However, we note that there have been many cases of previous mis-identifications of SN progenitors (even using adaptive optics imaging, e.g. \citealp{Maund15}). The proximity of SN~2016adj to the edge of the NACO FOV in the pre-explosion images is perhaps a  cause for concern, as there may be some additional geometric distortion. However, this should be accounted for in the transformation with an additional polynomial term. In any case, the ultimate test is to obtain new, deep AO images of Centaurus~A to test if the source to the N is still present after the SN has faded.




\subsection{HST imaging}

\begin{table}
\centering
\caption{HST pre-explosion observations covering the site of SN~2016adj.} 
\label{tab:HSTobs}
\begin{tabular}{lllc}
\hline\hline      
Instrument &
Filter &
Data &
Exptime (s)\\
\hline
WFC3/UVIS & F225W & 2010-07-17 & 2520.0 \\
WFC3/UVIS & F336W & 2010-07-17 & 2120.0 \\
WFC3/UVIS & F438W & 2010-07-17 & 1605.0 \\
WFC3/UVIS & F547M & 2010-07-06 & 1250.0 \\
WFC3/UVIS & F814W & 2010-07-06 & 1240.0 \\
WFC3/IR      & F160W & 2010-07-17 & 1797.7 \\
\hline
\end{tabular}
\end{table}

Centaurus~A has been a frequent target of HST, and a number of pre-explosion images available in the HST archive cover the site of SN~2016adj. Some of these data were taken with older generation instruments (i.e., WFPC2) and have relatively low spatial resolution (0.1\arcsec pixels). However, Centaurus~A was also observed with WFC3, and we list these pre-explosion observations in Table~\ref{tab:HSTobs}.

To locate the position of SN~2016adj within the pre-explosion HST images, we used a set of HST (+WFC3) F814W images taken on 2016 February 22.04 UT. Four 40~s exposures were taken in order to avoid saturating on the bright SN, and we co-added these to produce a single image with total exposure time 160~s. This image was aligned to the pipeline-drizzled WFC3 F814W image taken on 2010 July 6.61 UT. This pre-explosion image has a pixel scale of 0\farcs04.

Twenty-five sources common to both images were identified, and using the matched list of pixel coordinates in the pre- and post-explosion frames,  a geometric transformation was determined. The transformation allows for translation, rotation and a magnification factor in both x and y, as well as a polynomial term to account for any residual distortion. The rms uncertainty in the transformation is 0.11, 0.16 pixels in x and y, corresponding to 4 and 6 mas respectively.

\begin{figure*}
\centering
\resizebox{0.7\hsize}{!}
{\includegraphics{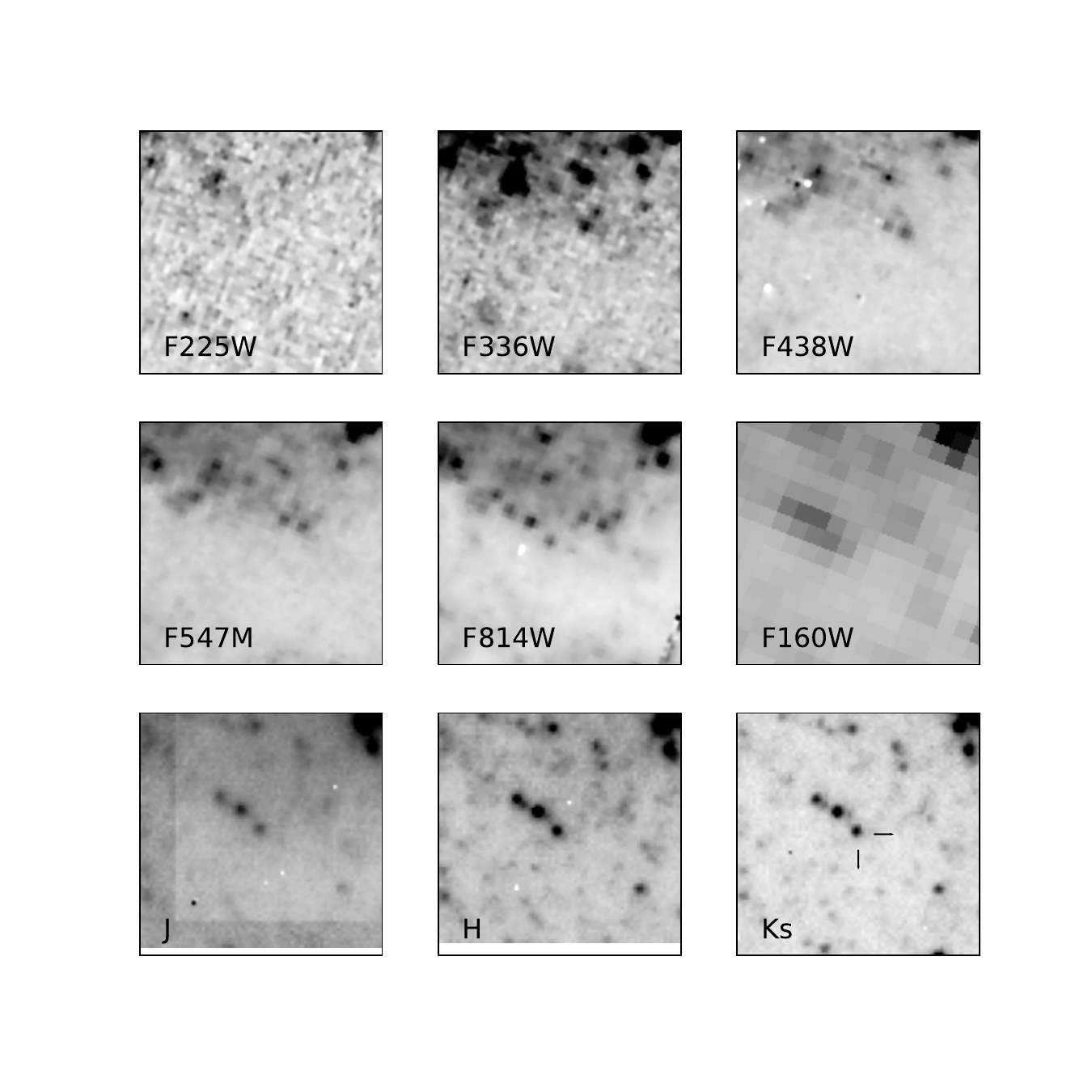}}
   \caption{Cutouts ($2\farcs0 \times 2\farcs0$) of pre-explosion images, centered on the location of SN~2016adj. The top two rows consist of HST (+WFC3) images and the bottom row show VLT (+NACO) images. In the lower right panel the location of SN is indicated with cross hairs. Its position is significantly offset from the nearby bright source that is seen in all filters redder than F814W.}
   \label{fig:cutouts}
\end{figure*}

We see a point source in the F814W image that lies slightly to the N of the transformed position of SN~2016adj, as shown in Fig. \ref{fig:cutouts}. This nearby source is 1.6 pixels distant, or 64 ~mas, which is much greater than the uncertainty in our transformation, and consistent with the offset seen in the VLT+NACO data. We conclude that the nearby source is likely unrelated to SN 2016adj, and that the progenitor is not detected in these images.

\subsection{Placing upper limits on the progenitor magnitude}
\label{sec:upperlimits}

While no progenitor candidate is identified in either HST or VLT pre-explosion images, we can still use these data to place limits on the magnitude of the progenitor. Using {\sc dolphot},  sources were injected into the images close to the transformed position of SN~2016adj. Fifty Monte Carlo trials were conducted for a range of injected source magnitudes, varying the pixel coordinates slightly each time. We define the limiting magnitude as the magnitude at which 50\% of these sources are recovered within 0.2~mag of their injected magnitude. To determine this, a sigmoid function is fit to the recovered fraction as a function of magnitude. Resulting, robust limiting magnitudes are computed for the WFC3 $F814W$ and F547M bands of $>$26.4 and $>$25.9 mag, respectively, while the bluer filters do not provide a robust constraint, nor does the F160W filter which is less sensitive than the VLT+NACO data. 

We also used artificial star tests to determine a limiting magnitude for the NACO data.
 Here, we were presented with a number of challenges. The extremely small FOV of NACO meant that there were no bright, isolated sources within the FOV that could be used to determine a zeropoint (the nearby star at $K_s = 9.2$ mag is too bright). Moreover, the AO images have a spatially varying PSF across the FOV. So, we built a model of the PSF from nearby sources, and injected progressively fainter sources until they could no longer be comfortably recovered by eye. 
 We determined the zeropoint for each image using standard stars observed on the previous or subsequent nights. The ESO weather archive suggests that all of these nights were photometric, and so we averaged each measurement of the zeropoint for each filter, rejecting points that lie $>3\sigma$ from the instrumental zeropoint as given by the NACO webpages. We were left with final limiting magnitudes of $J>21.6$, $H>20.9$ and $K_s>21.1$~mag.

\begin{figure*}
\centering
\resizebox{0.7\hsize}{!}
{\includegraphics{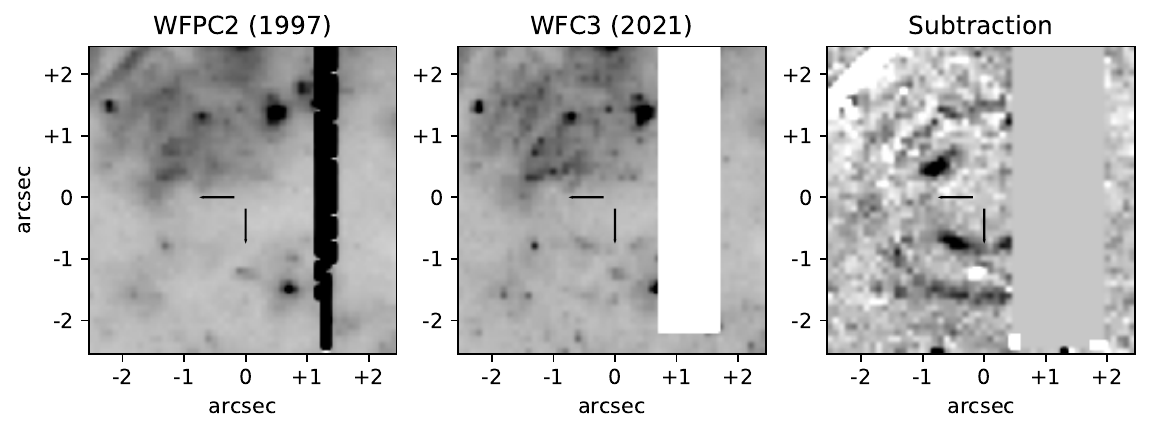}}
   \caption{Subtraction between pre-explosion WFPC2+F555W filter image, and late time WFC3+F555W image, covering the site of SN~2016adj. The WFC3 image has been transformed to match the pixel scale of the WFPC2 data. A diffraction spike from the nearby bright star in the field is visible in the WFPC2 image, and is masked in the WFC3 image. No disappearing progenitor can be seen in the subtracted image, although signatures of light echo emission are clearly visible.}
   \label{fig:f555W_sub}
\end{figure*}

\subsection{Luminosity limits on the progenitor of SN~2016adj}

To summarise the preceding sections, we see a source {\it close to, but significantly offset from} the position of SN~2016adj in both HST and VLT+NACO pre-explosion images. Independent analyses using either the HST or NACO data each suggest that this is not the progenitor at greater than 6$\sigma$ confidence.

We are hence left with a set of upper limits on the absolute magnitude of the progenitor of $F547M>25.9$, $F814W>26.4$, $J>21.6$, $H>20.9$, and $K_s>21.1$~mag. Given our adopted distance and reddening towards SN~2016adj (Sect. \ref{sec:distance}, \ref{sec:reddening}), the absolute magnitude of the progenitor of SN~2016adj must be $F547M>-8.4$, $F814W>-5.7$, $J>-8.2$, $H>-8.2$, $Ks>-7.5$~mag.

Even for our deepest optical limit ($F814W>-5.7$ mag), only a handful of the brightest WR stars would be detected \citep{Eldridge13,Massey02}. If the progenitor of SN~2016adj were a lower mass (and lower luminosity) star that had been stripped in a binary, then the optical limits are even less constraining. 

While WR stars emit most of their luminosity at UV wavelengths, if they are surrounded by significant CSM it is possible that some of this UV radiation is reprocessed to the infrared.
 To establish whether the NIR limits are useful, we compare to the catalog of MW WR stars in \cite{Rate20}. The WC stars in this catalog have $K_s$ absolute magnitudes ranging from $\sim-3.5$ (for early WC stars) to $\sim-7$ for late type WC stars. Only the {\it brightest} WC star in their sample had a $K_s$-band magnitude sufficient that we would see it in our NACO data. As for the optical, we hence conclude that we cannot use these data to place meaningful limits on the progenitor.

\subsection{Searching for a progenitor in F555W through  host-galaxy subtraction}

While no obvious candidate is identified in the pre-explosion F555W images, the availability of late-time HST data after the SN has faded allowed us to make a final test. We took the 2000s F555W pre-explosion image from 1997 August 01.88 UT, and aligned  it to the late time HST (+ WFC3) UVIS image taken on 2021 July 28.6 UT. After re-sampling the latter to match the pixel scale and orientation of the former,  the {\sc hotpants} package was used to perform difference imaging between the two images. The results of this subtraction can be seen in Fig.~\ref{fig:f555W_sub}; while the light echo emission  around SN~2016adj is clearly visible in the subtracted late phase image \citep[see also][]{Stritzinger2022},  there is no sign of a change in flux at the position of SN~2016adj.

\clearpage 
\section{CO modeling details}

\begin{figure}
\centering
\resizebox{\hsize}{!}
{\includegraphics{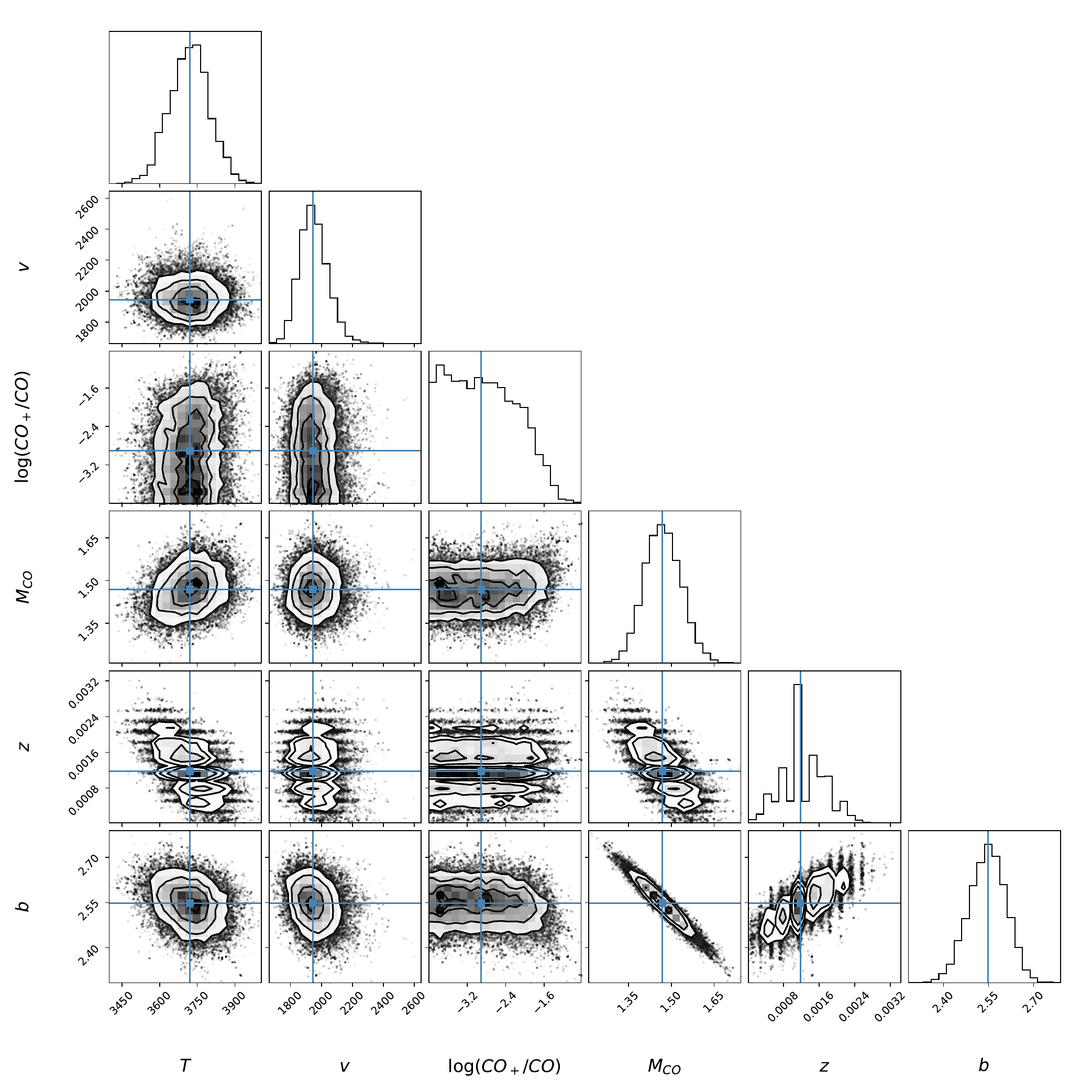}}
\caption{Posterior distribution of the MCMC  parameters used to determine the best fit.}
\label{fig:cornerplot}
\end{figure}

\FloatBarrier
 Fig.~\ref{fig:cornerplot} displays the corner plot associated with our MCMC calculations that led us to determine the best CO model fit discussed in Sect.~\ref{sec:COmodeling}. The corner plot displays the PDF (posterior probability functions) computed for the six model-fit parameters. 
\FloatBarrier

\end{appendix}

\end{document}